\documentclass{article}
\usepackage[margin=2cm]{geometry}
\usepackage{graphicx}
\usepackage{siunitx}
\usepackage{multirow}
\usepackage[section]{placeins}

\begin{document}

\title{A comprehensive design and performance analysis of LEO satellite quantum communication}
\author{J-P Bourgoin$^1$, E Meyer-Scott$^1$, B~L Higgins$^1$, B Helou$^1$, \\C Erven$^1$, H {H\"ubel}$^1$, B Kumar$^2$, D Hudson$^2$, I D'Souza$^2$, \\R Girard$^3$, R Laflamme$^1$ and T Jennewein$^1$
\\{\small$^1$Institute for Quantum Computing, University of Waterloo, Waterloo, ON, N2L 3G1, Canada}
\\{\small$^2$COM DEV Canada, Cambridge, ON, N1R 7H6, Canada}
\\{\small$^3$Canadian Space Agency, Saint-Hubert, QC, J3Y 8Y9, Canada}}

\maketitle
\begin{abstract}

Optical quantum communication utilizing satellite platforms has the potential to extend the reach of quantum key distribution (QKD) from terrestrial limits of ${\sim}200$~km to global scales. We have developed a thorough numerical simulation using realistic simulated orbits and incorporating the effects of pointing error, diffraction, atmosphere and telescope design, to obtain  estimates of the loss and background noise which a satellite-based system would experience. Combining with quantum optics simulations of sources and detection, we determine the length of secure key for QKD, as well as entanglement visibility and achievable distances for fundamental experiments. We analyze the performance of a low Earth orbit (LEO) satellite for downlink and uplink scenarios of the quantum optical signals. We argue that the advantages of locating the quantum source on the ground justify a greater scientific interest in an uplink as compared to a downlink. An uplink with a ground transmitter of at least 25~cm diameter and a 30~cm receiver telescope on the satellite could be used to successfully perform QKD multiple times per week with either an entangled photon source or with a weak coherent pulse source, as well as perform long-distance Bell tests and quantum teleportation. Our model helps to resolve important design considerations such as operating wavelength, type and specifications of sources and detectors, telescope designs, specific orbits and ground station locations, in view of anticipated overall system performance.

\end{abstract}

\section{Introduction} \label{sec.Intro}

Popular classical encryption methodologies rely on assumptions that certain mathematical operations are, and will forever remain, difficult to invert. In contrast, quantum cryptography techniques such as quantum key distribution (QKD) exploit the fundamentally quantum mechanical nature of reality to offer long-term security that is not based on computational assumptions, thereby preserving the security of communications regardless of any advancement of computational power~\cite{BB84,SBCDLP09}. Although QKD has reached the level of maturity sufficient for commercial implementation~\cite{idQuantique,MagiQ}, current transmission distances remain limited to the order of 200~km in fibre~\cite{TNZHHTY07, SWVTGZGTT09, ZHHJIJ12} and in free space~\cite{ UTSWSLBJPTOFMRSBWZ07}. Quantum repeaters have been suggested to overcome this limitation~\cite{BDCZ98}, but for such devices to be implemented efficiently, quantum memories are vital. However, this technology is not yet ready for application to global distances~\cite{SSRG11}. The use of a free-space link to an orbiting satellite for QKD can overcome this limitation~\cite{BHLMNP00, GH00, NHMPW02, RTGK02, UTSWSLBJPTOFMRSBWZ07, MHZG08, TBDNV11, EAWANS11, NMRFHW12, WYLZSHWYTZLLHHRPYJCPP12},  and several implementations are being pursued~\cite{NHMPW02, UJKPCMAVSFABBBCCGLHLLMPRRRSSTTTOPVWWWZZ08, MYMBHJ11, X11, TTTKA11, HBGMYJ12} with projected launch dates as early as 2017~\cite{X11}. Continued advances in technology and the growing concern for more reliable data transmission security protocols have contributed to making satellite QKD an attractive and feasible proposition.

Using QKD, two distant parties, Alice and Bob, can generate a cryptographic key, random and secret but known to both, through the use of a quantum channel (which might be accessible to an eavesdropper, Eve) and an authenticated public classical channel. Here we focus on QKD schemes which transmit quantum information encoded in photon polarization, which has successfully been demonstrated in long-distance free-space links. There are two main types of QKD schemes: prepare-and-measure schemes, such as the seminal BB84 protocol~\cite{BB84}---e.g.\ using a weak coherent pulse (WCP) source with a decoy state method~\cite{Hwa03}---and entanglement-based schemes, such as the BBM92 protocol~\cite{BBM92}, where the trustworthiness of the source can be determined by assessing the strength of measured correlations via a Bell test~\cite{Horodecki2009}. In either protocol, Eve is unable to gain any information about the key without leaving telltale signs that she has been listening, as a consequence of the no-cloning theorem~\cite{WZ82}.

Accordingly, there are two approaches for using satellites to establish long-distance QKD links. One approach utilizes the verifiability of entanglement correlations in schemes such as BBM92 with an entangled source, or entangling Bell state measurement~\cite{LCQ12}, on the satellite. These schemes are challenging as they require a satellite payload with two telescopes capable of establishing two independent quantum links simultaneously. This imposes an additional restriction on the attainable distance between the receivers on Earth, as low elevation angles would experience too much atmospheric loss to be useful.

We focus on the other, simpler, approach in which the satellite is used as a trusted node. The satellite establishes a key first with one ground station and then orbits over a second ground station to establish a second key. The first key is then encrypted with the second key, and sent to a ground station. This allows the ground station to decrypt and obtain the second key. In this way, the two ground stations then share a secret key with which they can securely communicate. Because this approach only requires a single quantum link at a time, it is technically simpler, more cost-effective, and therefore faster to deploy than the untrusted satellite schemes. The drawback of this approach is that the satellite also possesses the secure key with which the ground stations communicate, and thus we must mandate (or assume) that an eavesdropper is incapable of interrogating the key storage on the satellite.  

Here we present results obtained from a detailed link analysis model constructed as a part of feasibility studies for the Canadian Quantum Encryption and Science Satellite (QEYSSat) mission concept. Encompassing downlink and uplink scenarios for each of WCP (BB84) and entangled photon (BBM92) sources, our analysis includes full loss and background calculations for one year of satellite passes determined according to precise orbit estimations. Quantum bit error ratios (QBERs) and sifted key rates are calculated and used to determine the length of secure keys, incorporating finite size effects, obtainable from each pass. We consider operation in a low Earth orbit (LEO, altitude up to the order of 1000~km) which has the advantages of having less optical loss and being less costly to attain and successfully operate than higher orbits, making it the most favourable scenario in the near term. To reduce background noise, the QKD link will operate at nighttime. We thus consider a circular sun-synchronous noon/midnight orbit, crossing the equator at approximately noon or midnight local time, thereby optimizing the amount of time spent in Earth's shadow. From these we assess the feasibility of a trusted-node satellite platform utilizing current technologies to achieve global-scale QKD. 

In addition, we also investigate the capability of this satellite platform to perform long-distance Bell tests~\cite{Bell64} and quantum teleportation experiments~\cite{PhysRevLett.70.1895}. Bell tests, which are fundamental tests of the underlying nature of reality, have to date been performed up to 144~km~\cite{SUKRMHRFLJZ10}. A quantum satellite platform has the potential to greatly extend this distance, thereby testing the validity of quantum mechanics in a new regime~\cite{RJADHKKLMMMMMSSST12}. Quantum teleportation, a precursor to entanglement swapping, allows one to transfer a quantum state from one location to another given a source of shared entanglement. This type of experiment has been performed in full (with remote preparation and feed-forward) up to 550~m~\cite{LHBZG07}, and simplified (with local preparation and remote feed-forward) up to 143~km~\cite{YRLCYWLLZJCXPJHYWCPP12, MHSWKNWMKAMJUZ12}.

\section{Detailed simulation of photonic quantum communication}\label{QuantOpt}

\begin{figure}[tb]
\centering
\includegraphics[width=0.8\linewidth]{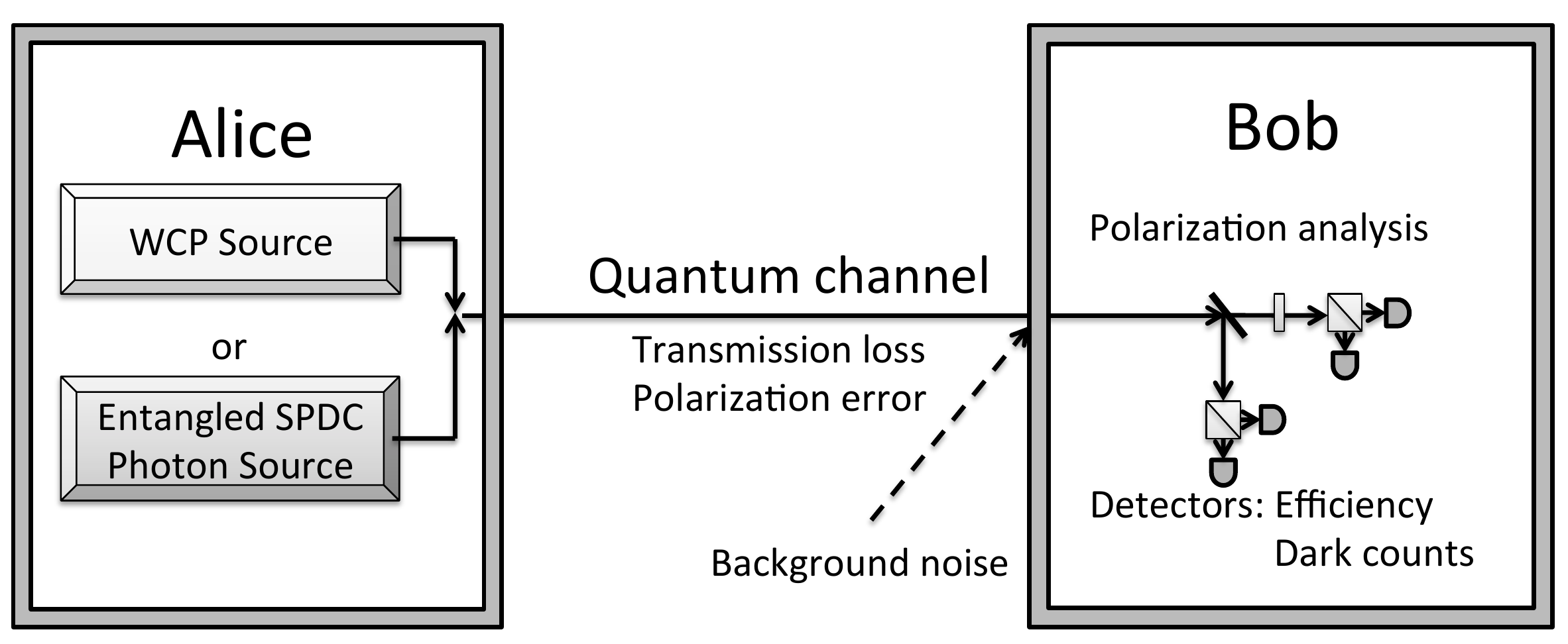}
\caption[Quantum optics simulation]{Devices considered in the quantum optics simulation. Each source is modelled separately using the appropriate quantum unitary operator. Optical losses are added in the quantum channel, accounting for atmospheric losses, and finally the polarization state of the photons is analysed and the photon detection probability evaluated. For the WCP source, Alice chooses the polarization to send for each pulse, whereas for the entangled photon source, Alice measures one photon of the pair to determine its polarization state. Bob's polarization analysis consists of four detectors in a passive polarization analysis apparatus arranged for QKD states.} \label{Simulation_diagram}
\end{figure}

To comprehensively determine the capabilities and limitations of a quantum communication link between the ground and a satellite, we first need to construct a realistic numerical quantum optics simulation capable of predicting photon rates, entanglement visibilities, and secure key rates~\cite{JBW11}. The system we simulate is illustrated in Figure~\ref{Simulation_diagram}, consisting of source, quantum channel, and detection. The key stages for this simulation are:
\begin{enumerate}
\item Photons are created in a mode of a Fock space of finite dimension. We chose dimension 7, allowing us to perform the simulation with 0--6 photons in each mode considered. Doing so we can incorporate the effects of multi-pair emission in down-conversion (for BBM92, Bell tests, and teleportation) and Poissonian statistics in weak coherent pulses (for BB84) to high accuracy.
\item Photons undergo a rotation to simulate imperfectly aligned polarization optics, and appropriate losses are applied to the quantum channel.
\item Photons are measured resulting in count rate statistics, with added noise accounting for background light and detector dark counts. A realistic detector efficiency is used.
\item These statistics are taken in various loss and background count rate regimes to assess optimal and typical expected performance.
\end{enumerate}

Details of the simulations are given in Appendix~\ref{app.QKD.ent} for entanglement-based QKD, and Appendix~\ref{app.QKD.WCP} for WCP QKD. In both cases, the results of the simulation are inserted into an appropriate key-rate formula to predict the number of key bits extractable under given loss and background noise conditions, including fluctuations caused by the finite length of the received key.

In addition, we simulate conditions for Bell tests~\cite{PhysRevLett.81.5039} and quantum teleportation~\cite{Jin2010Experime} between ground and satellite (details in Appendix~\ref{app.Bell}). Both experiments utilize an entangled photon source. The criterion for success for the Bell test is to violate the CHSH inequality~\cite{PhysRevLett.23.880} with $3\sigma$ certainty, i.e.\ three standard deviations above the classical limit. For teleportation, success is defined as violation of the cloning limit for polarization visibility~\cite{BDEFMS98}, again with $3\sigma$ certainty. With our models, we determine the maximum ground-satellite distance over which these experiments are possible for various telescope sizes.

\section{Challenges of quantum transmission in space}

Several effects contribute to deterioration of the optical transmission as it propagates between the satellite and ground-based platforms. The mean optical beam will be broadened by diffraction, systematic pointing error, and atmospheric turbulence, all of which lead to transmission losses. In addition, atmospheric absorption introduces losses dependent on wavelength and atmospheric constituents. We take each of these effects into account in our simulation, as discussed below.

Beam diffraction from the transmitter output depends on wavelength, telescope size and design, and the outgoing beam spatial mode. Although it can be mitigated by using bigger telescopes or shorter photon wavelengths, a compromise between telescope sizes and costs, and subsequent free-space transmission performance is necessary to ensure acceptable outcomes. We find that diffraction is the major loss contributor in a downlink satellite transmission. Diffraction also influences uplink transmission, but we find that mitigation of this by increasing telescope sizes is limited by atmospheric turbulence (see below).

Atmospheric turbulence arises due to local refractive index fluctuations caused by temperature variations. This effect leads to a broadening of the beam as well as beam wander over time (which for our purposes can be modelled as further broadening). Turbulence has no negative effect on polarization-based QKD beyond additional loss~\cite{S67}---indeed, sophisticated filtering techniques can aid in key generation~\cite{SV09, SV10, VSV12, EHMBLWJ12}. Turbulence predominately occurs in the lower 20~km of the atmosphere and is strongest near the surface of the Earth~\cite{AS93}, thus its impact can be neglected in the case of a downlink as the effects occur only near the end of the transmission path. For uplink transmissions, however, turbulence is important, affecting the beginning of propagation. Turbulence-induced short-term transmission jitter (of order 10--100~ms) is significant in certain applications, but for QKD we need only consider the total number of photons received, proportional to time-averaged loss. We find this from the long-term averaged beam size in the presence of turbulence. The total beam size then comes from the combination of diffraction and turbulence. Because turbulence does not depend on the transmitter aperture, this imposes an inherent limitation on the performance improvement obtainable by increasing transmitter size. For visible wavelengths, turbulence will typically dominate diffraction when using a transmitting telescope of more than 25--50~cm diameter. 
The effect of turbulence can be mitigated by choosing a good ground station. An adaptive optics system could also be used to compensate the effects of turbulence.

Jitter in the telescopes and imprecision in the tracking system will cause misalignment with a time scale typically ${\sim}0.1$--1~s. As with turbulence fluctuations, this systematic pointing error can be averaged over time as additional beam broadening. Controlling for jitter is more challenging on a satellite, thus a downlink will be more vulnerable to this effect. The pointing accuracy must be better than the combined beam waist from diffraction and turbulence to avoid becoming a dominant source of loss.

\begin{figure}
  \centering
  \includegraphics[width=0.49\linewidth]{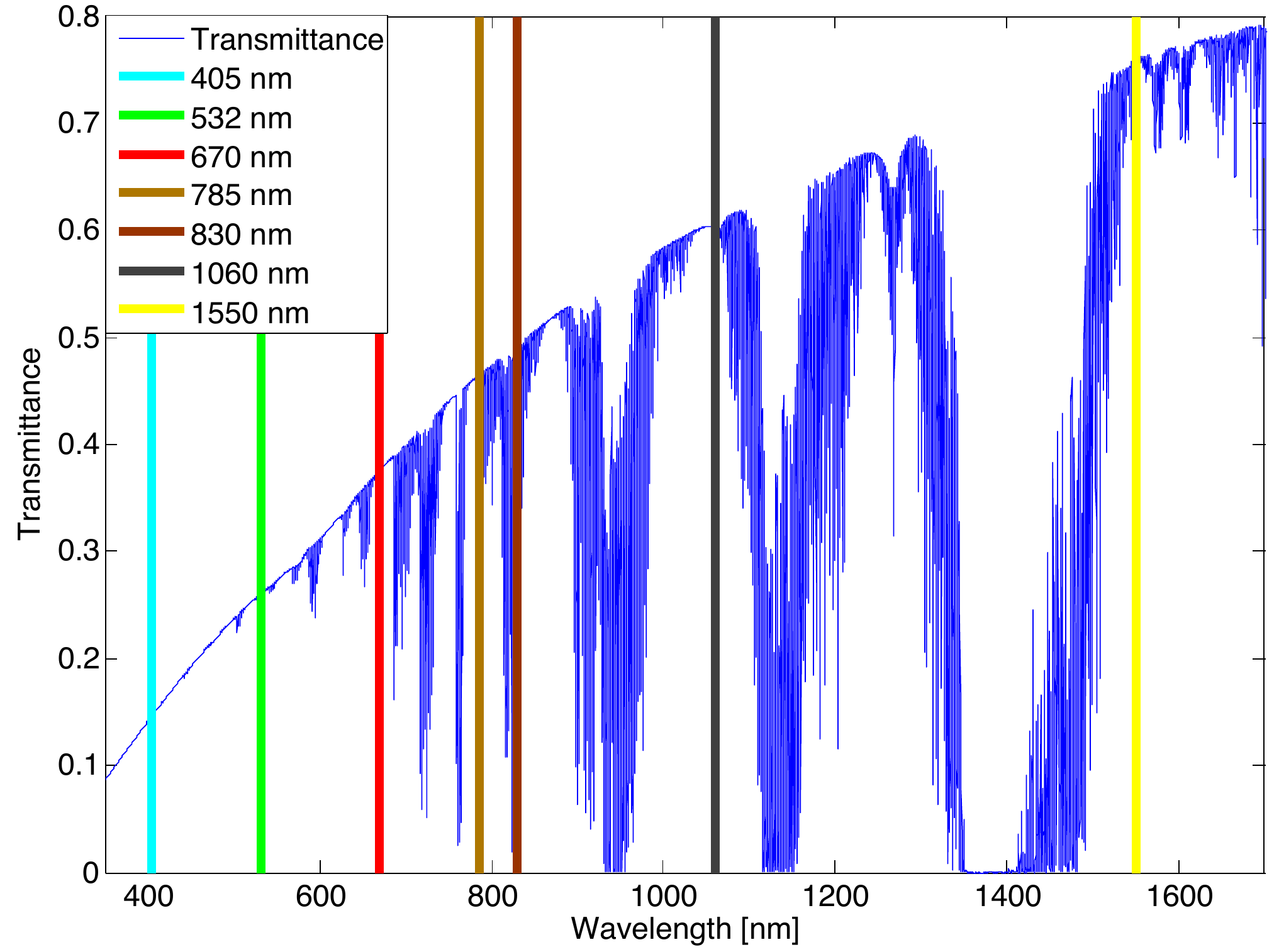}
  \includegraphics[width=0.455\linewidth]{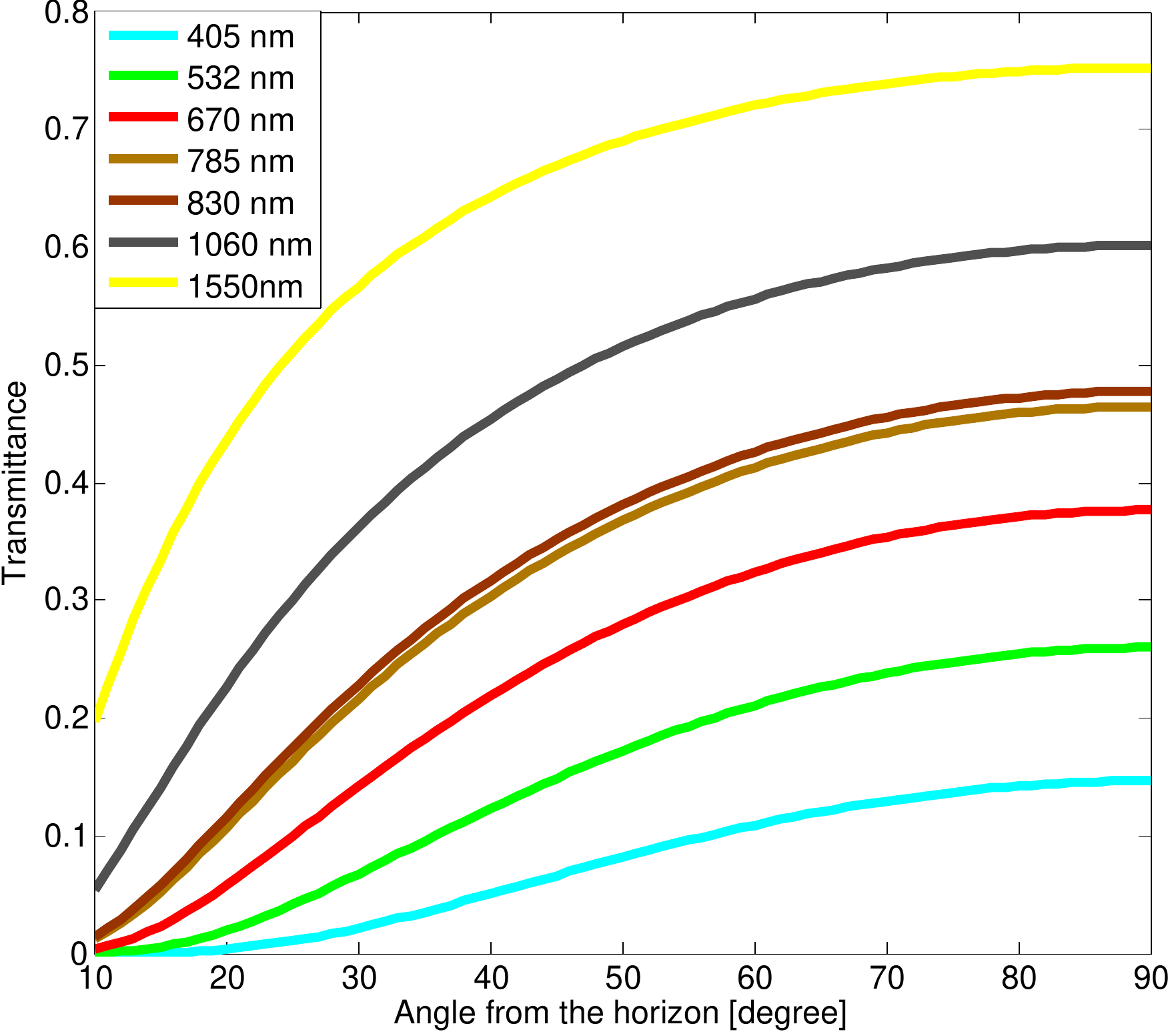}
  \caption[Simulated atmospheric transmittance]{Simulated atmospheric transmittance at a typical rural location, for propagation at zenith (left) and for different elevation angles (right). Coloured lines represent wavelengths of commercially available laser systems. Several transmission windows are evident, within which optical transmission would experience low loss. Generally, the transmission tends to be better at higher wavelengths, but other factors (e.g.\ diffraction, sources, detectors) must be taken into account to properly determine the best wavelength choice.}
  \label{fig.AtmTrans}
\end{figure}

Transmittance through atmosphere is dependent on both wavelength (see Figure~\ref{fig.AtmTrans}, left) and angle (Figure~\ref{fig.AtmTrans}, right), and is a result of the types and concentrations of molecules and particles that are present. Several low-loss transmission windows can be found---most notable are those at 665--685~nm, 775--785~nm, 1000--1070~nm, and 1540--1680~nm, all of which support wavelengths of commercial laser diodes. Using MODTRAN~5~\cite{Mod5}, we model atmospheric transmittance of a rural sea-level location with a visibility of 5~km, chosen to approximate ground stations near large cities (such stations could be utilized to connect city-wide QKD networks globally).

Our numerical model incorporates all the aforementioned loss contributions, as well as (wavelength dependent) scattering and absorption losses due to receiver optical components and detectors, to determine the expected key rates. Diffraction is simulated by discretising the transmission beam intensity profile into a $50\times50$ grid. A radial intensity profile comprising 5000 samples spanning 50~m from the centre of the receiver is then calculated following Rayleigh-Sommerfeld diffraction as propagated from each point of the transmission profile grid. This discretisation allows us to model our final beam profile for a wide range of beam waists, shapes, and telescope designs. Pointing error and atmospheric turbulence (in the case of an uplink) is added using a 2-dimensional convolution between the calculated diffraction profile and the Gaussian distribution of pointing error and turbulence. The final intensity profile is integrated over the receiving area to determine the received power (proportional to the probability of receiving each photon). We then add the remaining loss contributions (atmospheric and optical transmissions, detector efficiency). For details, see Appendix~\ref{app.LossCalcs}.

\section{Beam waist, pointing, and telescope design}

\begin{figure}[tbp]
  \centering
  \includegraphics[width=0.49\linewidth]{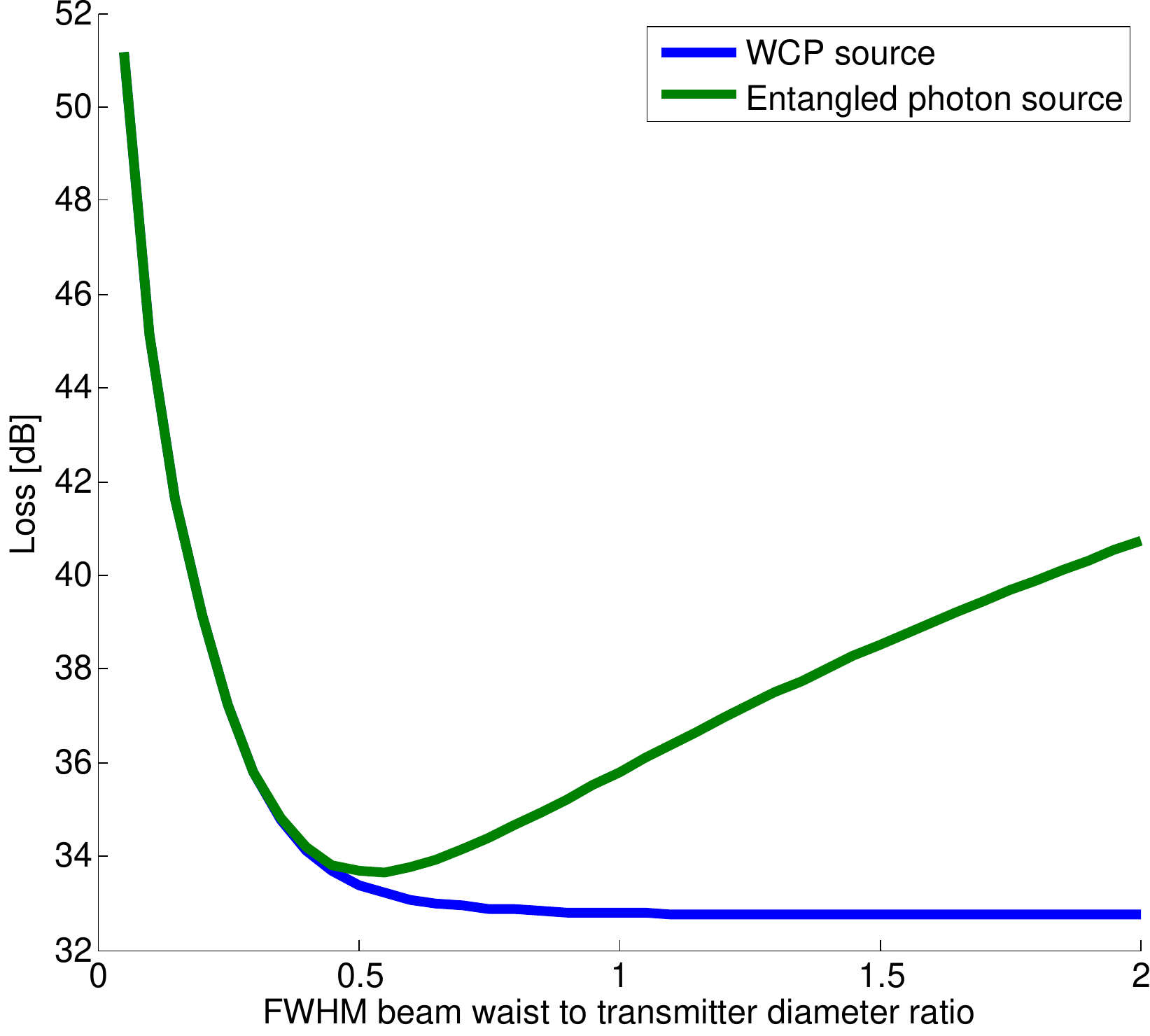}
  \includegraphics[width=0.49\linewidth]{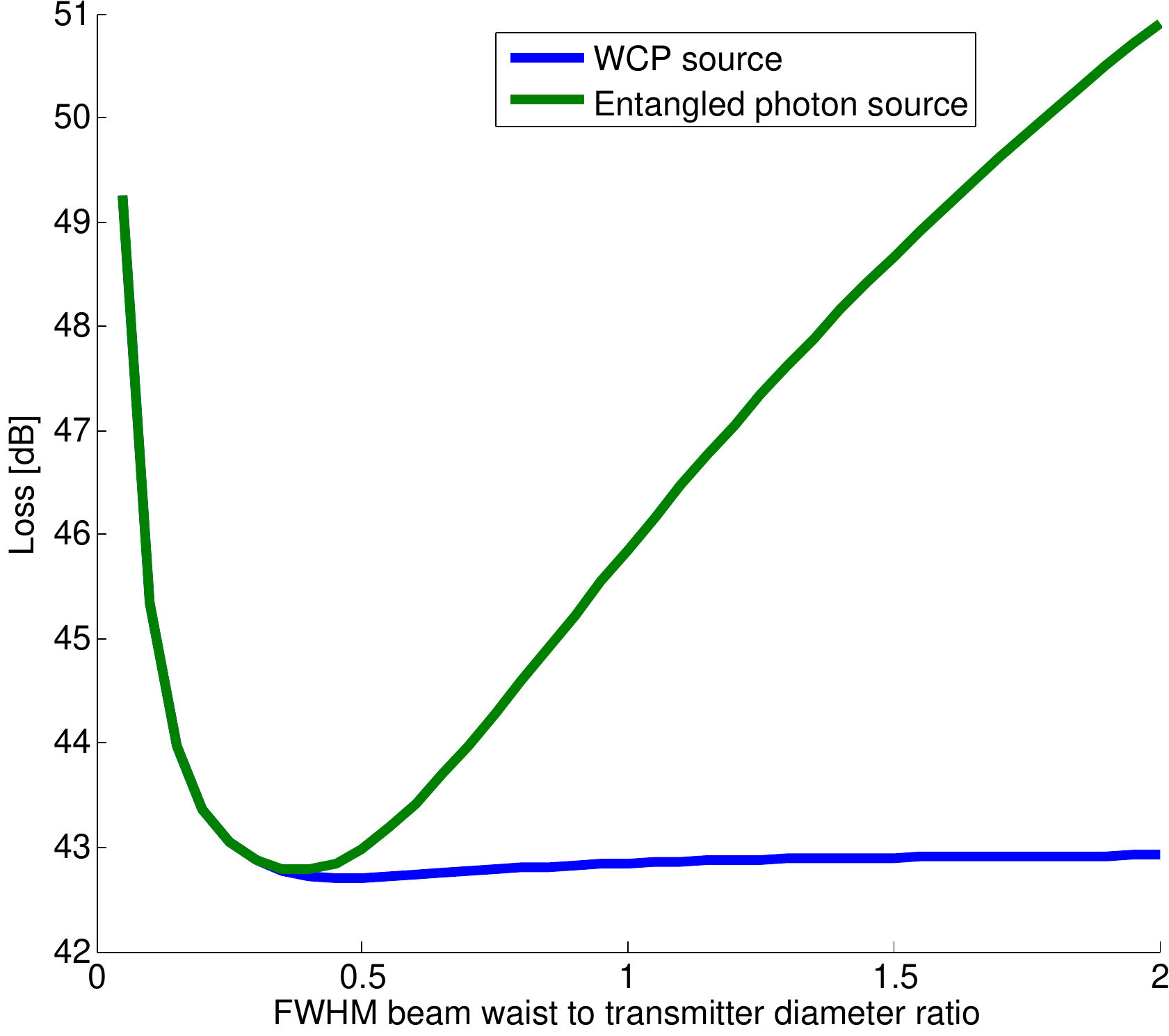}
  \caption[loss vs FWHM 600~km]{Loss at 40$^\circ$ from zenith as a function of the outgoing beam waist (FWHM) for a downlink (left) and an uplink (right). The WCP performs better than the entangled source at large beam waist because the loss from clipping can be included in realizing the required attenuation. The optimum for a downlink is to have a beam waist as large as possible for a WCP source and a beam waist of half the telescope diameter for an entangled photon source. In an uplink, the best beam waist for both sources is smaller then their corresponding value for a downlink because of atmospheric turbulence effects. For downlink: wavelength, 670~nm; satellite transmitter diameter, 10~cm; ground receiver, 50~cm. For	 uplink: wavelength, 785~nm; satellite receiver, 30~cm; ground transmitter, 25~cm. (Different wavelengths minimize losses---see Sections~\ref{sec.wavelengths} and~\ref{sec.results}.) In both cases, orbit altitude is 600~km with no pointing error. Atmosphere is rural sea level.}
  \label{fig.loss_vs_FWHM_600km}
\end{figure}

The optical beam waist (here assumed Gaussian), measured at the transmitting aperture, can be engineered by changing the curvature of the lenses/mirrors of the telescope. Our simulations show that for a downlink (see Figure~\ref{fig.loss_vs_FWHM_600km}, left), an entangled photon source has optimal loss with a full width at half maximum (FWHM) beam waist of half the diameter of the transmitter (consistent with existing literature~\cite{ST98}). This is because a beam that is too small causes exaggerated diffraction, while a beam that is too large will be clipped to the size of the transmitter telescope. We also find that the loss due to beam waist for a WCP source becomes effectively constant for any FWHM beam waist greater than the transmitter telescope diameter. The reason is that the WCP source must emit less than one photon per pulse (on average), and the loss from clipping the outer portion of the beam can be utilized as attenuation towards this end. Therefore, Alice can compensate further clipping losses at the telescope by increasing her source intensity. The beam waist may be made so large, whilst increasing source intensity to compensate, that it essentially becomes a plane wave where diffraction is entirely due to the transmitter's size.

In an uplink, the turbulence broadening dominates over diffraction, reducing the advantage of a larger transmitting telescope. In Figure~\ref{fig.loss_vs_FWHM_600km} (right), the optimal beam waist reflects the size of the beam where diffraction becomes negligible compared to turbulence, and increasing the beam size further has almost no effect on the final beam broadening from all sources. Because of this, it is actually better to keep the beam waist smaller, with less clipping, even if doing so increases diffraction.
Because the influence of atmospheric turbulence also depends on the propagation angle through the atmosphere, the optimal ratio of the beam waist to transmitter size is a function of angle. Continuous readjustment of the beam waist throughout a satellite pass is a significant complication and unlikely to return major improvements.
Thus, for our calculations we use the same FWHM beam waist as with the downlink case so that the diffraction remains based on the telescope size. For all other figures presented here, the FWHM beam waist has been fixed to the transmitter's diameter for the WCP source, and to half the transmitter's diameter for the entangled photon source.

\begin{figure}[tbp]
  \centering
  \includegraphics[width=0.49\linewidth]{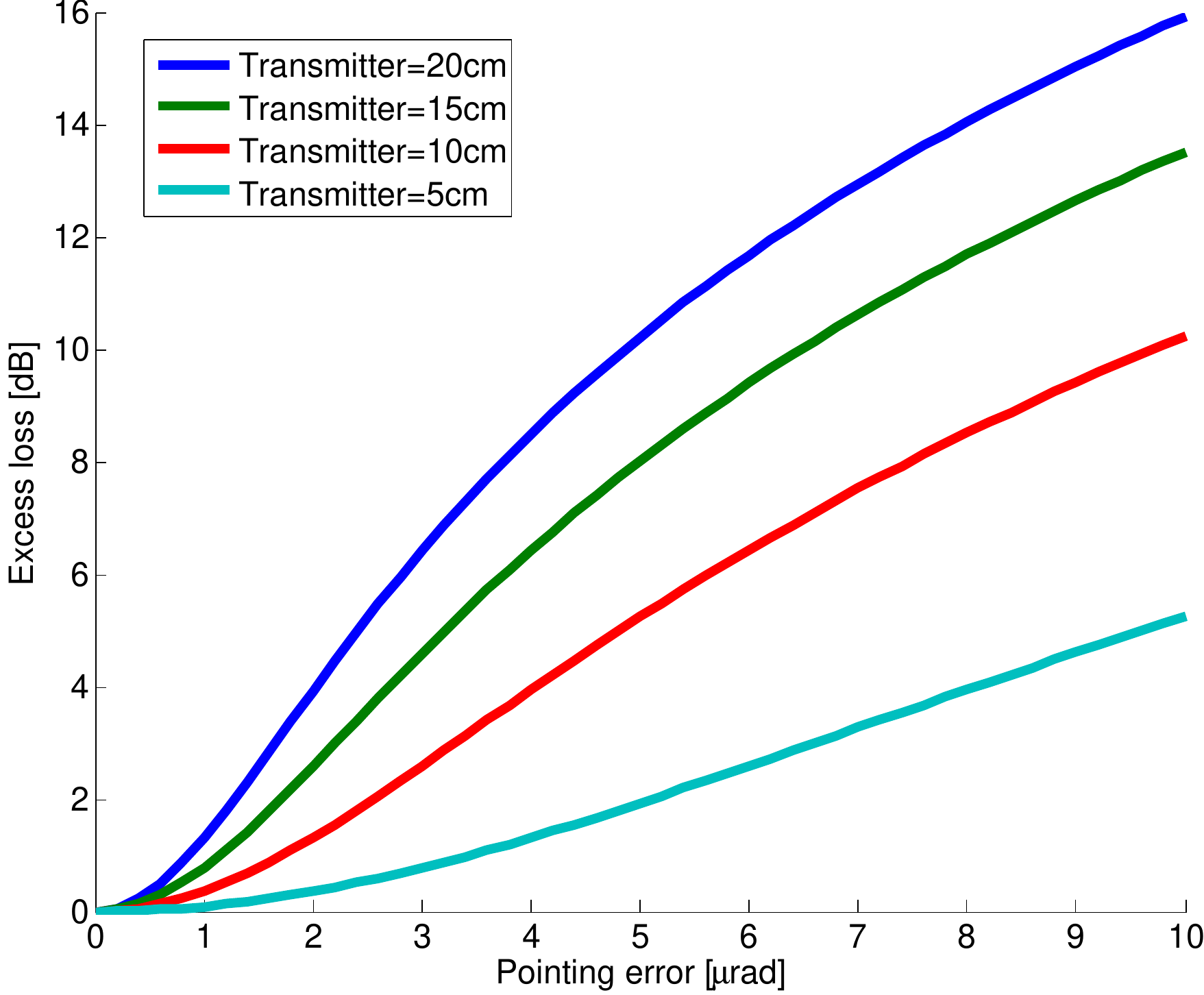}
  \includegraphics[width=0.49\linewidth]{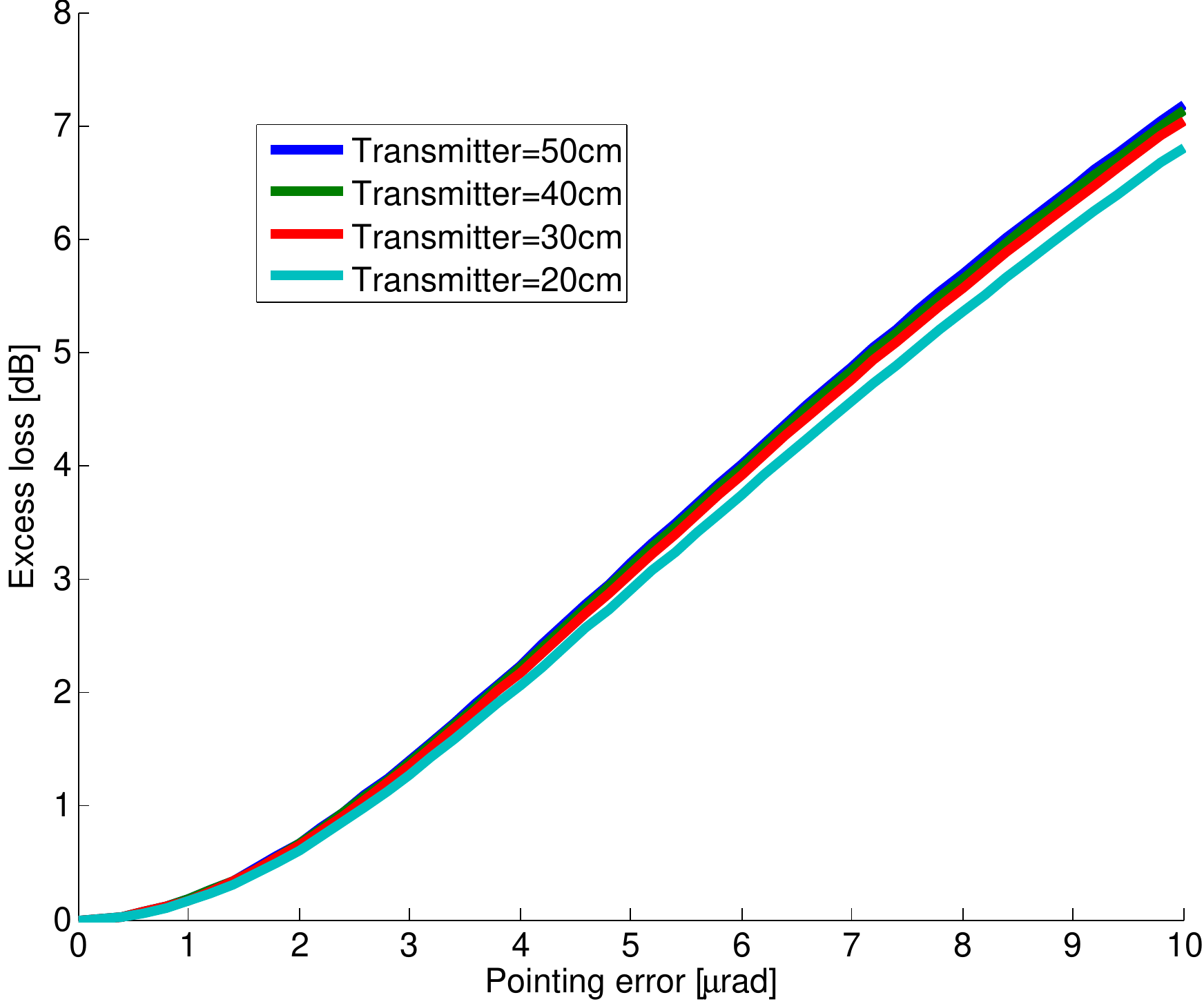}
  \caption[loss vs pointing 600~km]{Excess loss due to systematic pointing error of the transmitter for various transmitter sizes at 40$^\circ$ from zenith in a downlink (left) and in an uplink (right) assuming a two-dimensional Gaussian distribution of the pointing error. For our performance analysis we assumed a pointing error of 2~$\mu$rad, inducing only up to 4~dB of loss in a downlink and less than 1~dB of loss in an uplink. For downlink: wavelength, 670~nm, ground receiver diameter, 50~cm. For uplink: wavelength, 785~nm; satellite receiver, 30~cm. In both cases, the orbit altitude is 600~km and the atmosphere is rural sea level.}
  \label{fig.loss_vs_pointing_600km}
\end{figure}

Again modelled as time-averaged Gaussian beam broadening, the excess loss due to transmitter systematic pointing error is shown in Figure~\ref{fig.loss_vs_pointing_600km}. For a downlink (left), the impact depends strongly on the transmitter size, due to diffraction. To minimize loss, it is sufficient to reduce pointing error such that diffraction becomes the dominant source of broadening. For an uplink, the transmitter size (for transmitters of 20~cm or more) has little impact as atmospheric turbulence dominates, rather than diffraction. The goal is then to reduce pointing error below the influence of atmospheric turbulence. Under our model, pointing accuracies of better than 2~$\mu$rad root mean square (RMS), as demonstrated in previous satellite experiments~\cite{TTTSKTKKTK08}, would cause 1--4~dB of loss in a downlink for up to a 20~cm transmitter, and less than 1~dB of loss in an uplink for all transmitter sizes. We apply this value for the transmitter in our analysis. The receiver only needs to point to an accuracy within its field of view---we assume 50~$\mu$rad, for which sufficient pointing can be easily achieved.

\begin{figure}[tbp]
  \centering
  \includegraphics[width=0.49\linewidth]{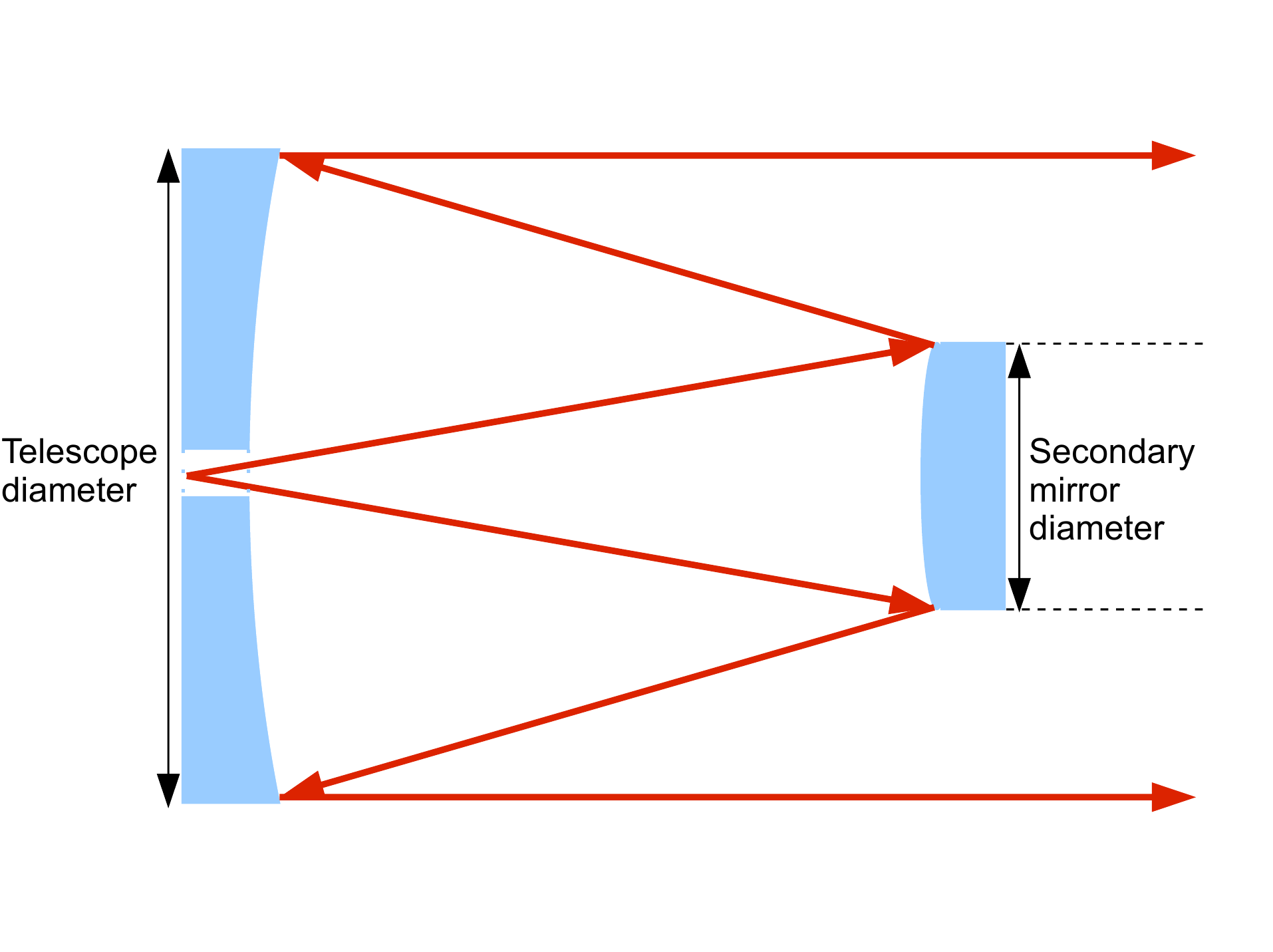}
  \includegraphics[width=0.49\linewidth]{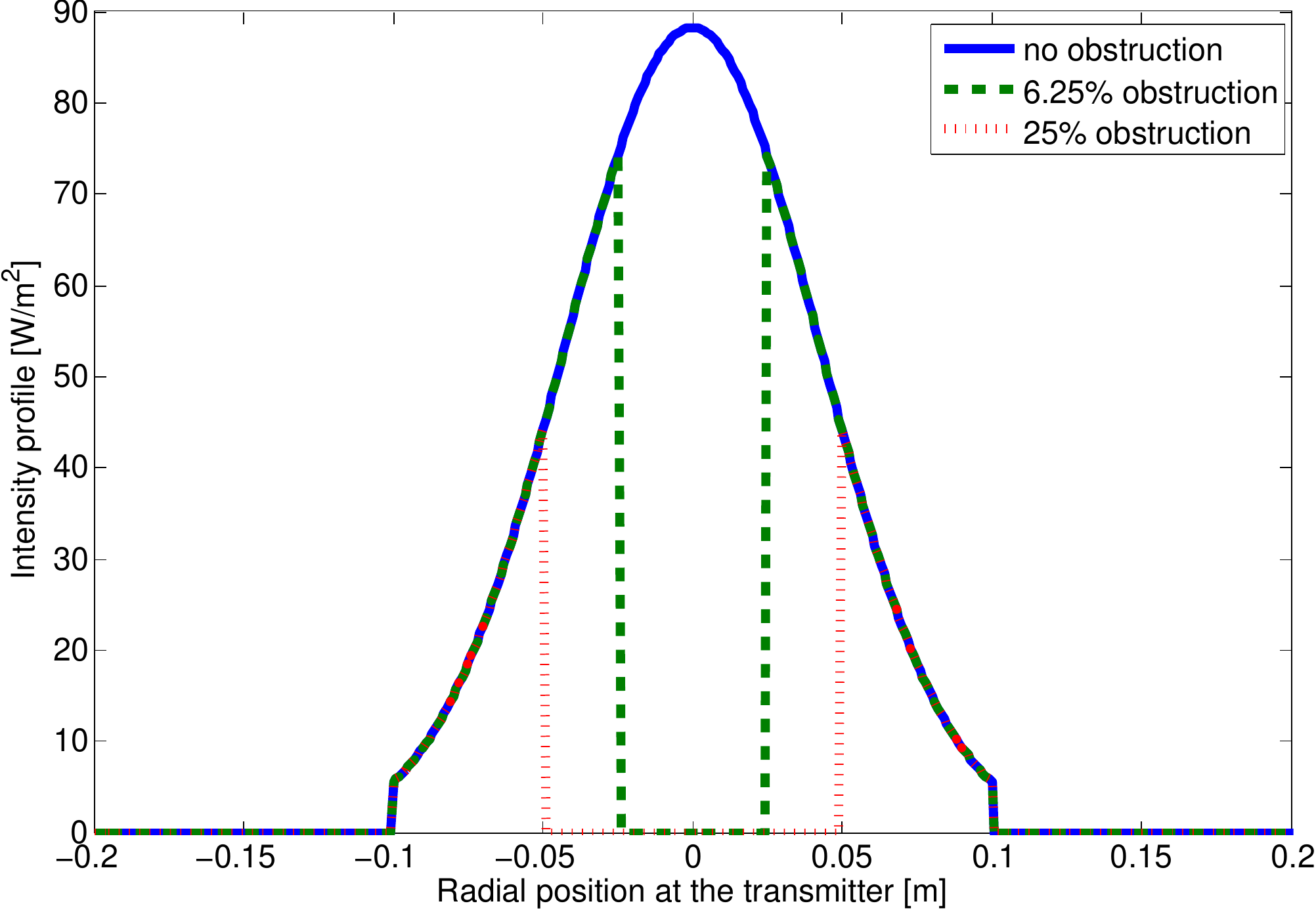}
  \includegraphics[width=0.49\linewidth]{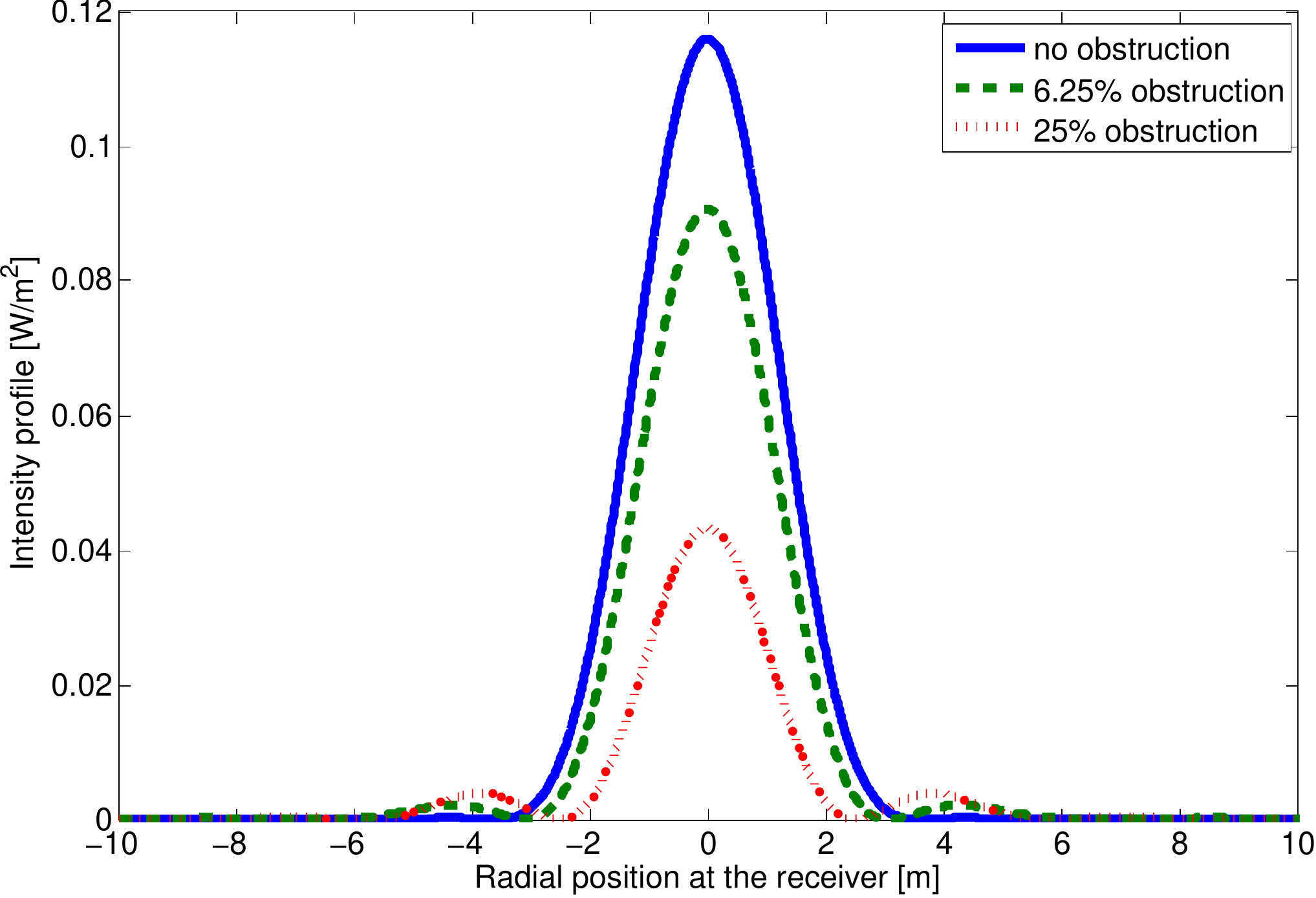}
 \includegraphics[width=0.49\linewidth]{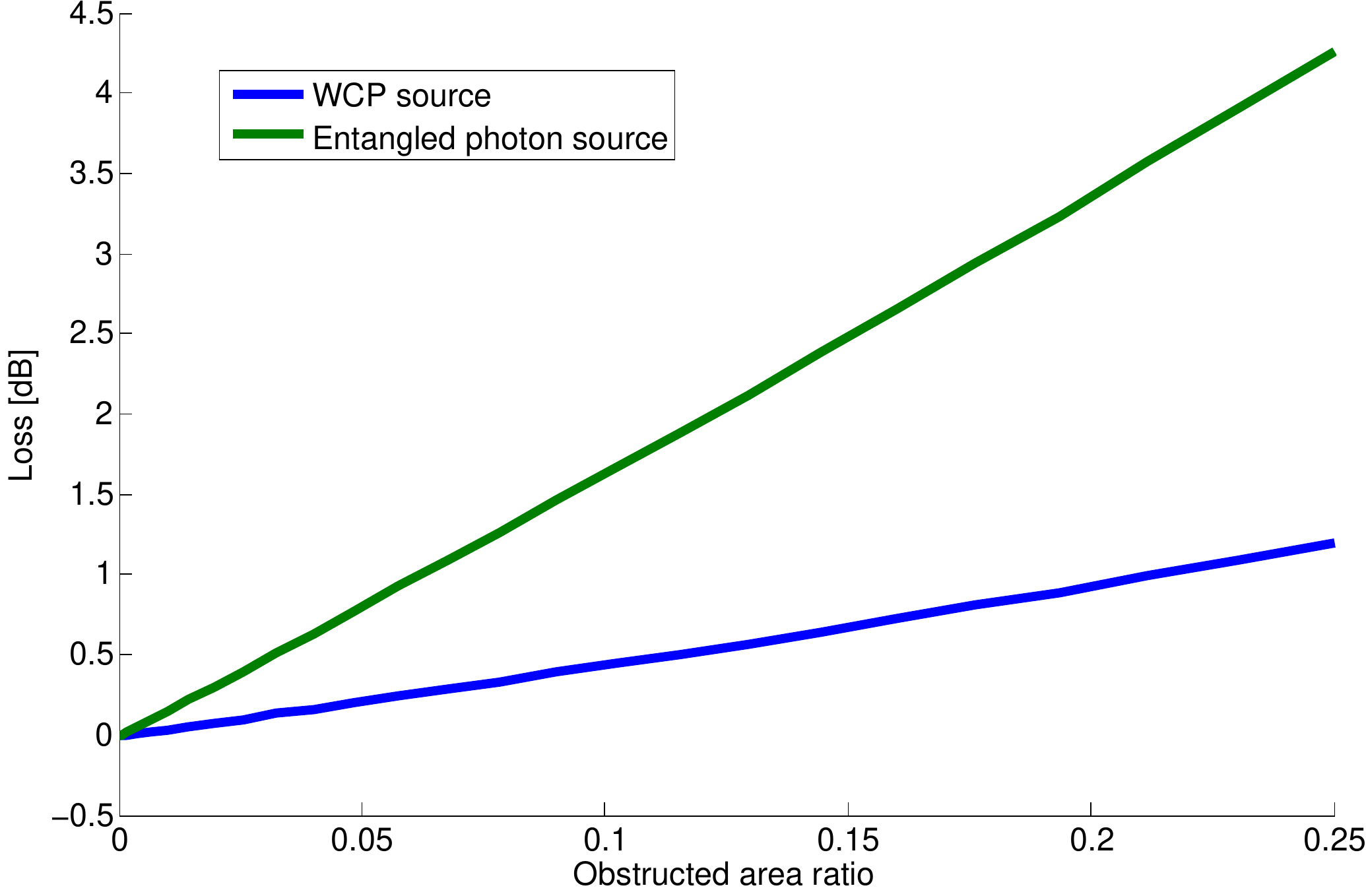}
  \caption[Cassegrain_transmitter]{Beam profile from a transmitter with a central secondary mirror blocking a portion of the outgoing beam (top-left) and the accompanying intensity profile at the transmitter (top-right). The intensity profile at the receiving telescope was computed (bottom-left) and the additional loss due to this type of transmitting telescope was evaluated for both WCP and entangled photon sources (bottom-right). The impact of this design is less than 1~dB for an obstruction of up to 6.2\% of the area (i.e., a secondary mirror with a diameter of up to 25\% the diameter of the primary mirror).}
  \label{fig.Cassegrain}
\end{figure}

Modern large telescopes are typically of a reflective design with a secondary mirror placed in the path of the beam, therefore blocking part of it (see Figure~\ref{fig.Cassegrain}, top-left). We incorporate in our analysis the loss due to a secondary mirror centred in the optical beam path, such as a symmetrical Gregorian, Newtonian or Cassegrain design. Because we assume a Gaussian beam, this design represents a worst case scenario, providing a lower bound on the expected link transmission when using a reflective design.

The secondary mirror has two effects: it blocks a portion of the beam (Figure~\ref{fig.Cassegrain}, top-right) and it alters the diffraction (Figure~\ref{fig.Cassegrain}, bottom-left). For both, the additional loss depends only on the ratio of primary and secondary mirror diameters when considering the far field. Figure~\ref{fig.Cassegrain} (bottom-right) shows that such a design has little impact for reasonable primary/secondary mirror ratios. Like beam waist, the effect of blocking the central part of the beam is lesser for a WCP source than an entangled source because the transmission power can be adjusted to counteract the obstruction loss. On the receiving end, the size of the beam---typically on the order of 10~m---is already much larger than any receiving telescope we consider. The loss due to the receiver design is therefore almost entirely dependent on the area of the telescope. A reflective design simply reduces the receiving area proportionally.

\section{Wavelength considerations}\label{sec.wavelengths}

Diffraction losses decrease for shorter wavelengths, whereas atmospheric transmittance and turbulence losses reduce for longer wavelengths. As the magnitudes of these depend on transmitter and receiver telescope sizes, so too will the wavelength that minimizes the total loss owing to both effects. For a satellite incorporating a downlink, it is preferable to use a robust and space-qualified laser either as a WCP source or as part of an entangled photon source. Ideally this would consist of a diode laser module, capable of producing a certain wavelength or wavelengths within a very small bandwidth range. However, diode lasers exist only for certain specific wavelengths, limiting the choices available (wavelengths we considered are shown in Figure~\ref{fig.AtmTrans}).

Commercially available detector technology typically fits into two categories. Loosely speaking, these are visible (specifically, 400--1000~nm) and infrared (950--1650~nm) wavelengths. Visible wavelengths are typically serviced by Si avalanche photodiode (APD) technology capable of ${>}50\%$ detection efficiency with low dark counts and maximum count rates in the MHz range~\cite{SBCDLP09}. Infrared wavelengths are serviced by either InGaAs APDs or, recently, by superconducting single photon detectors. Currently, InGaAs APDs suffer from lower detection efficiencies, higher dark count rates, and low repetition rates~\cite{SBCDLP09} limiting their usefulness for satellite missions.\footnote{Notably, new techniques such as self-differencing~\cite{YKSS07} are improving InGaAs detectors such that they may yet become practical for satellite missions.} Superconducting single photon detectors have made considerable progress over the last few years~\cite{GOCLSSVDWS01, MNMS03, MNMS03, RLMN05, KMMMRCSVGSPVS05, RYDKAVGB06, LMN08, BLVN11, MNDBHCMB11}, reaching high efficiency, low dark counts, and broad-spectrum sensitivity. While promising, current superconducting detectors are in the research stage, and all such devices require cryogenic cooling to operate~\cite{SBCDLP09}, making them impractical for low cost satellite missions (especially those incorporating uplinks).

Si-APDs are a mature technology with low technological requirements for a satellite mission, supporting wavelengths of multiple free-space transmission windows (see Figure~\ref{fig.AtmTrans}). For these reasons, we focus on these wavelengths, and calculate losses for commercially-available laser wavelengths in this range for various possible telescope diameters. Two types of Si-APDs were studied: thin APD (from Micro Photon Devices) detector efficiencies are used for wavelengths below 532~nm~\cite{MPD}, and thick APD (from Excelitas Technologies, formerly part of Perkin Elmer) efficiencies for 532~nm and above~\cite{Excelitas}. We also assumed a detector dark count rate of 20~cps, in line with the capabilities of these detectors.

\section{Results summary}\label{sec.results}

We calculate the total loss, along with projected background counts, for one year's worth of nighttime satellite passes. To determine the background contribution (see Appendix~\ref{app.BackgroundCalcs}) we assume a half-moon at $45^\circ$ elevation. The location used for calculating the artificial light pollution was 20~km from the city of Ottawa, Canada, with a latitude of approximately $45^\circ$ North. This location, at the the edge of the city, represents a scenario where the ground station may be linked to a city's ground-based secure QKD network, with the satellite acting as a trusted node to establish global quantum-secured links.

We consider a 600~km circular sun-synchronous noon/midnight low Earth orbit. The orbit is modelled using Systems Tool Kit (STK) 9 from Analytical Graphics, Inc.\ (AGI)~\cite{AGISTK9}. We incorporate only satellite elevations greater than $10^\circ$ above the horizon, encompassing a ``usable'' portion of a pass, as lower angles typically exhibit losses too high to constructively add to any of the considered schemes. Under this condition we find 713 usable passes over one year, or about 2 passes per night. Notably, the results of the simulations are largely insensitive to the selection of orbit height---e.g.\ lowering the orbit to 500~km does improve the signal-to-noise ratio, but this effect is muted by the reduced contact time to the ground station.

\begin{figure}[tbp]
  \centering
  \includegraphics[width=0.49\linewidth]{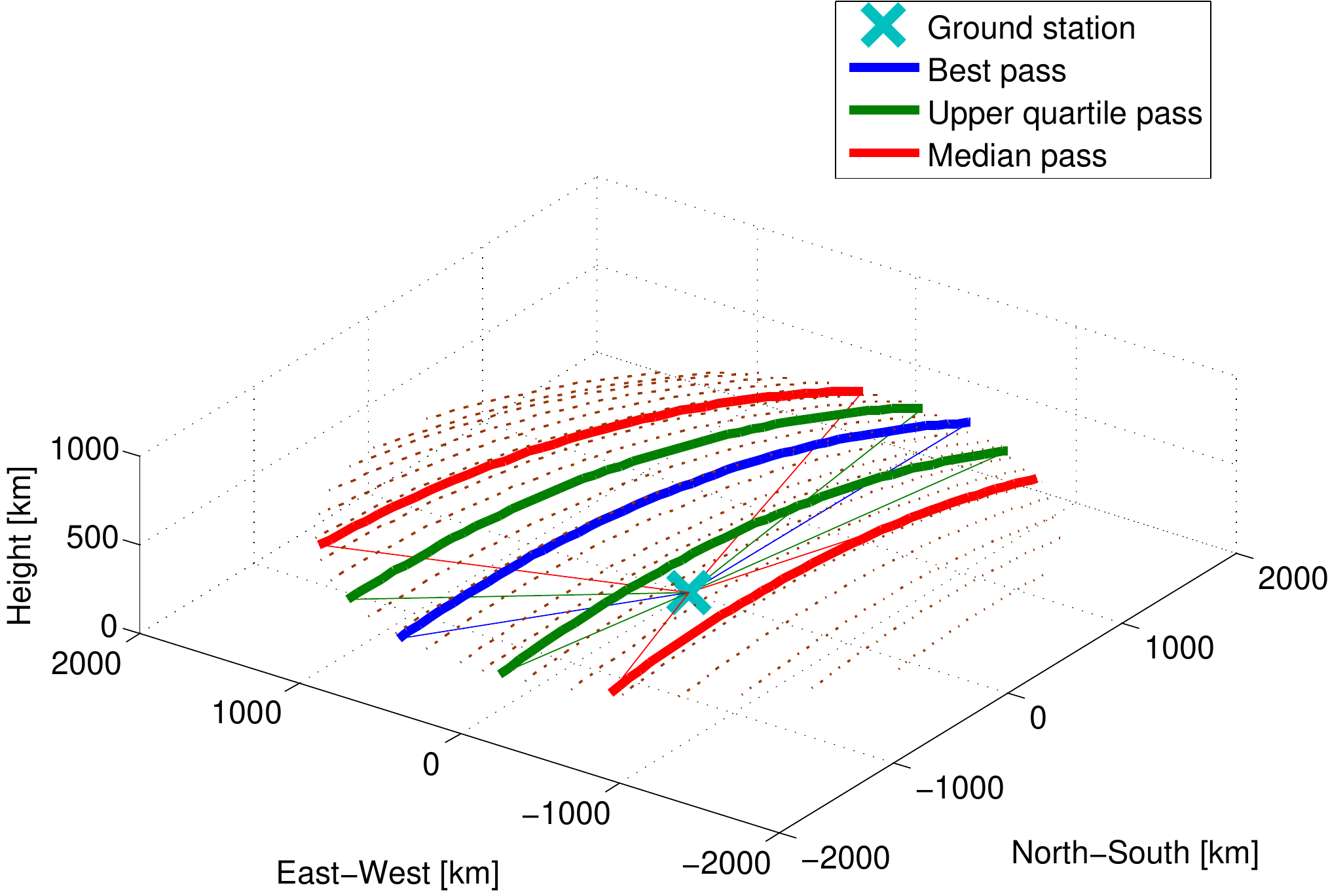}  
  \includegraphics[width=0.49\linewidth]{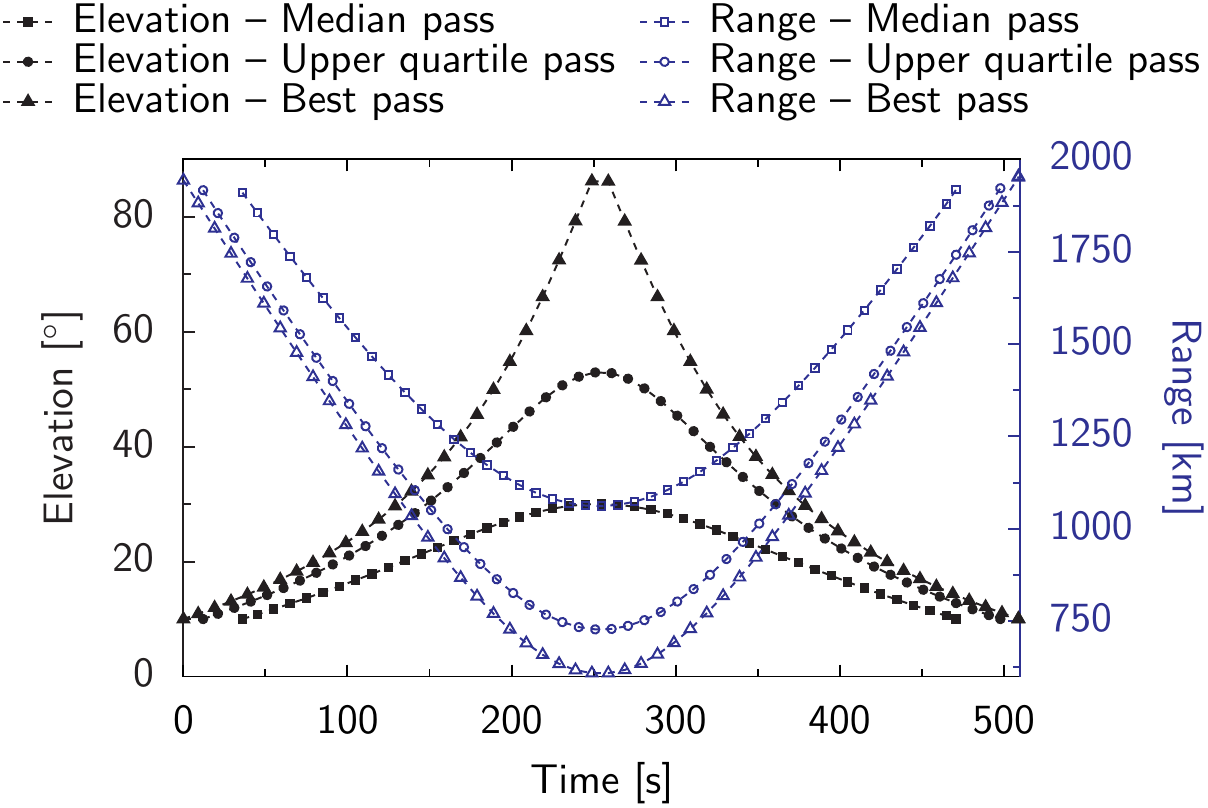}
  \caption[passes]{Illustration of satellite passes (left) and range and elevation angle of the satellite (right) relative to the ground location during the best pass, upper quartile pass, and median pass, i.e.\ having longest, upper quartile, and median usable duration, respectively, of all passes over one year. Left: Best, upper quartile, and median passes are shown as thick blue, green, and red lines, respectively (thin lines connect the ground station with the link termination points for these passes), with 20 additional example passes (brown dotted lines). The best pass transits directly over the ground station (i.e.\ reaching $90^\circ$ elevation), while other passes fall to either side. Right: Half of all passes have a duration of at least 450~s, only 80~s less than the best pass, yet less than 25\% of passes reach $53^\circ$ of elevation from zenith.}
  \label{fig.range_elevation}
\end{figure}

\begin{figure}[tbp]
  \centering
  \includegraphics[width=0.49\linewidth]{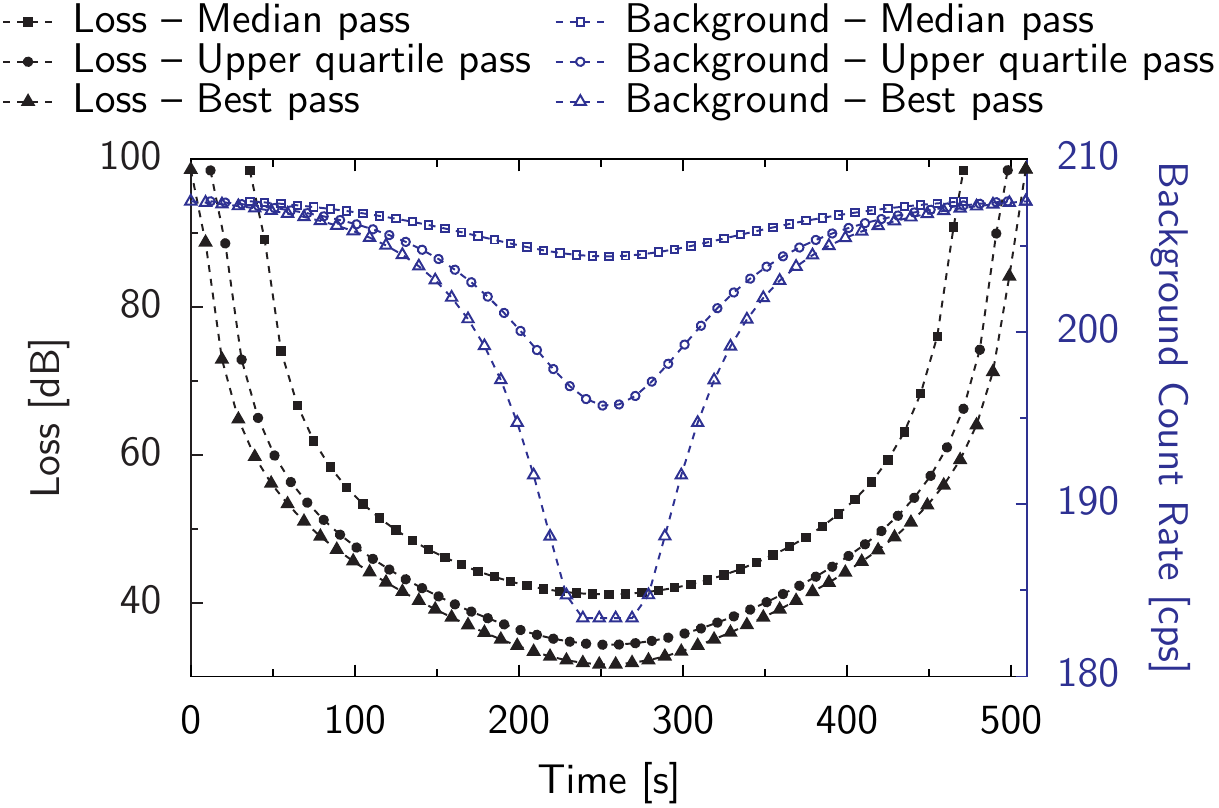}
  \includegraphics[width=0.49\linewidth]{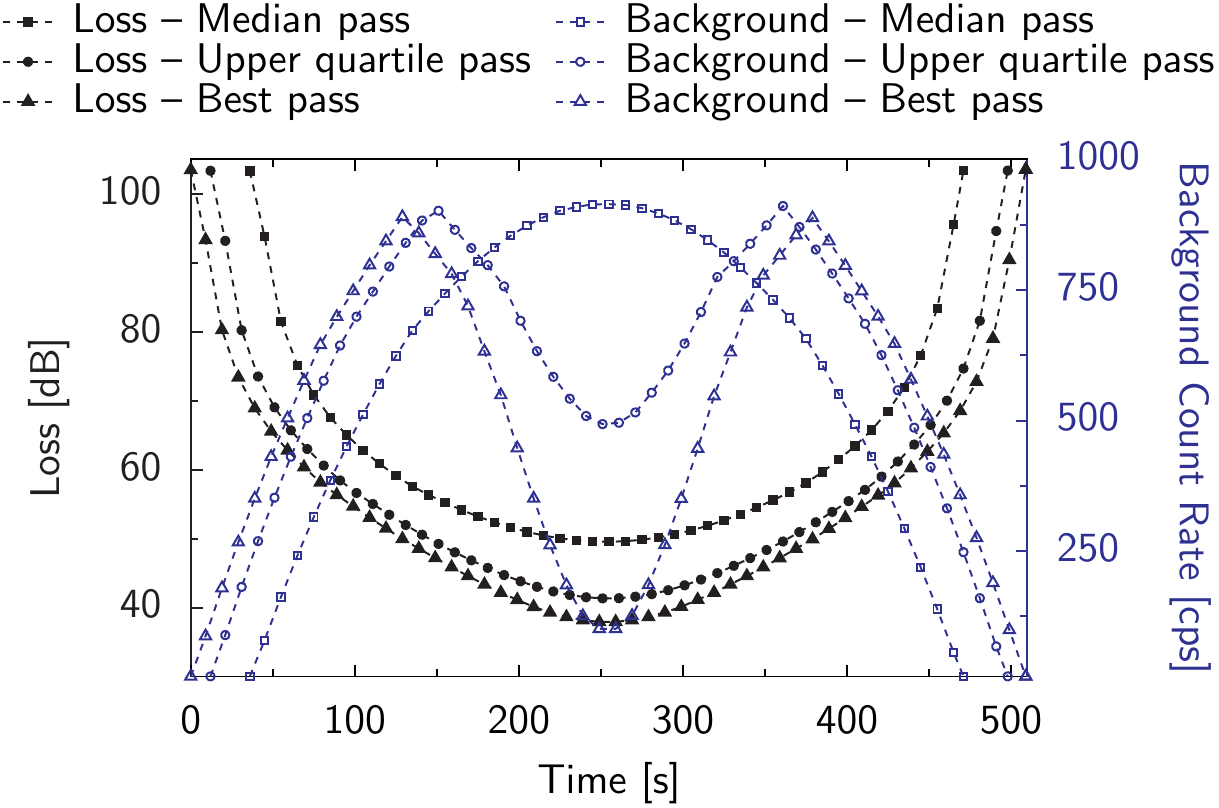}
  \caption[passes]{Loss and detected background count rate during the best pass, upper quartile pass, and median pass for a downlink (left) and an uplink (right). The uplink background is mainly due to artificial light and is lower at high elevation angle, when the satellite has a smaller field of view area on the ground. Orbit altitude is 600~km. For downlink, wavelength is 670~nm and satellite transmitter telescope diameter is 10~cm. For uplink, wavelength is 785~nm and satellite receiver telescope diameter is 30~cm. In both cases, the receiver applies an optical filter with 1~nm bandwidth. Ground telescope is 50~cm with pointing error of 2~$\mu$rad. Atmosphere is rural sea level.}
  \label{fig.loss_background}
\end{figure}

Figure~\ref{fig.range_elevation} shows examples of passes for the 600~km orbit, including the best, upper quartile and median pass. The best pass is the pass possessing the maximum usable duration, the upper quartile pass is the pass for which 25\% of all satellite passes have longer usable duration, and the median pass is the pass for which 50\% of all satellite passes have longer usable portions (and 50\% have shorter usable portions). Figure~\ref{fig.loss_background} shows the loss and background count rates for these passes. We see that an uplink experiences more loss (by ${\approx}5$~dB) than a downlink with a satellite telescope of $1/3$ the diameter of the uplink's satellite telescope, due to the atmospheric turbulence. Additionally, the background count rate for an uplink is almost an order of magnitude higher than a downlink, owing to artificial light pollution emitted upward. Atmospheric transmission loss decreases as the satellite approaches zenith. For an uplink, this also means lower loss for background photons, however the ground area imaged by the satellite also reduces near zenith, reducing the background photons collected overall.

To properly determine the optimal wavelength, we examine the secure key length that can be obtained during an upper quartile pass. The results (Table~\ref{tab.loss_vs_wavelength_600km}) suggest that the common 670~nm laser line is a highly suitable wavelength for a downlink, be this using a WCP or an entangled photon source. The optimal wavelength for the uplink is higher, closer to the laser line at 785~nm, owing to the reduction of atmospheric turbulence. Table~\ref{tab.loss_vs_wavelength_600km} also shows that ${\approx}800$~nm, typical of spontaneous parametric down-conversion entangled photon sources, would work well in both cases (justifying our choice of wavelength in the examples presented above).

\begin{table}[hbt]
\caption{Calculated length of distributed cryptographic key for various wavelengths with a WCP (left) and an entangled photon (right) source. Of the laser-line wavelengths studied, 670~nm produces the longest key for a downlink, while 785~nm produces the longest key for the uplink. Downlink is with a 10~cm transmitter and a 50~cm receiver; uplink is with a 50~cm transmitter and a 30~cm receiver.  Simulations are of the upper quartile satellite pass (in terms of pass duration) with a 600~km orbit, pointing error of 2~$\mu$rad, and rural atmosphere (5~km visibility) at sea level. Source rate: 300~MHz for WCP and 100~MHz for entangled photon source; detector dark count rate: 20~cps; detection time window: 0.5~ns.}\label{tab.loss_vs_wavelength_600km}
\centering\begin{tabular}{>{\centering}p{2cm}S[table-format=3.2]S[table-format=3.2]S[table-format=3.2]S[table-format=3.2]} 
\hline
\multicolumn{1}{l}{} &\multicolumn{4}{>{\centering}p{10.2cm}}{Secure key length obtained for the upper quartile satellite pass [kbit]}\\
\cline{2-5}
Wavelength [nm] & \multicolumn{1}{>{\centering}p{2cm}}{Downlink, WCP source} & \multicolumn{1}{>{\centering}p{2cm}}{Uplink, WCP source} &  \multicolumn{1}{>{\centering}p{2.6cm}}{Downlink, entangled photon source} &  \multicolumn{1}{>{\centering}p{2.6cm}}{Uplink, entangled photon source}  \\
\hline
405 & 68.5 & 3.5 & 6.2 & 0 \\
532 & 264.5 & 33.1 & 119.3 & 12.1  \\
670 & 465.6 & 87.7 & 324.7 & 67.4 \\
785 & 458.3 & 111.3 & 272.9 & 75.7  \\
830 & 317.3 & 82.1 & 136.1 & 39.7  \\
1060 & 175.4 & 67.6 & 21.8 & 8.1  \\
1550 & 123.9 & 94.8 & 12.8 & 14.4  \\
\hline
\end{tabular}
\end{table}

We accumulate key rate statistics for the full one-year set of satellite passes, with various transmitter and receiver telescope sizes, to determine the expected number of secure key bits generated each month. Each pass generates a secure key independently---gradual accumulation of cryptographic key bits from each satellite pass ensures these bits are available for use when required.\footnote{One could obtain a larger monthly secure key by combining the raw keys of several passes, thereby reducing finite size effects, at the cost of a reduction in the frequency of key accumulation/usage.}

We further assume that only half of the nights have clear skies, automatically rendering half the passes unusable due to cloud coverage. Actual cloud coverage will depend on the ground station location ultimately chosen. The average global cloud coverage on land is between 50--90\%, with over 25\% of clouds having a thin density~\cite{WJMB05}. Many areas, particularly in drier or more elevated regions, experience less than 20\% cloud cover, some having near 0\% cloud cover~\cite{MSMS00}. A location with 50\% cloud coverage would likely represent a worst case of any site that would be reasonably considered.

The results show that a downlink can generate more secure key bits than an uplink for the same ground and satellite telescopes. Furthermore, the WCP source outperforms the entangled photon source, due in part to the higher source rate for WCP, and in part to the inefficiency of detecting the transmitter's heralding photon in the entangled source. A downlink with a satellite transmitter telescope as small as 10~cm and a receiver of 50~cm could be used to successfully exchange a key of 4.5~Mbit per month with an entangled photon source, and 25~Mbit per month with a WCP source. In an uplink, a 30~cm receiver telescope on the satellite and a ground transmitter of at least 25~cm could produce 0.4~Mbit key per month with an entangled photon source and 3~Mbit per month with a WCP source.

Interestingly, for an uplink, varying the size of the ground transmitter telescope has little effect on the number of key bits generated. This is because, for a transmitter telescope of 25~cm or more, turbulence dominates the beam divergence, limiting any gains that could otherwise be found by reducing diffraction via increasing the transmitter telescope diameter.

We also determine the long-distance performance of two other important quantum experiments: Bell tests and quantum teleportation. For both experiments, we analyse each satellite pass independently to determine which pass can perform a successful Bell test or teleportation with $3\sigma$ certainty. Since data from an entire pass is needed for success, we calculate the minimum ground-satellite distance of each successful pass. Finding the greatest of these minima from all passes gives the longest distance test achievable with our parameters. That is, at least one pass from our simulated year of orbits will be capable of performing the experiment with a $3\sigma$ violation while maintaining, for the entire experiment, a distance at least the ``maximum distance'' reported.

A downlink reaches a distance of 1700~km in a Bell test and 1050~km for teleportation for a satellite transmitter telescope of 10~cm and a receiver of 50~cm. An uplink with a 30~cm receiver telescope on the satellite and a ground transmitter of 25~cm would be capable of performing a Bell test at 1350~km and teleportation at over 675~km. Both are significantly beyond that which can be achieved on the ground alone.

For an uplink, exposure to radiation in the space environment is expected to lead to an increase in the dark count rate over time~\cite{SKAS04, MMMLFPBBGRGGSRCM11}, the degree to which can be mitigated by appropriate shielding. We performed a preliminary analysis (see Table~\ref{tab.detector_degradation}) that showed a dark count rate on the order of 1000~cps would cause only a minor degradation of performance for QKD and Bell tests. Teleportation, however, requires stricter control over radiation-induced dark counts. These results provide important guidelines for determining the requirement of radiation hardening.

\begin{table}[hbt]
\caption{Effect of higher dark counts in the detectors on the key generation and on the maximum distances of fundamental experiments. QKD (with either a WCP source or an entangled photon source) and Bell tests are resistant to detector dark count rate increase of up to 1000~cps, and both QKD with an entangled photon source and Bell tests can still be performed (with reduced efficiency) at 10000~cps. Teleportation is only able to accept a dark count rate increase on the order of 100~cps. Wavelength 785~nm, 50~cm transmitter and a 30~cm receiver. Orbit 600~km, pointing error 2~$\mu$rad, and rural atmosphere (5~km visibility) at sea level. Source rate: 300~MHz for WCP and 100~MHz for entangled photon source; detection time window: 0.5~ns.}\label{tab.detector_degradation}
\centering\begin{tabular}{c S[table-format=2.4]S[table-format=2.4]S[table-format=6.2]S[table-format=6.2]} 
\hline
\multicolumn{1}{>{\centering}p{1.8cm}}{Detector dark [cps]} & \multicolumn{1}{>{\centering}p{2.3cm}}{WCP source key [Mbit/month]} & \multicolumn{1}{>{\centering}p{2.2cm}}{Entangled source key [Mbit/month]} & \multicolumn{1}{>{\centering}p{2.3cm}}{Max.\ Bell test distance [km]} & \multicolumn{1}{>{\centering}p{2.4cm}}{Max.\ teleportation distance [km]}  \\
\hline
20 & 3.222 & 0.440 & 1359 &  683 \\
100 & 3.109 & 0.426 & 1283 & 660 \\
1000 & 2.292 & 0.322 & 1039 & 0 \\
10000 & 0 & 0.012 & 660 & 0 \\
\hline
\end{tabular}
\end{table}

Complete results of our quantum satellite performance analysis are given in Appendix~\ref{app.Comp_perf}. It is evident from our analysis that a downlink outperforms an uplink in general, and would thus be the preferred option for global QKD implementation. However, it should be recalled that a downlink 
introduces a number of technical challenges in addition to those of an uplink.
An uplink also allows the source to be easily interchangeable, permitting a wider range of experiments and tests. With these considerations in mind, a strong argument can be made that an uplink is the better choice to scientifically study global-scale QKD implementations (and other experiments) prior to implementing a full-scale (downlink) global QKD system.

\section{Conclusion}

We have presented an in-depth analysis of satellite QKD possible with current technologies. To do so, we constructed a detailed link analysis to determine various optimal or near-optimal link parameters and to estimate the length of the secure key that would be obtained. We found that the optimal beam waist has a FWHM equal to the transmitter diameter for a WCP source, and a FWHM of half the transmitter diameter for an entangled photon source. From the available laser diode wavelengths and given the technological and logistical limitations of current photon detectors, the laser line wavelength of 670~nm would be well suited for a downlink using a 600~km noon/midnight orbiting satellite, and an uplink at the same orbit altitude would benefit more from 785~nm. We also considered the effect of pointing error and telescope design, finding that a pointing error of 2~$\mu$rad or less and a central secondary mirror at up to a quarter of the telescope's diameter would each induce no more than a 1--4~dB increase in the overall loss.

Finite-size secure key analysis was used to determine that a reasonable satellite QKD system could generate on the order of megabits of secure key per month, and successfully perform fundamental Bell tests and teleportation experiments at distances on the order of 1000~km. Preliminary analysis of radiation degradation of detectors suggests that up to two orders of magnitude increase in dark counts could be tolerated. Many of the key assumptions (e.g.\ atmospheric conditions, systematic pointing error, cloud coverage, artificial background counts, moon location and illumination) are on the conservative side and therefore strengthen our confidence in the performance estimates. Additional technologies, such as adaptive optics (for which our own analysis is in preparation), could be used to further improve performance. Multiple passes (that may not be usable on their own) could also be combined to increase the usability and performance. We note that, despite better performance from a downlink, an uplink can provide reasonable key rates for an initial demonstration mission, with the additional benefit of more flexibility of a ground based source, as well as a significantly simpler satellite payload.

The model we developed allows us to include realistic designs of satellite quantum optics experiments to thoroughly predict their performance. This model can be applied to various scenarios beyond the single link LEO: other orbits (medium Earth orbits, geostationary orbits, etc.), double links including inter-satellite links, and various other scenarios such as using a reflector on a satellite.

\section{Acknowledgements}

The authors would like to thank A. Koujelev (CSA), J.-F. Lavigne, F. Ch\^ateauneuf and A. Foug\`eres (INO), N. {L\"utkenhaus}, X. Ma and Z. Yan (IQC), and  M. Toyoshima (NICT) for helpful discussions, as well as C. Holloway (IQC) for help in speeding up the numerical simulations. Support for this work by CSA, DRDC, NSERC, CIFAR, CFI and Ontario's ERA program is gratefully acknowledged.

\appendix
\section{Numerically modelling realistic loss}\label{app.LossCalcs}

\begin{figure}[tbp]
  \centering
  \includegraphics[width=0.46\linewidth]{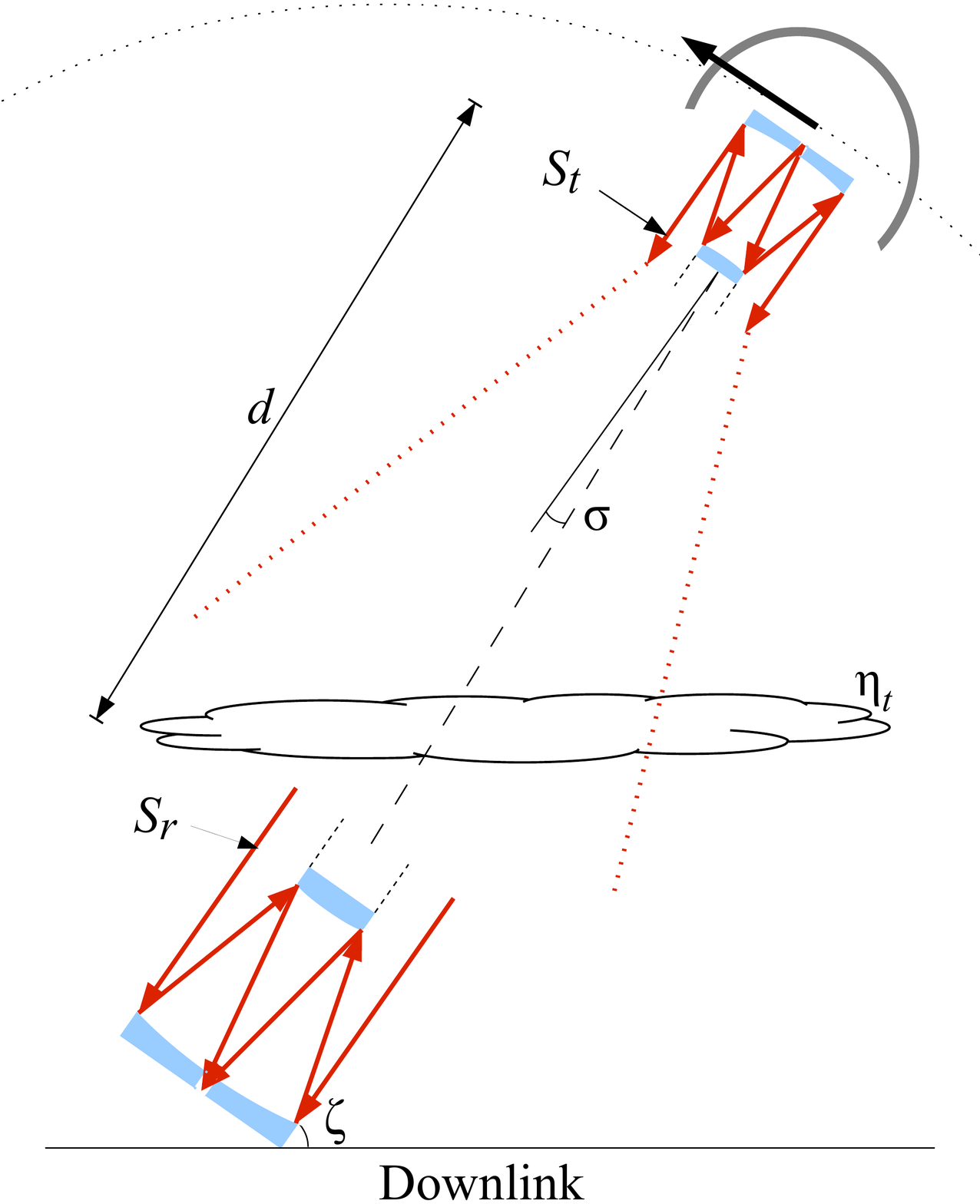}
\hspace{0.05\linewidth}
  \includegraphics[width=0.46\linewidth]{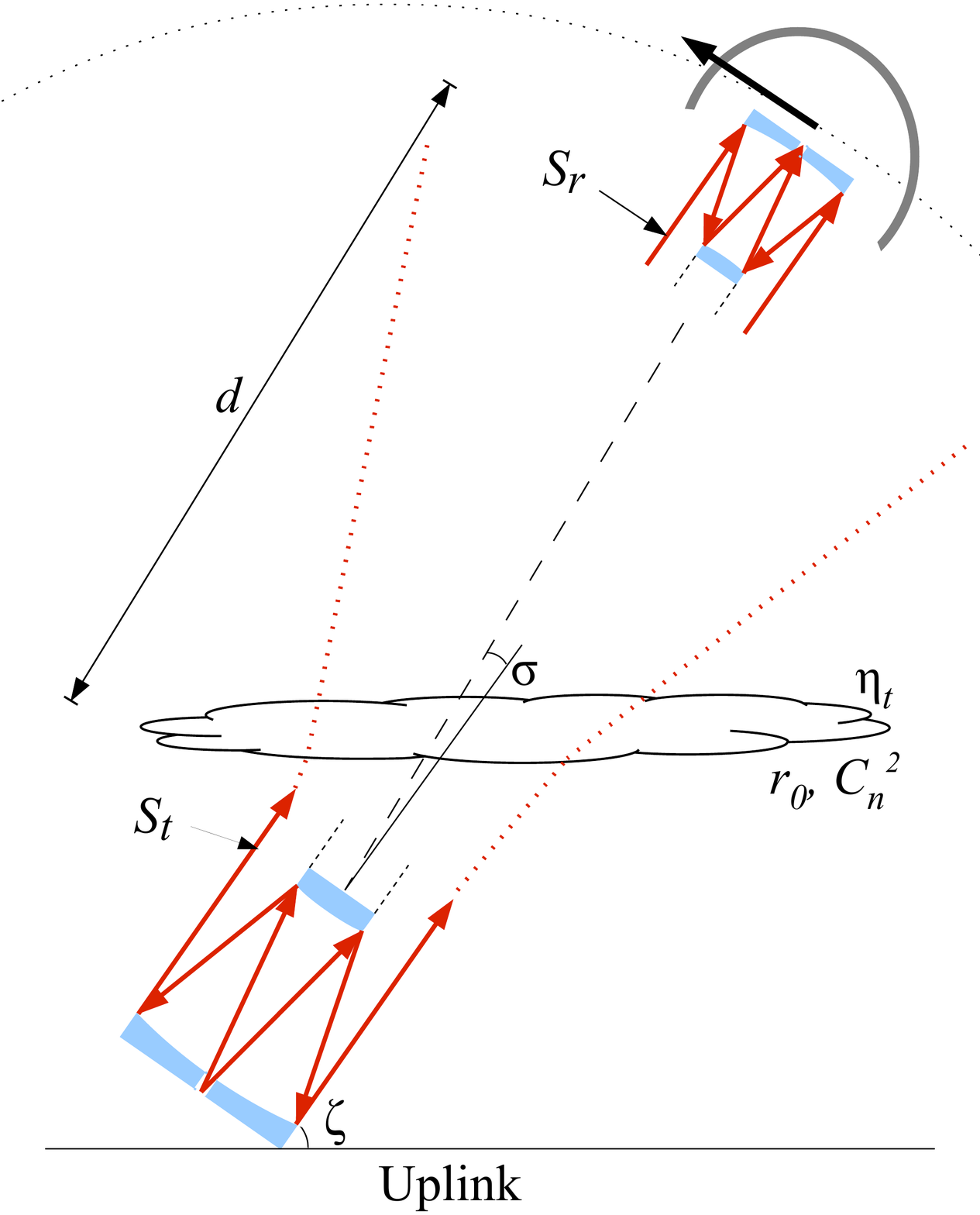}
  \caption[Link sketch]{Sketch of the downlink (left) and uplink (right) transmission from a satellite. The variables follow those used in the text: $S_\text{t}$ is the transmitter's surface, $S_\text{r}$ is the receiver's surface, $\sigma$ is the pointing error, $d$ is the distance from the satellite to the ground station, and $\zeta$ is the elevation angle from ground. $\eta_\text{t}$ is the atmospheric transmittance, and the atmospheric turbulence is characterized by $r_0$, the transverse coherence length, and $C_\text{n}^2(z)$, the refractive-index structure constant.}
  \label{fig.link_sketch}
\end{figure}

To determine the feasibility of satellite QKD, a numerical model was developed to calculate the total loss using realistic parameters (Figure~\ref{fig.link_sketch}). The loss from diffraction is calculated using the Rayleigh-Sommerfeld diffraction~\cite{G96}:
 \begin{equation}
I_1(\vec{v}) = \left|\frac{d^2}{\lambda^2} \int\!\!\!\int_{S_\text{t}} \frac{\sqrt{I_0(\vec{v}')}}{|\vec{v} - \vec{v}'|^2} \exp \left(\frac{2i\pi|\vec{v} - \vec{v}'|}{\lambda}\right) d x d y \right|^2,
\label{equ.diffraction}
\end{equation}
where $I$ is the intensity at the receiver, $\vec{v}$ is the location at the receiver, $\vec{v}'$ is the location at the transmitter, and we integrate over the surface of the transmitter, $S_\text{t}$. $I_0$ is the intensity at the transmitter, $\lambda$ is the beam's wavelength, and $d$ is the distance from the satellite to the ground station. By using the fact that the beam's profile has circular symmetry, we need only calculate the intensity at $y=0$ (or $x=0$) and determine $I(r)$ where $r$ is the radius from the centre of the beam.

The loss from the pointing error is calculated by first determining the distribution over time of the beam centre at the receiver. We assume a two-dimensional Gaussian distribution of pointing, given by
\begin{equation}
g(r)=\frac{1}{2\pi\sigma^2}\exp\left(-\frac{r^{2}}{2\sigma^{2}}\right),
\label{eq.pointing}
\end{equation}
where $\sigma$ is the standard deviation caused by pointing error. The beam profile with pointing error $I_2(\vec{v})$ is then obtained by taking a two-dimensional convolution~\cite{B09} of the beam after diffraction with the distribution of the receiver due to pointing error:
\begin{equation}
I_2(\vec{v})=(I_1 * g)(r, \theta) = \int_0^{2\pi} d\theta' \int_{-\infty}^{\infty} I_1(r')g(r-r') d r'.
\label{eq.convolution}
\end{equation}

For an uplink, we must also incorporate the broadening of the beam due to atmospheric turbulence, on a time scale of order 10--100~ms. Averaging over a period significantly longer than this time scale results in a Gaussian distribution. From the Hufnagel-Valley model of atmospheric turbulence~\cite{AS93,BTDNV07} the waist $w$ of the distribution at the receiver is
\begin{equation}
w_2 = \frac{2\sqrt{2}d \lambda}{\pi r_0},
\label{eq.turbulence_waist}
\end{equation}
where $r_0$ is the transverse coherence length,
\begin{equation}
r_0 = \left[1.46\sec(\frac{\pi}{2} - \zeta) \left(\frac{2\pi}{\lambda}\right)^2\int_{0}^{h} C_\text{n}^2 (z)\left(1-\frac{z}{h}\right)^{\frac{5}{3}} d z\right]^{-\frac{3}{5}},
\label{eq.coherence_length}
\end{equation}
with $\zeta$ the elevation angle of the satellite from the ground, $h$ the altitude of the receiver, and $C_\text{n}^2(z)$ the refractive-index structure constant, given by
\begin{equation}
C_\text{n}^2(z) = 0.005\,94(v/27)^2(z\cdot 10^{-5})^{10}e^{-z/1000}+2.7\cdot 10^{-16}e^{-z/1500}+Ae^{-z/100}.
\label{eq.refractive_index_struct}
\end{equation}
The parameters $v$ and $A$ depend on the atmospheric conditions. For this paper we used the values $A = 1.7\times10^{-14}$~m$^{-\frac{2}{3}}$ and $v = 21$~m/s, which are typical values at sea level during nighttime~\cite{TOV06}. Once we have the waist of the Gaussian distribution we can use a two-dimensional convolution with the intensity profile $I_2(\vec{v})$ that includes pointing error to obtain the overall beam profile $I_3(\vec{v})$.

Once we have determined the profile of the beam at the receiver, in either the uplink or downlink (where $I_3 = I_2$), we then integrate it over the receiving area to obtain the received optical power ($P$),
\begin{equation}
P = \int\!\!\!\int_{S_\text{r}} I_3(\vec{v}) d x d y,
\label{eq.power_int}
\end{equation}
where $S_\text{r}$ is the surface of the receiver. The resulting power is proportional to the average number of detected photons. 

The atmospheric transmission and the detector efficiency is then added by multiplying the power with the detector efficiency $\eta_\text{d}$ and atmospheric transmittance $\eta_\text{t}$ at the transmitting wavelength, modelled using MODTRAN~5 (Figure~\ref{fig.AtmTrans})~\cite{Mod5}. In this paper we used a sea-level rural atmosphere. The MODTRAN parameters we used are listed in Appendix~\ref{app.MODTRAN_parameters}. Finally the ratio of the final power to the initial power ($P_0$) is converted into loss in dB and a 3~dB loss is added to account for losses in the polarization analyser and coupling loss into the fiber to the detectors.
\begin{equation}
  L = -10 \log_{10} \left(\frac{\eta_\text{t}\eta_\text{d}P}{P_0}\right) + 3.
\label{eq.final_loss}
\end{equation}

\section{Estimating background light}\label{app.BackgroundCalcs}

\begin{figure}[tbph]
\centering
\includegraphics[width=\linewidth]{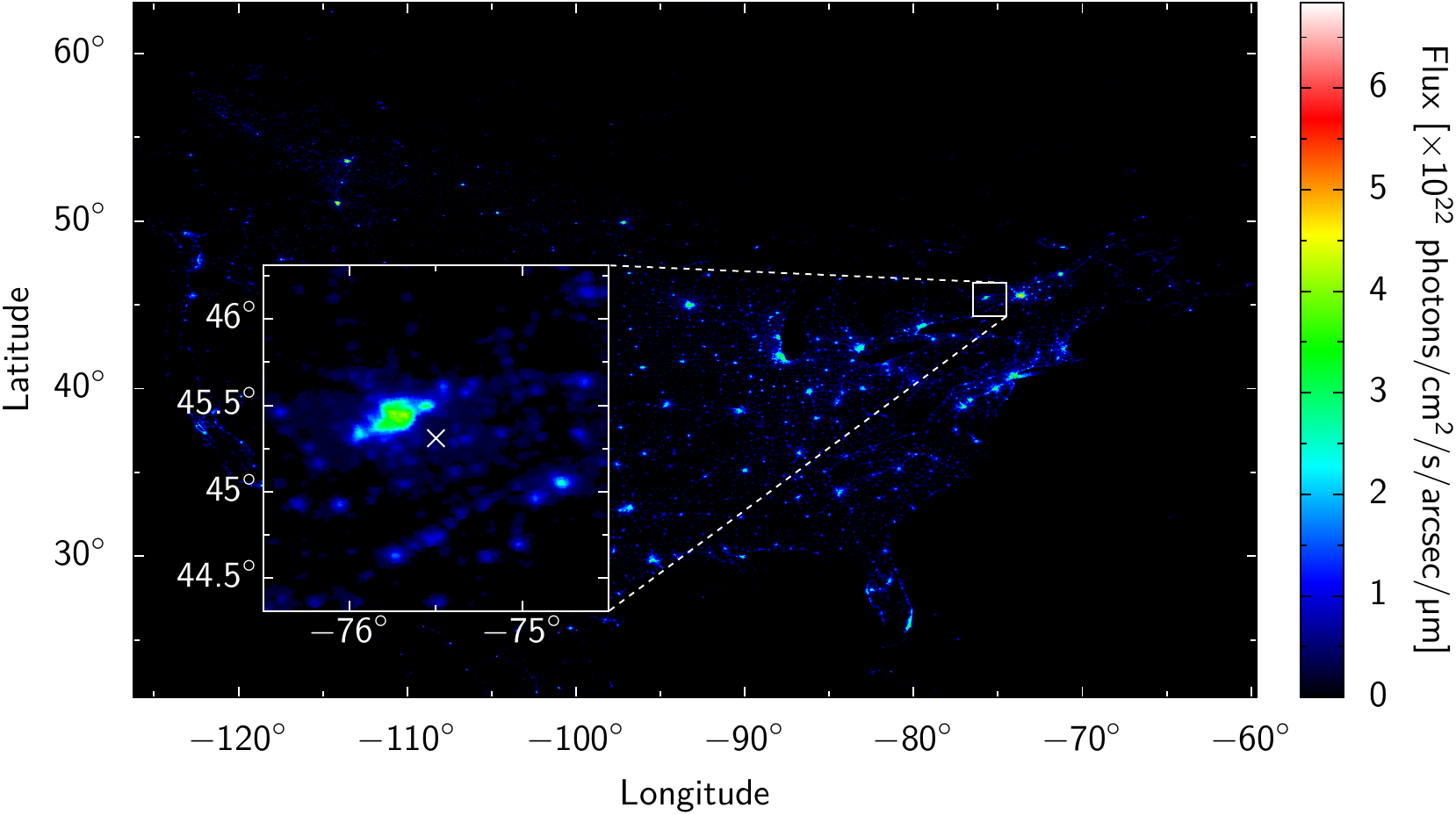}
\caption[Light pollution from human activities in North America]{Light pollution from human activities in North America, data from World Atlas of Artificial Sky Brightness (Ref.~\cite{CFE01}). The inset shows a closer view of the location of the simulated ground site, marked with a cross, approximately 20~km outside Ottawa.} \label{fig.global_northam}
\end{figure}

Background light originates from both natural and artificial sources. Natural sources, such as the Sun, Moon and stars, have been thoroughly characterized elsewhere. Artificial sources consist largely of light pollution from human activities. Shown in Figure~\ref{fig.global_northam}, this light pollution was characterized over the surface of the Earth during 1996 and 1997 by the Defence Meteorological Satellite Program's (DMSP) Operational Linescan System (OLS)~\cite{EBDBSK99}. We utilize this light pollution data for our calculations.

\subsection{Downlink}

A receiver which is located at a ground station will receive light from bright objects in the sky (e.g.\ stars) and from scattered light originally emitted by human activities. Astronomers have characterized the natural brightness of the night sky at different locations~\cite{KSTTGGKIC87, BE98, P08}. Theoretical models and computer algorithms to predict the night sky brightness also exist~\cite{B10}. The contribution of the Moon to the night sky brightness has been studied~\cite{KS91}.

The nighttime sky brightness due to light pollution can be calculated from the DMSP-OLS data which specifies the measured upward flux emitted at a given ground location~\cite{CFEB00}. Since the ground-based telescope is pointing towards a satellite, it can receive background counts from the Sun's light reflecting off the satellite and into the telescope. For our case we will assume the satellite is not illuminated by the Sun. (For our orbit, this will be valid for all nighttime passes at most ground station locations.) The overall sum of these contributions amounts to the total number of background counts per second:
\begin{equation}
N_\text{tot} = \frac{1}{E_{\nu_0}} \left\{ (H_\text{nat} + H_\text{art})\times\pi(\text{FOV})^{2}\times\pi r^{2}\right\}, 
\end{equation}
where $\nu_0$ is the mean frequency of the laser emitted towards the receiver, $E_{\nu_\text{0}}$ is the energy of a single photon of frequency $\nu_\text{0}$, $r$ is the telescope's radius, $\text{FOV}$ is the angular field of view of the receiving telescope, and $H_\text{nat}$ and $H_\text{art}$ are the natural and artificial sky brightness, respectively, listed in~\cite{KS91, CFEB00, B10}. 

\subsection{Uplink}

A receiver located on a satellite is directed towards the surface of the Earth, and thus receives light from natural and artificial sources. The primary natural light source is the light from the Sun reflected first off the Moon and then off the surface of the Earth towards the satellite. Reflected light from other natural sources is negligible. To obtain the background count rates from natural sources, we begin by considering Planck's law for blackbody radiation~\cite{B99},
\begin{equation}
I(\nu, T) = \frac{2h\nu^{3}}{c^{2}}\left(\exp(h\nu/kT)- 1\right)^{-1} \label{eq:PlancksLaw},
\end{equation}
where $\nu$ is the frequency of the emitted radiation, $h$ is the Planck constant, $c$ is the speed of light, $k$ is the Boltzmann constant, and $T$ is the temperature of the emitter. Using the equation at temperatures equal to that of the surface of the Sun (about 5778~K on average~\cite{B99}) gives an accurate estimate of the spectrum of emitted light. The average Earth albedo (quantifying how strongly its surface reflects light) is 30\%~\cite{B99}. The Moon's albedo depends primarily on the lunar phase. We used empirical data to obtain the Moon's albedo at a certain lunar phase. We assume Lambertian diffusion~\cite{BTDNV07}, i.e.\ the radiance of reflected light is independent of angle. Consequently, the number of photons reflected by the Moon $N_\text{M}$ is given by
\begin{equation}
N_\text{Moon} = a_\text{Moon}\frac{I\left(\nu_0, T_\text{Sun}\right)}{E_{\nu_0}}\pi R_\text{Moon}^{2},
\end{equation}
where $a_\text{Moon}$ is the Moon's albedo, $T_\text{Sun}$ is the Sun's temperature and $R_\text{Moon}$ is the Moon's radius. If the Moon is at normal incidence, the solid angle to the area on Earth $\Lambda$ seen from the Moon is $\Lambda/d_\text{EM}^{2}$, where $d_\text{EM}$ is the distance between the Earth and the Moon. Consequently, the number of background photons reaching the telescope after Lambertian reflection from the surface of the Earth is
\begin{equation}
N_\text{Sun} = e\left[a_\text{Earth}N_\text{Moon}\left(\frac{\Lambda}{d_\text{EM}^{2}}\right)\Omega\right],
\label{eq:SunContribution}
\end{equation}
where $e$ is the extinction coefficient of the atmosphere, $a_\text{Earth}$ is the Earth's albedo and $\Omega$ is the solid angle from which the telescope can be seen from Earth. $e$ takes into account the traversal of light through the atmosphere twice: First, light reflected from the Moon reaches the surface of the Earth. Then, this light is reflected into the receiving telescope.

We use the DMSP-OLS data to obtain the number of background counts due to light pollution~\cite{EBDBSK99}. The DMSP-OLS data takes the form of a high-resolution image. The pixel coordinates of this image correspond to physical locations on the surface of the Earth, and the pixel values denote the nighttime radiance. This allows us to obtain the average radiance $\bar{L}$ emitted by a certain location due to nighttime activities. We can then directly obtain light pollution emitted into the receiver:
\begin{equation}
N_\text{night} = e_\text{n}\left(\frac{\bar{L}}{E_{\nu_\text{0}}}\Lambda\Omega\right), \label{eq:nightcounts}
\end{equation}
where $e_\text{n}$ is the extinction coefficient due to light traversing the atmosphere to reach the satellite's receiver.

We obtain the total number of background counts by summing the contributions from all sources:
\begin{equation}
N_\text{BG} = N_\text{Sun} + N_\text{night}. \label{eq:totalsat}
\end{equation}
One caveat with this result is that (\ref{eq:totalsat}), as we calculated, is necessarily an approximation based on data taken almost a decade ago, and its accuracy varies seasonally due to changes in composition of the atmosphere. In addition, we assume that the radiance $\bar{L}$ is constant at all frequencies because there is little data on the composition of the types of lamps used in a certain region. Some information about the composition of lighting types is expected to arrive in the future when the Nightsat mission becomes operational~\cite{ECPASSNLRSWE07}. (\ref{eq:totalsat}) is nevertheless accurate enough to give a reasonable estimate of the expected magnitude of background counts. The background is then used, along with the loss, in the quantum optics simulations described in Section~\ref{QuantOpt} to estimate the QBER, raw key rate, and final secure key length.

\section{Simulating photonic quantum communication}\label{app.Detailed_QOsim}

\subsection{QKD with an entangled photon source}\label{app.QKD.ent}

To simulate entanglement, a model of a spontaneous parametric down-conversion (SPDC) source is used to produce correlated photons, which are then permuted to introduce entanglement. The squeezing operator for the down-conversion process is
\begin{equation}
S(\varepsilon) = \exp{\left(\varepsilon(a^\dagger_1 a^\dagger_2 - a_1 a_2)\right)},
 \label{eqn.Hspdc}
\end{equation}
where $a^\dagger_m$ and $a_m$ are the creation and annihilation operators respectively for mode $m$, and $\varepsilon$ contains the pump power and probability of down-conversion. With this operator applied to a two-mode vacuum, photons in modes 1 and 2 are always created in pairs. However, as for a real SPDC source, multiple uncorrelated photon pairs are also created simultaneously,  leading to errors. This model applies to both a pulsed or continuous wave pumping scheme, with the former having detection probabilities defined per pulse, and the latter per coincident detection window of Alice and Bob's detectors.

The generated two-photon state, initially of the form $\vert HH \rangle$ (with each $H$ inside the ket indicating a horizontally polarized photon; $V$, vertically polarized), is permuted to construct an entangled state: $\vert \Psi^-\rangle = \frac{1}{\sqrt2}\left(\vert HV \rangle - \vert VH \rangle \right)$. The results obtained in this paper do not depend on the specific maximally-entangled bipartite qubit state chosen. In modern SPDC sources of entangled photon pairs,  the entangled state is created directly, but this permutation trick achieves the same results in a way that is computationally much simpler.

The maximally entangled state has visibility $V=100\%$, where $V$ is defined as 
\begin{equation}
V=\frac{C_\text{E}-C_\text{U}}{C_\text{E}+C_\text{U}}.
\label{eqn.V}
\end{equation}
Here $C_\text{E}$ is the coincident photon counts detected possessing the expected polarization (i.e., anti-correlated polarizations for the $\vert \Psi^- \rangle$ state), and $C_\text{U}$ is the coincident counts detected with unexpected (correlated) polarization.

Next, we simulate realistic degradation in entanglement visibility due to polarization misalignment between source and receiver. We apply a unitary rotation to Bob's photon (but not Alice's), leading to some ``unexpected'' coincident counts and hence to degraded visibility. We chose an entanglement visibility of 98\%, a pessimistic case, as better alignments are readily achieved experimentally~\cite{EMLW09}.

After traversing the quantum channel, the photon state is detected. The total loss of the channel, comprising link efficiency, receiver optics efficiency, and detector efficiency, is incorporated at this point as an overall efficiency of each detector, i.e.\ $\eta_\text{total}=\eta_\text{d}\eta_\text{optics}\eta_\text{link}$ (where $\eta_\text{d}$ is the intrinsic efficiency of the detector at the relevant wavelength, $\eta_\text{optics}$ is the efficiency of Bob's optics, and $\eta_\text{link}$ is the efficiency of the channel between Alice's and Bob's telescopes). We model our detectors as bucket detectors, producing a classical signal with non-zero probability if one or more photons enters the device. Our simulated detectors do not resolve photon number, in line with current commercial detectors. We also include a realistic dark count rate of 20~darks/s per detector.

\begin{figure}[tb]
\centering
\includegraphics[width=0.7\linewidth]{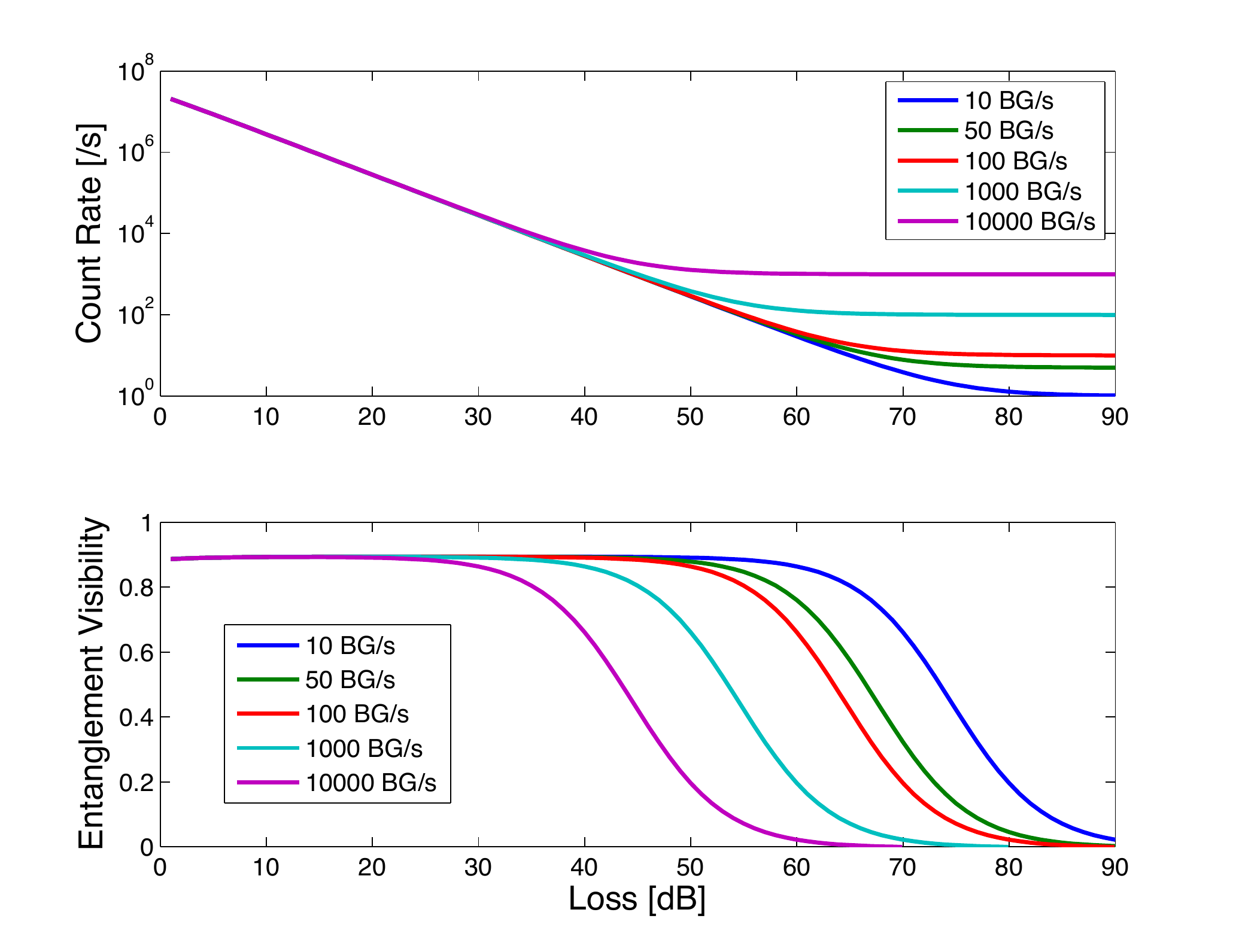}
\caption[Entangled photon visibility and count rate]{Entangled photon visibility and coincidence count rate as functions of loss in Bob's channel, for various background count rates (BG/s). The entangled source operates at a pair production rate of 100~MHz; double-pair emissions and the slight polarization misalignment lead to imperfect entanglement visibility at low loss. At high loss, the visibility goes to zero, and the count rate asymptotes to the product of background count rate, the number of detectors (four), the ratio of the detection window (1.0~ns) to repetition period (10~ns), and Alice's detection efficiency (0.25, i.e.\ only background counts that arrive in coincidence with a photon detected by Alice are included in the rate).} \label{fig.vis_loss_entanglement}
\end{figure}

Due to noise in the detectors, multi-pair emission, and the slight polarization misalignment, the photon will no longer be perfectly entangled when it arrives, leading to an entanglement visibility $V$ less than 100\%. This visibility (or equivalently, the quantum bit error ratio, QBER, $1 - V/2$), along with count rates (detector events), is extracted for use in feasibility calculations. Examples of these visibilities and count rates for an entangled photon source are shown in Figure~\ref{fig.vis_loss_entanglement}. Where ideally we would like to possess both a high count rate and a low double-pair emission probability, there exists a trade-off between these two properties as an increased count rate leads to increased double-pair probability, and two photon pairs emitted simultaneously will be uncorrelated, leading to reduced entanglement visibility. In our simulations we use $\varepsilon=0.22$ (the strength of the SPDC operator in (\ref{eqn.Hspdc})), corresponding to an average number of pairs per pulse of 0.1~\cite{MFL07}.

Finally, we calculate a secure key rate based on the count rate and entanglement visibility, following~\cite{PhysRevLett.100.200501}. The final key rate per detected coincident pair is
\begin{equation}
 R = q\left[1-H_2(E+\xi)-f(E)H_2(E)-\Delta/N\right],
 \label{eqn.QKDEnt}
\end{equation}
where $q=1/2$ is the basis reconciliation (sifting) factor, $E$ is the QBER, $\xi$ is a security parameter from~\cite{PhysRevLett.100.200501}, $H_2(x)=-x\log_2(x)-(1-x)\log_2(1-x)$ is the binary entropy function, $f(E)=1.22$ is the error correction efficiency, and $N$ is the length of raw key. The security parameter $\Delta$ is given by $\Delta = 2\log_2 1/[2(\epsilon - \overline{\epsilon} - \epsilon_\text{EC})] + 7\sqrt{N\log_2[2/(\overline{\epsilon} - \overline{\epsilon}')]}$, where $\epsilon$ is the total allowable probability that the final key is insecure, chosen to be $\epsilon = 10^{-9}$, $\epsilon_\text{EC} = 10^{-10}$ is the error correction failure probability, and $\overline{\epsilon}$ and $\overline{\epsilon}'$ can be optimized numerically with the constraint $\epsilon - \epsilon_\text{EC} > \overline{\epsilon} > \overline{\epsilon}' \ge 0$. The key rate in~(\ref{eqn.QKDEnt}), when combined with the output of the link analysis, is the principal figure-of-merit for entanglement-based QKD.
 
\subsection{QKD with a WCP source}\label{app.QKD.WCP}

\begin{figure}[tb]
\centering
\includegraphics[width=0.7\linewidth]{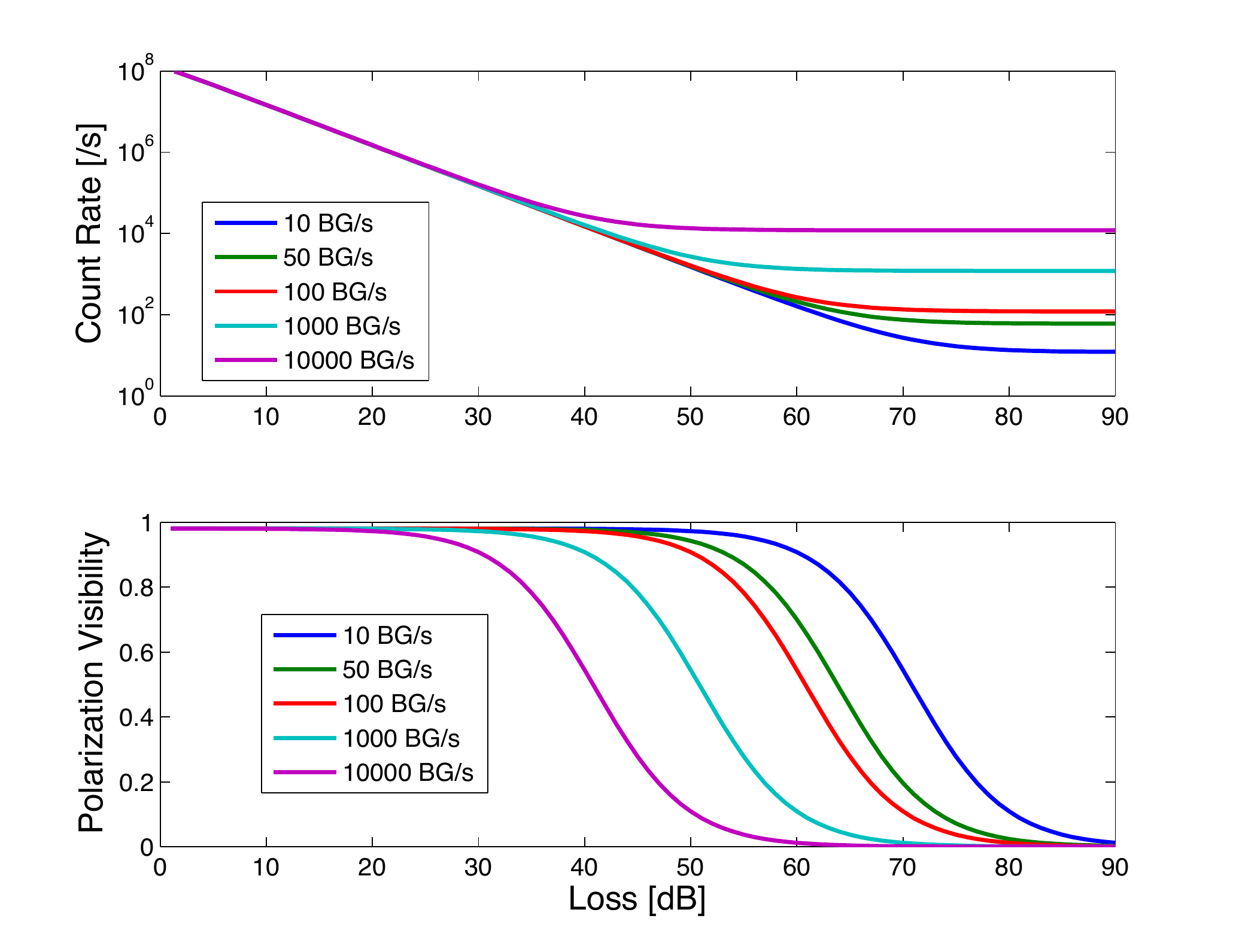}
\caption[WCP polarization visibility and count rate]{Weak coherent pulse photon polarization visibility and count rate as functions of channel loss in Bob's arm, for various background count rates (BG/s). The source operates at a repetition rate of 300~MHz, with an average photon number per pulse of 0.5. The count rate here includes only detections that arrive within 1~ns of an expected laser pulse from Alice. At high loss, the visibility goes to zero, and the count rate asymptotes to the product of background count rate, the number of detectors (four), and the ratio of the detection window (1.0~ns) to the repetition period (3.3~ns).} \label{fig.vis_loss_WCP}
\end{figure}

For QKD using a WCP source, the quantum mechanical description of photon creation is the displacement operator:
\begin{equation}
D(\alpha) = \exp(\alpha a^\dagger - \alpha^* a),
\label{eqn.Hwcp}
\end{equation}
where $|\alpha|^2 = \mu$ is the average photon number per pulse. The displacement operator models the output of a highly attenuated laser (for small $\mu$). As our detectors (and discrete-variable qubit protocols) are insensitive to the laser phase, the displacement operator also describes the incoherent Poissonian photon number distribution assumed in security proofs~\cite{GLLP04}. As above, polarization misalignment, noise, and loss are applied and the photons are measured, giving a count rate and QBER, which can be used in secure key rate calculations. For the unentangled photons of a WCP source, the polarization visibility is defined as 
\begin{equation}
V = \frac{N_\text{E} - N_\text{U}}{N_\text{E} + N_\text{U}}.
\label{eqn.VWCP}
\end{equation}
Here, $N_\text{E}$ is the number of detections with polarization parallel to the state that Alice sent (the expected counts), and $N_\text{U}$ is the number of detections with perpendicular polarization (the unexpected counts). Examples of polarization visibility and count rate for a WCP source are shown in Figure~\ref{fig.vis_loss_WCP}. It is notable that the WCP source is slightly less resilient against background noise than the entangled source (worse visibility at high loss). This is because a noise count must arrive in coincidence with a detection on Alice's side to be considered in an entangled scheme, whereas for a WCP source every noise count that arrives in coincidence with a laser pulse time-slice is accepted.

Because of the Poissonian statistics of photon number in laser pulses, some pulses will possess more than one photon, and thus may be subject to the photon number splitting attack~\cite{1367-2630-4-1-344} where an adversary Eve splits off one photon from the pulse and stores it (in a quantum memory) to measure only after Bob reveals his measurement basis. Eve then measures the stored photon in the same basis, thereby gaining full information about multiphoton pulses in an undetectable manner.

To combat this attack, the decoy pulse method was introduced, wherein Alice changes the average photon number $\mu$ of randomly interspersed pulses (decoys) which are not utilized in generating the secure key. Because Eve cannot know \emph{a priori} whether a given pulse is a signal or decoy, the decoy pulses enforce much stricter bounds on how much information Eve can gain from multiphoton signals, and thus how much privacy amplification must be performed. Using the decoy pulse method, the lower bound on asymptotic key rate per laser pulse is~\cite{PhysRevA.72.012326}
\begin{equation}
R \ge q \{-Q_\mu f(E_\mu) H_2(E_\mu) + Q_1 \left[1-H_2(E_1)\right]\},
\label{eqn.asymp}
\end{equation}
where $q=1/2$ is the basis reconciliation factor, $Q_\mu$ is the signal gain (i.e.\ the ratio of Bob's detections to pulses sent by Alice for average photon number $\mu$), $E_\mu$ is the quantum bit error rate for signal pulses, $f(E_\mu)=1.22$ is the error correction efficiency for practical error correction codes, $H_2(x)$ is the binary entropy function, and $Q_1$ and $E_1$ are the estimated gain and error rate for single photon pulses. The key rate is then the gain of qubit pulses, minus the information leaked from error correction on all signal pulses, minus the privacy amplification on qubit pulses. This key rate is multiplied by the system clock rate or laser pulse rate to obtain secure key bits per second. 

We consider the one-decoy protocol from~\cite{PhysRevA.72.012326}. In this protocol, Alice randomly chooses to send either a signal pulse with average photon number $\mu$ or a decoy pulse with average photon number $\nu < \mu$.  $Q_\mu$ and $Q_\nu$ are the gains of signal and decoy pulses respectively, and $Q_1$ and $E_1$ can be estimated from measurable quantities as
\begin{eqnarray}
Q_1&=&\frac{\mu^2 e^{-\mu}}{\mu\nu-\nu^2}\left(Q_\nu e^\nu-Q_\mu e^\mu \frac{\nu^2}{\mu^2}\right),\\
E_1&=&\frac{E_\nu Q_\nu}{Q_1}.
\end{eqnarray}

Equation~(\ref{eqn.asymp}) quantifies the asymptotic key rate over an infinitely long key generation time, but in order to generate a key on a single pass, finite size statistics of the observed parameters must be incorporated. A rigorous finite size analysis for WCP QKD is still incomplete, but an \emph{ad hoc} version can be developed as follows. Firstly, a factor $N_\mu/(N_\mu+N_\nu)$ must multiply the key rate since only signal pulses contribute to the final key, where $N_\mu$ ($N_\nu$) is the number of Bob's received signal (decoy) counts. We assume no bits are required for error rate estimation as the error correction algorithm identifies the number of errors precisely. Secondly, the parameters used to calculate $Q_1$ and $E_1$ must be modified to account for the chance of statistical fluctuation in their values. Specifically, 10 standard deviations~\cite{Sun2009Decoy-st} are incorporated into $Q_\mu$, $Q_\nu$, $E_\mu$ and $E_\nu$ such that the worst case scenario is considered, and the probability that the actual values fall outside this range is less than $10^{-25}$. Finally, the parameter $\Delta$ found in~(\ref{eqn.QKDEnt}) must also be added, with $N=N_\mu$.

The final key rate for a WCP source with finite size effects is then lower-bounded by
\begin{equation}
R \ge q\frac{N_\mu}{N_\mu+N_\nu} \{-Q_\mu f(E_\mu) H_2(E_\mu) + Q_1 \left[1-H_2(E_1)\right]-Q_\mu\Delta/N_\mu\}.
\end{equation}
Given the known $\mu$, $\nu$, $N_\mu$, $N_\nu$, $\epsilon$, $\epsilon_\text{EC}$, bounded $Q_\mu$, $Q_\nu$, $E_\mu$, $E_\nu$ and estimated $Q_1$, $E_1$ from our quantum optics simulations, a secure key rate $R$ can be calculated for each satellite passage. In our simulations we used $\mu=0.5$ and $\nu=0.1$.

\subsection{Bell tests and quantum teleportation}\label{app.Bell}

For the Bell test, one photon of the entangled pair is measured locally at the source, while the other is sent to the satellite. With sufficient photons possessing strong entanglement correlations, the CHSH inequality~\cite{PhysRevLett.23.880} is violated, thereby excluding models of reality which preserve intuitive notions of locality and realism. We take as the measure of a successful Bell test the violation of the CHSH inequality, with certainty of $3\sigma$ (i.e.\ three standard deviations above the classical limit), within the time of a single satellite passage. The CHSH inequality is given by
\begin{equation}
S_\text{CHSH} = \vert E(\phi_\text{A}, \phi_\text{B}) - E(\phi_\text{A}, \phi_\text{B}') \vert + \vert E(\phi_\text{A}', \phi_\text{B}) + E(\phi_\text{A}',\phi_\text{B}') \vert \leq 2,
\label{eqn.CHSH}
\end{equation}
where $E(\phi_\text{A}, \phi_\text{B}) = (N_{++} - N_{+-} - N_{-+} + N_{--})/N_\text{total}$ is the joint correlation at Alice and Bob's measurement angles $\phi_A$ and $\phi_B$ respectively. The $N_{ij}$ are the number of coincident counts between Alice's $i$ and Bob's $j$ detectors, where $i,j \in \{+,-\}$. The simulation begins as with entanglement QKD with the production of entangled photons via~(\ref{eqn.Hwcp}) and channel transmission loss; the only difference is the angles at which the detection polarisers are set. From the simulated detection probabilities, we find the value of the CHSH parameter and its uncertainty, and thereby determine if the inequality is violated by at least 3$\sigma$.

Quantum teleportation simulations proceed similarly, except that as part of the teleportation protocol, Bob supplies an additional photon that is interfered with one of the photons of the entangled pair (generated by Alice) on a beam splitter. This input photon has an arbitrary polarization state, which is subsequently transferred to the distant entangled photon through the teleportation process. Here we consider an entangled photon state generated from SPDC as above, with one of the entangled photons sent over the long-distance link to Bob. Bob possesses a WCP source, supplying the additional input photon which interferes with the entangled photon arriving from the SPDC source. A joint measurement is performed on these two modes, teleporting the WCP polarization state onto the other photon of the entangled pair that Alice produced. This measurement requires the two photons to be indistinguishable in all modes (spatial, timing, spectral). This mode overlap is achievable but can be challenging in real world implementation as it requires very precise alignment and filtering. In the interest of finding the limit of quantum teleportation to a satellite with current technologies, we focus on the performance with perfect mode overlap.

In the case of a downlink, Bob is located on the ground while the satellite provides entangled photon pairs, acting as Alice. In an uplink, the situation is reversed, with the entangled photon source/Alice on the ground and Bob on the satellite.  The hallmark of teleportation is that the original photon state is destroyed as it is transferred to the new photon. Thus the visibility of the final state should be higher than that possible with an optimal quantum cloner, i.e.\ $V > V_\text{cloner} = 2/3$~\cite{BDEFMS98}, again with $3\sigma$ certainty. Here $V$ is defined as in~(\ref{eqn.VWCP}) with $N_\text{E}$ being the number of detections with polarization parallel to Bob's WCP input polarization, and $N_\text{U}$ being the number of detections perpendicular.

\begin{figure}[tb]
\centering
\includegraphics[width=0.7\linewidth]{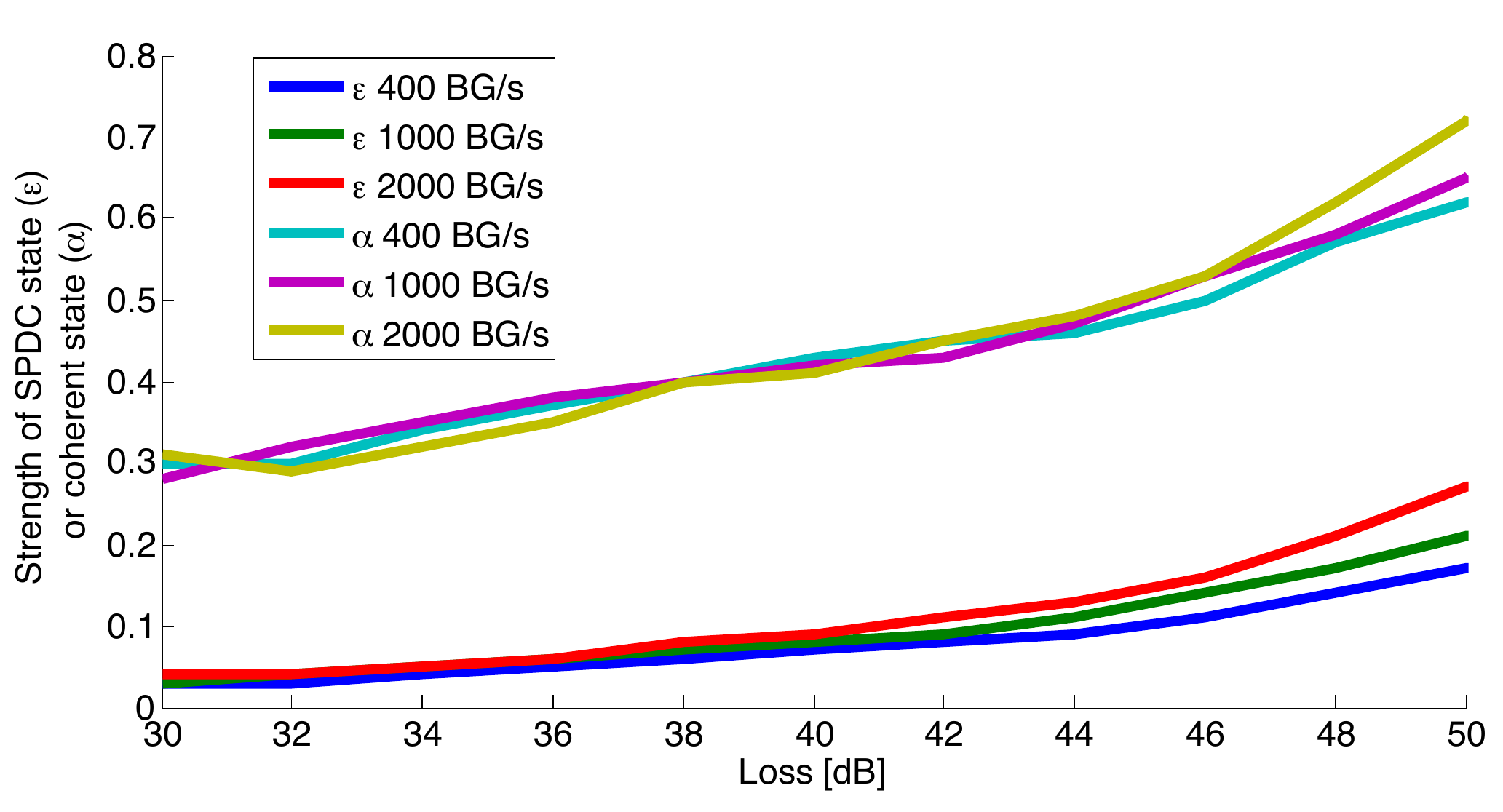}
\caption[Optimized parameters for teleportation]{Optimized parameters $\varepsilon$ and $\alpha$ for teleportation as a functions of channel loss in Bob's arm, for various background counts per second per detector (BG/s). Here $\alpha$ is the strength of the coherent state, (\ref{eqn.Hwcp}), whose polarization is teleported, with average photon number $|\alpha|^2$. Similarly, $\varepsilon$ is the strength of the entangled photon state from SPDC, (\ref{eqn.Hspdc}), with average number of pairs per pulse $2\sinh^2\varepsilon$. As the loss increases, the optimal value of both parameters tends to also increase, in order to counteract the reduction in signal-to-noise at the detectors.} \label{fig.Param_telep}
\end{figure}

We also perform numerical optimizations to determine the strengths of the SPDC ($\varepsilon$) and WCP ($\alpha$) states most suitable for teleportation. Figure~\ref{fig.Param_telep} shows that the optimal value of both parameters tends to increase with loss, with $\varepsilon$ always greater than $\alpha$. Points where the optimal $\varepsilon$ decreases with loss and the crossing of the curves are most likely due to imprecision in the numerical simulation. For our simulations, two representative sets of parameters were chosen: $\varepsilon=0.41$ and $\alpha=0.07$ for downlink simulations (where the usable part of a pass is typically around 35--40~dB of total loss with 180--200 background counts) and $\varepsilon=0.55$ and $\alpha=0.14$ for uplink simulations (usable part possessing around 45--50~dB loss and 250--750 background counts).

\section{Comprehensive performance analysis}\label{app.Comp_perf}

\begin{figure}[tbp]
  \centering
  \includegraphics[width=0.49\linewidth]{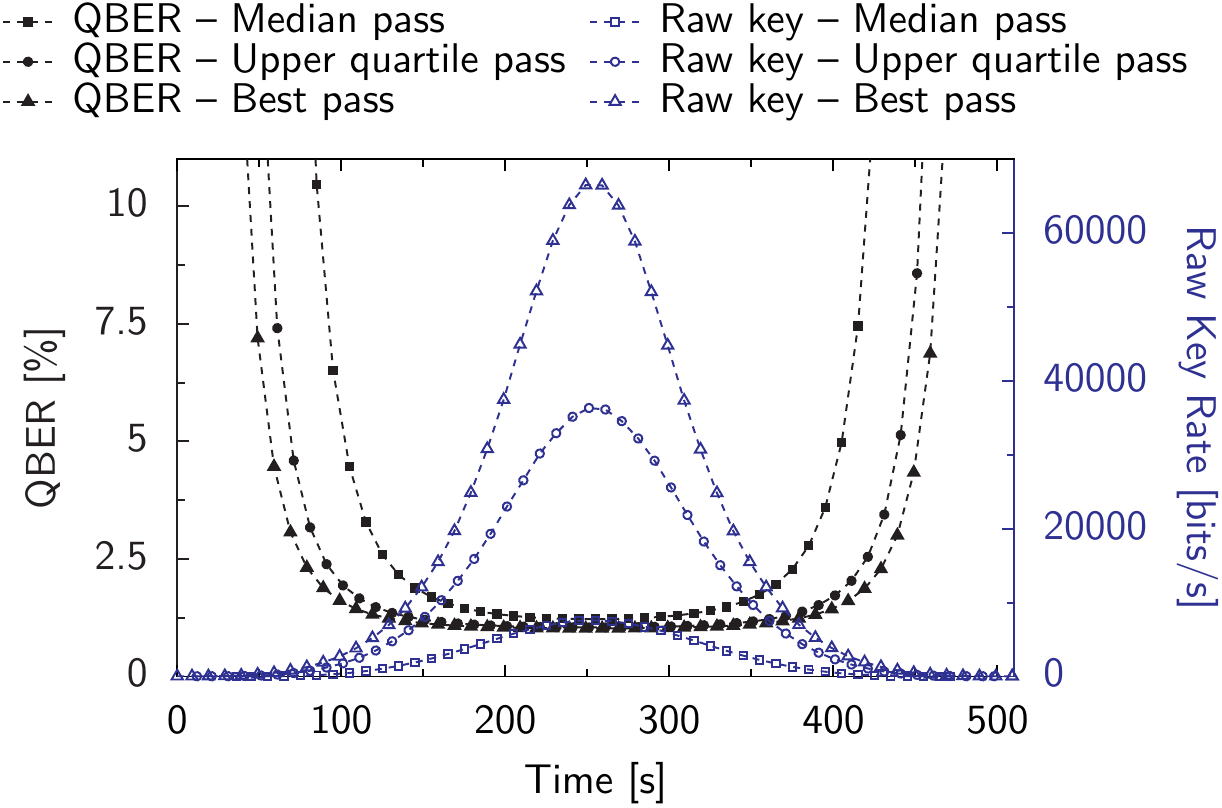}
  \includegraphics[width=0.49\linewidth]{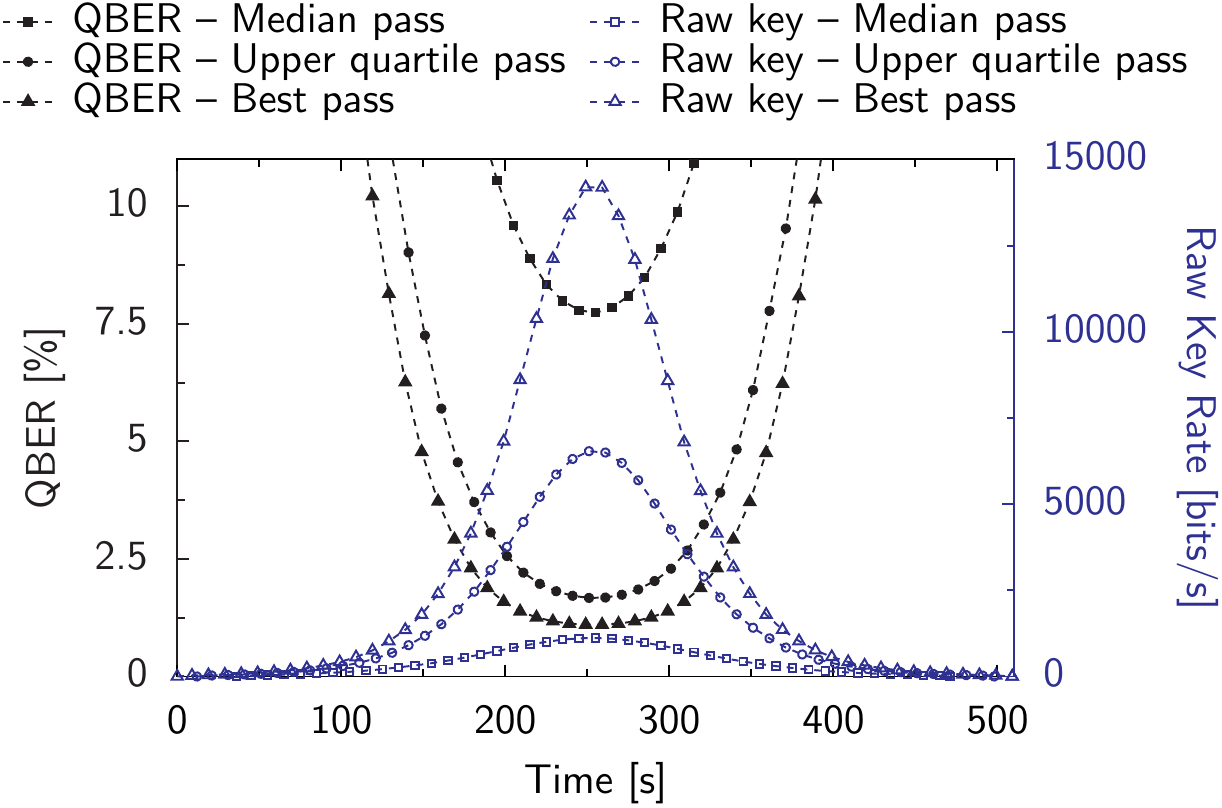}
  \caption[passes]{QBER and raw key rate during the best pass, upper quartile pass, and median pass for a downlink (left) and an uplink (right) utilizing a WCP source. The QBER is significantly higher at low elevations, preventing the generation of secure key from the raw key for most protocols when the QBER is above 11\%. Altitude, wavelength, telescope and atmospheric conditions follow Figure~\ref{fig.loss_background}. Source rate: 300~MHz; detector dark count rate: 20~cps; detection time window: 0.5~ns.}
  \label{fig.QBER_key_WCP}
\end{figure}

\begin{figure}[tbp]
  \centering
  \includegraphics[width=0.49\linewidth]{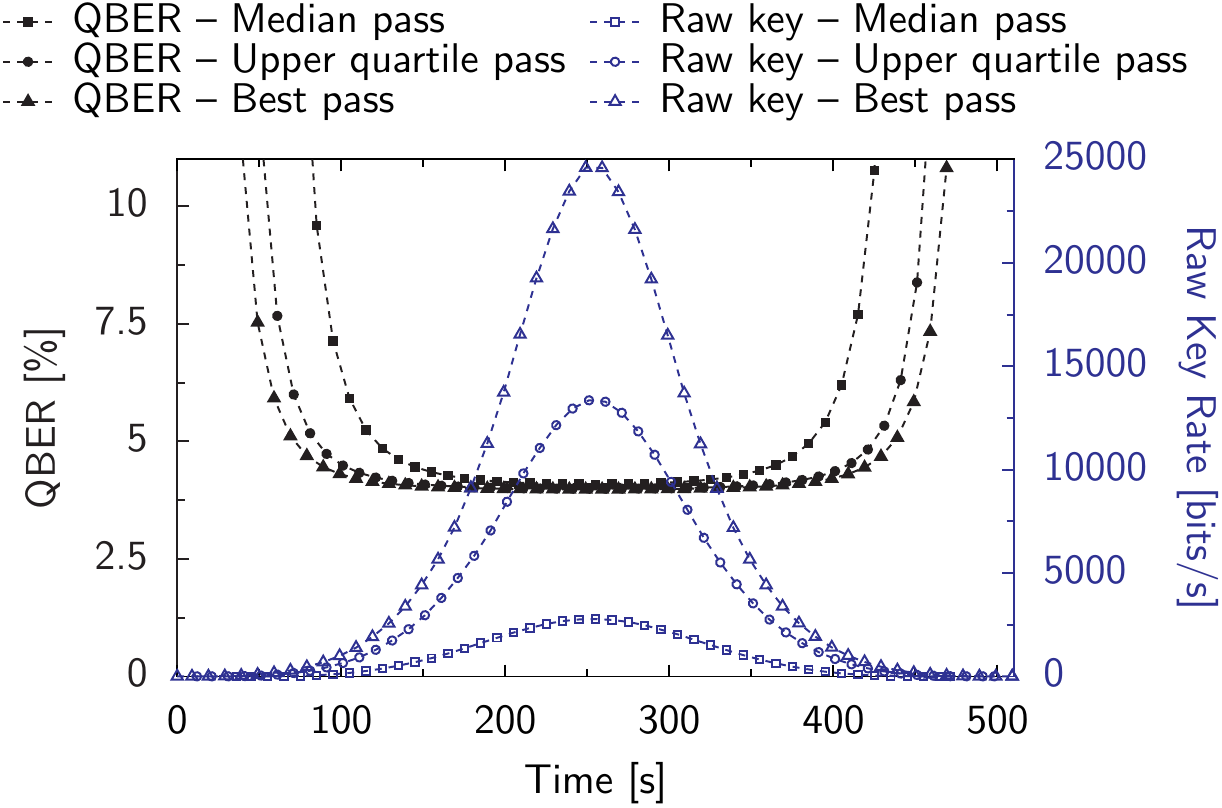}
  \includegraphics[width=0.49\linewidth]{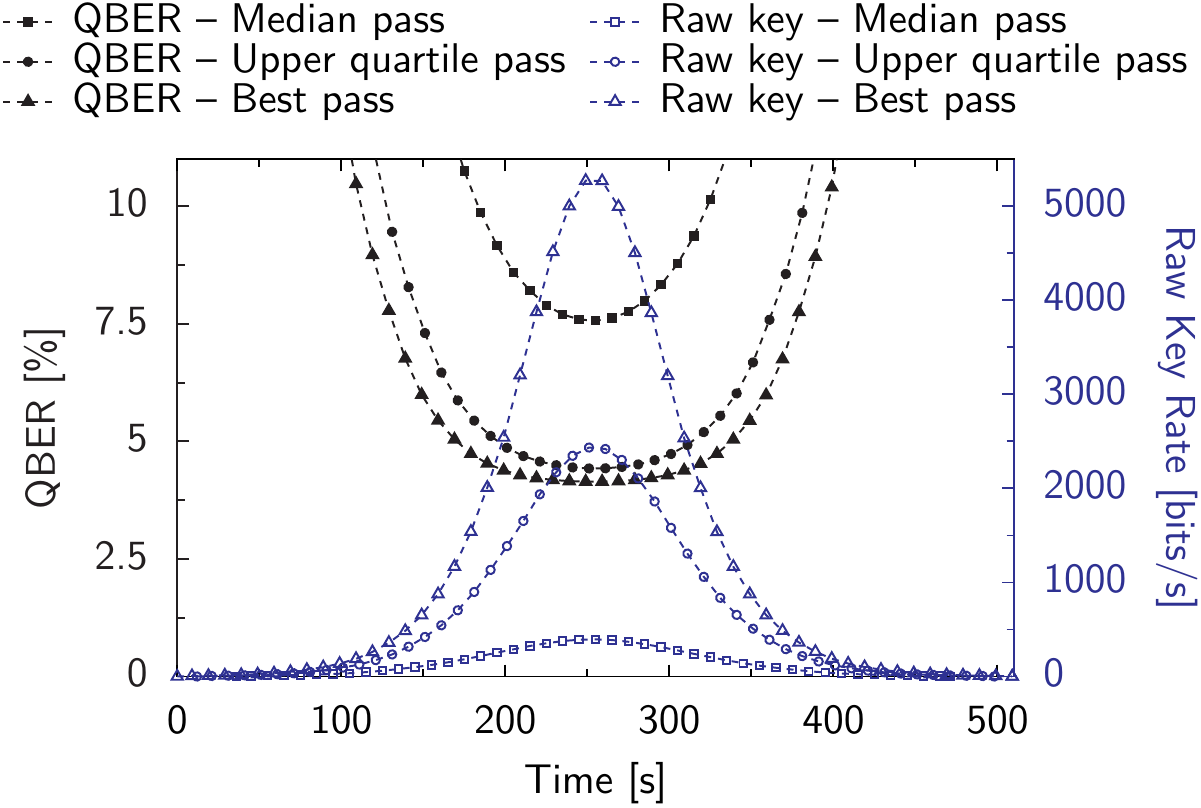}
  \caption[passes]{QBER and raw key rate during the best pass, upper quartile pass, and median pass for a downlink (left) and an uplink (right) utilizing an entangled photon source. The entangled photon source has a higher intrinsic QBER than the WCP source, primarily because of multi-pair emissions and a lower source rate. Altitude, wavelength, telescope and atmospheric conditions follow Figure~\ref{fig.loss_background}. Source rate: 100~MHz; detector dark count rate: 20~cps; detection time window: 0.5~ns.}
  \label{fig.QBER_key_Entangled}
\end{figure}

\begin{figure}[tbp]
  \centering
 \includegraphics[width=0.49\linewidth]{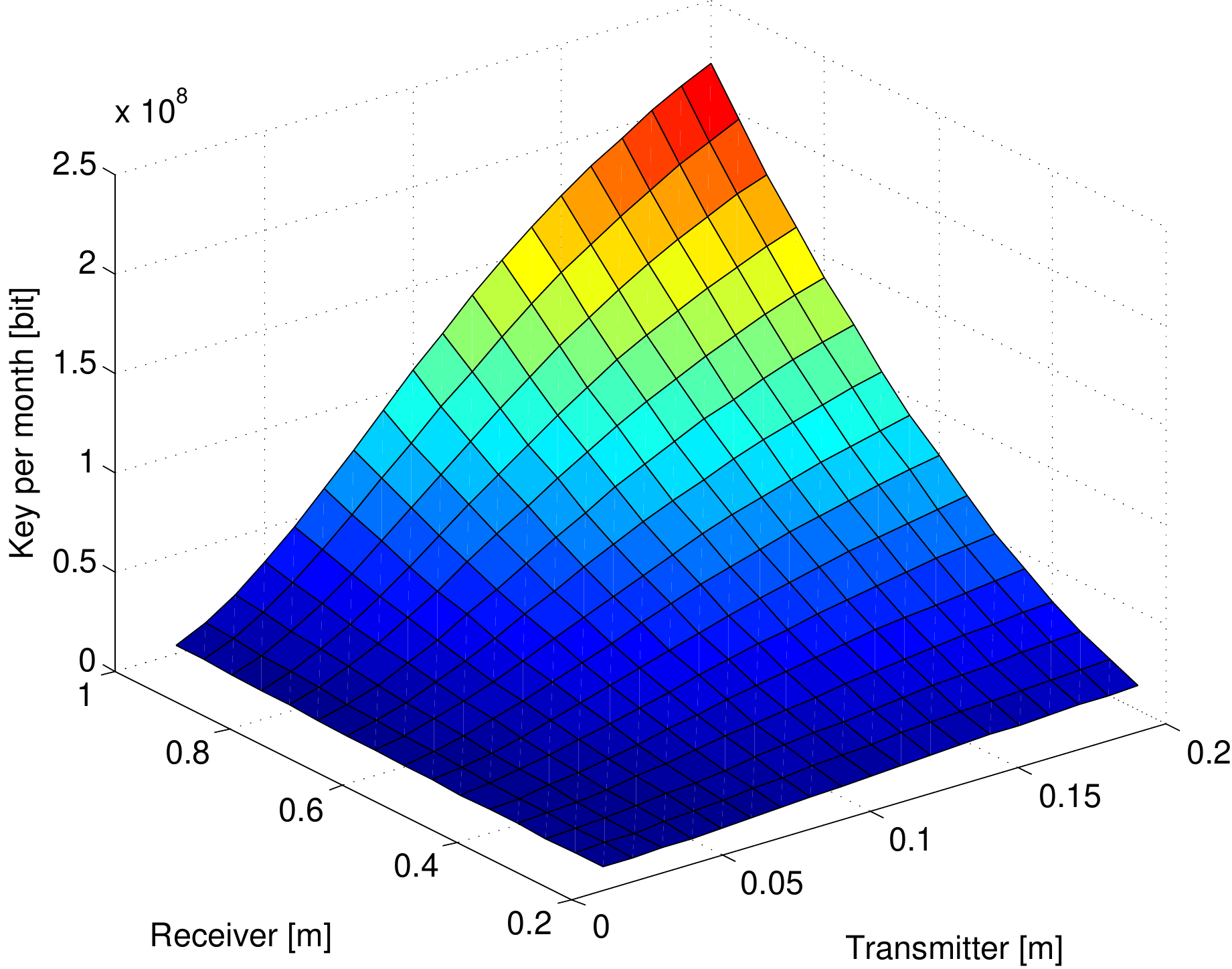}  
 \includegraphics[width=0.49\linewidth]{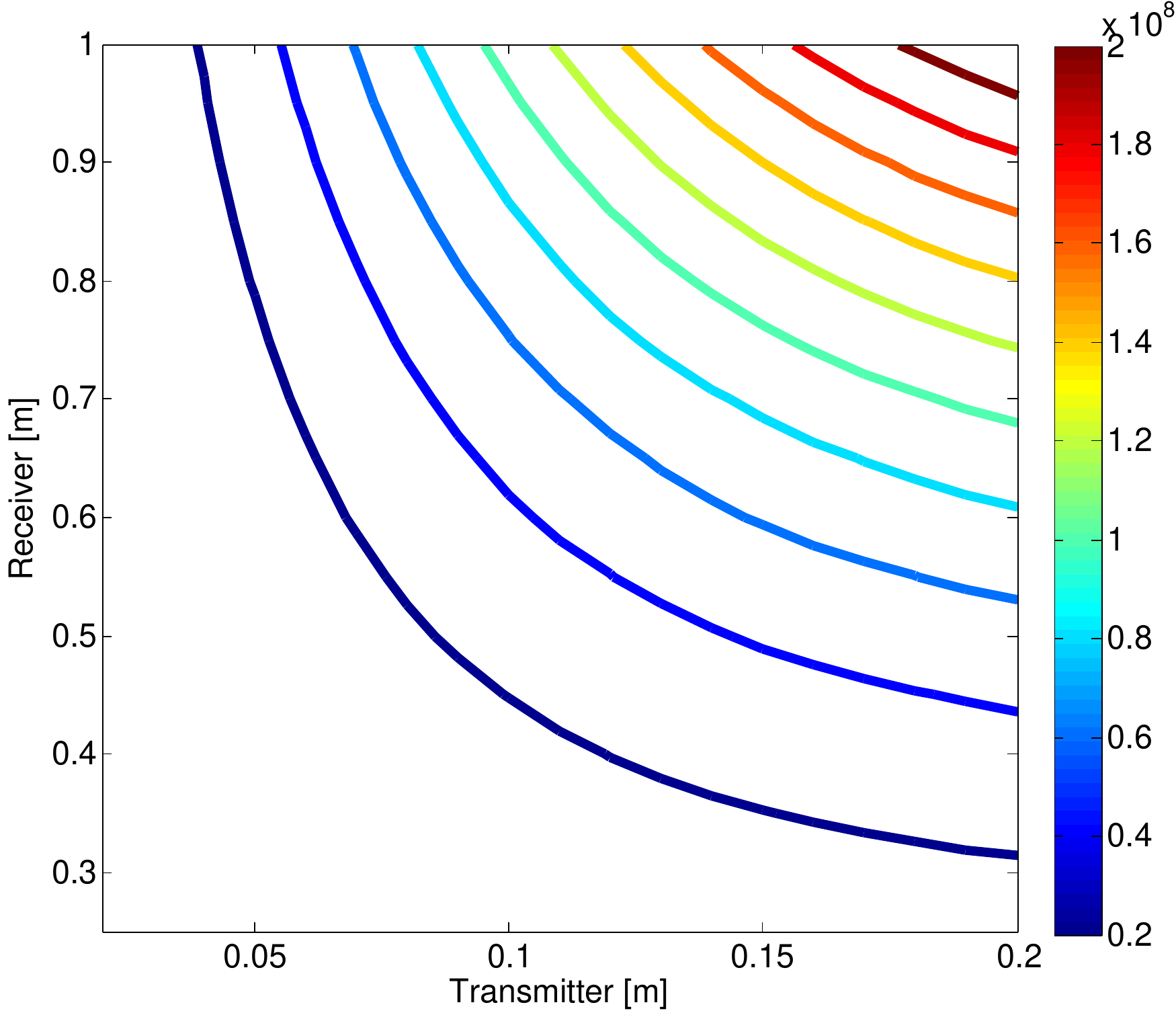} \\
  \includegraphics[width=0.49\linewidth]{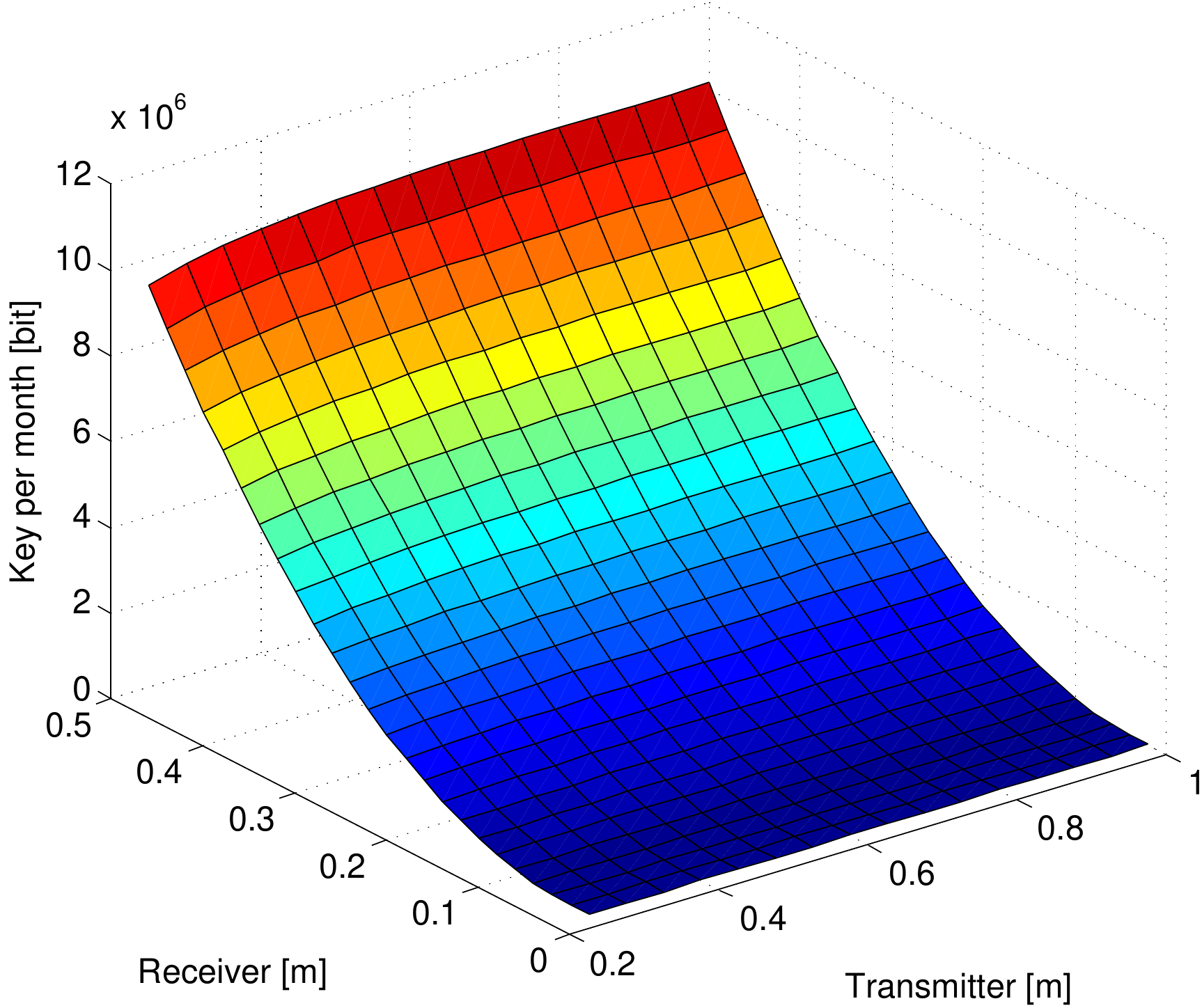}  
 \includegraphics[width=0.49\linewidth]{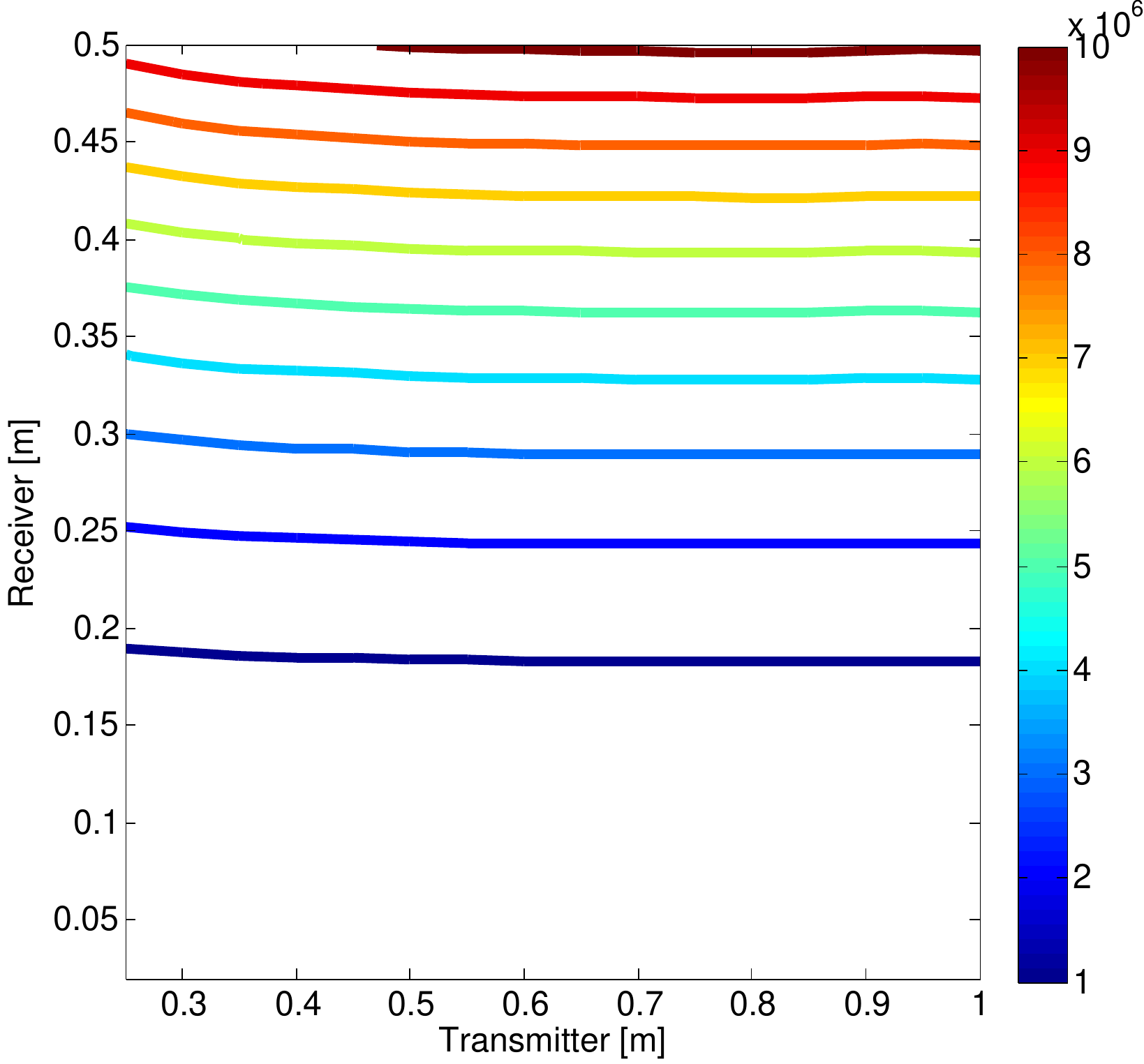}
 \caption[key WCP]{Estimated key per month with a WCP source for various telescope sizes, assuming half the passes are unobstructed by cloud cover. Top: downlink; bottom: uplink. A downlink with a satellite transmitter telescope of 10~cm and a receiver of 50~cm could be used to successfully exchange a key of 25~Mbit per month, while an uplink with a 30~cm receiver telescope on the satellite and a ground transmitter of 25~cm could produce 3~Mbit per month. In an uplink, the size of the ground transmitter has little importance because atmospheric turbulence dominates diffraction. Conditions are as in previous figures, with a downlink wavelength of 670~nm, uplink wavelength of 785~nm, and source rate of 300~MHz.}
  \label{fig.key_WCP}
\end{figure}

\begin{figure}[tbp]
  \centering
  \includegraphics[width=0.49\linewidth]{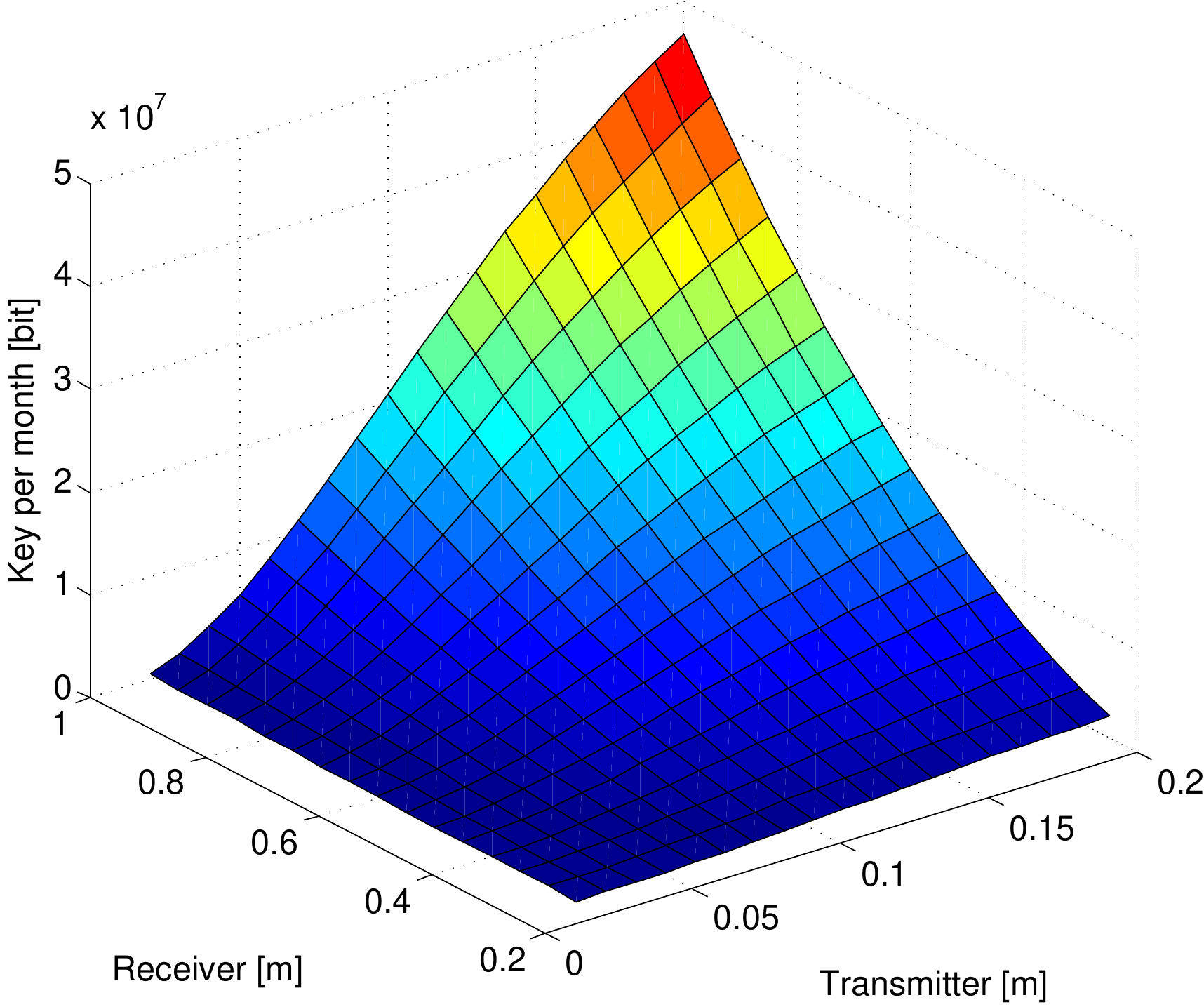} 
 \includegraphics[width=0.49\linewidth]{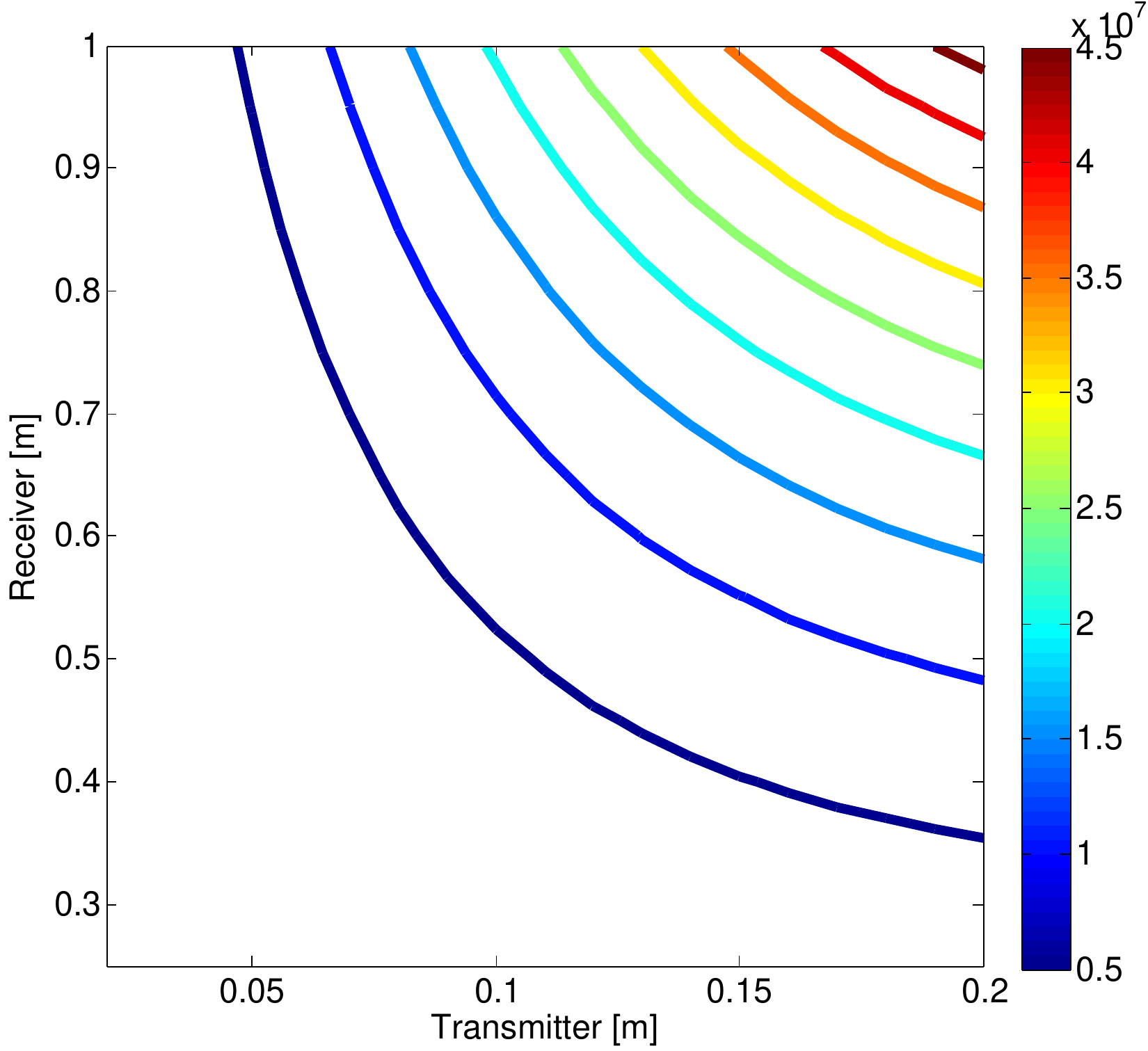} \\
\includegraphics[width=0.49\linewidth]{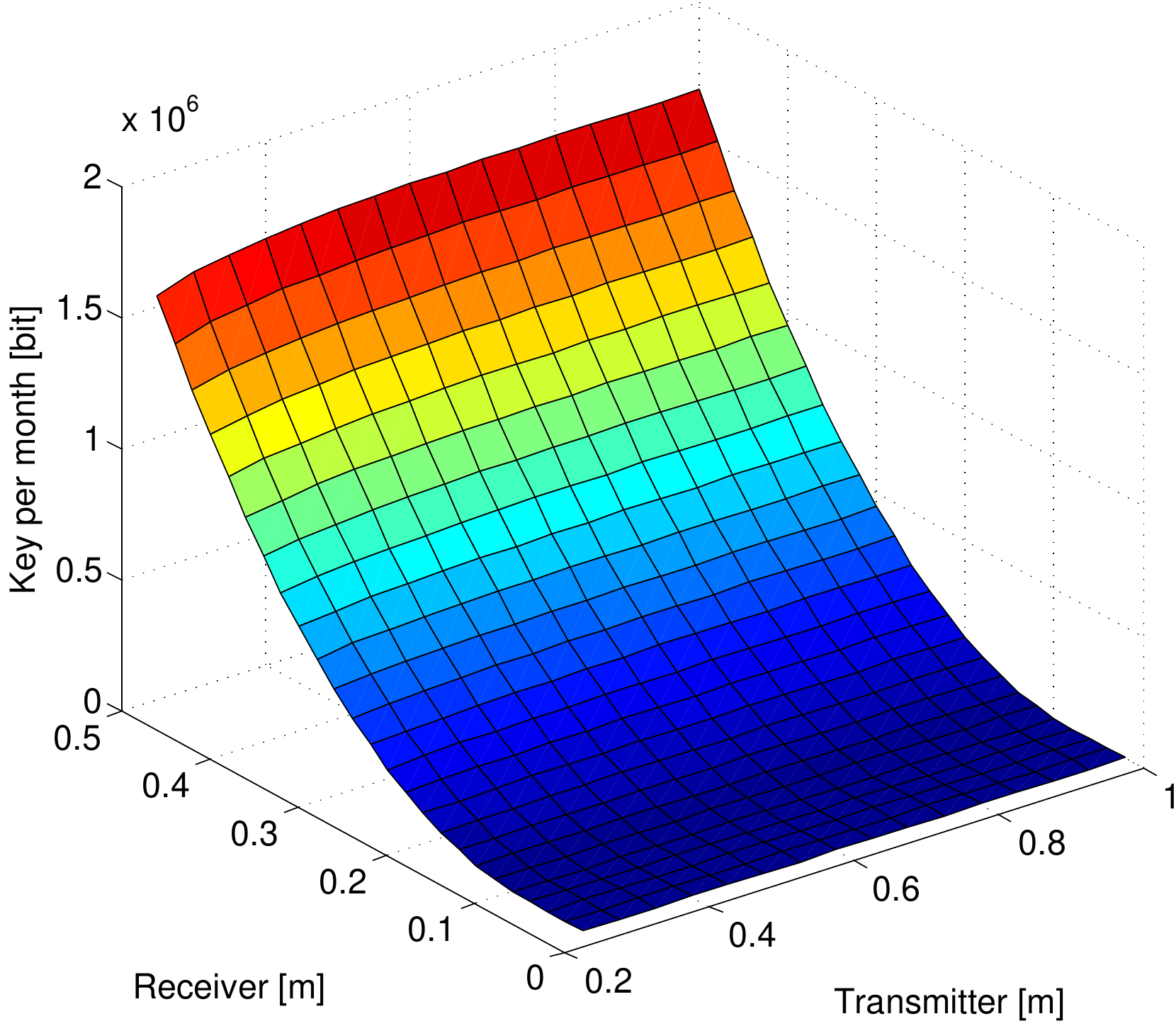} 
 \includegraphics[width=0.49\linewidth]{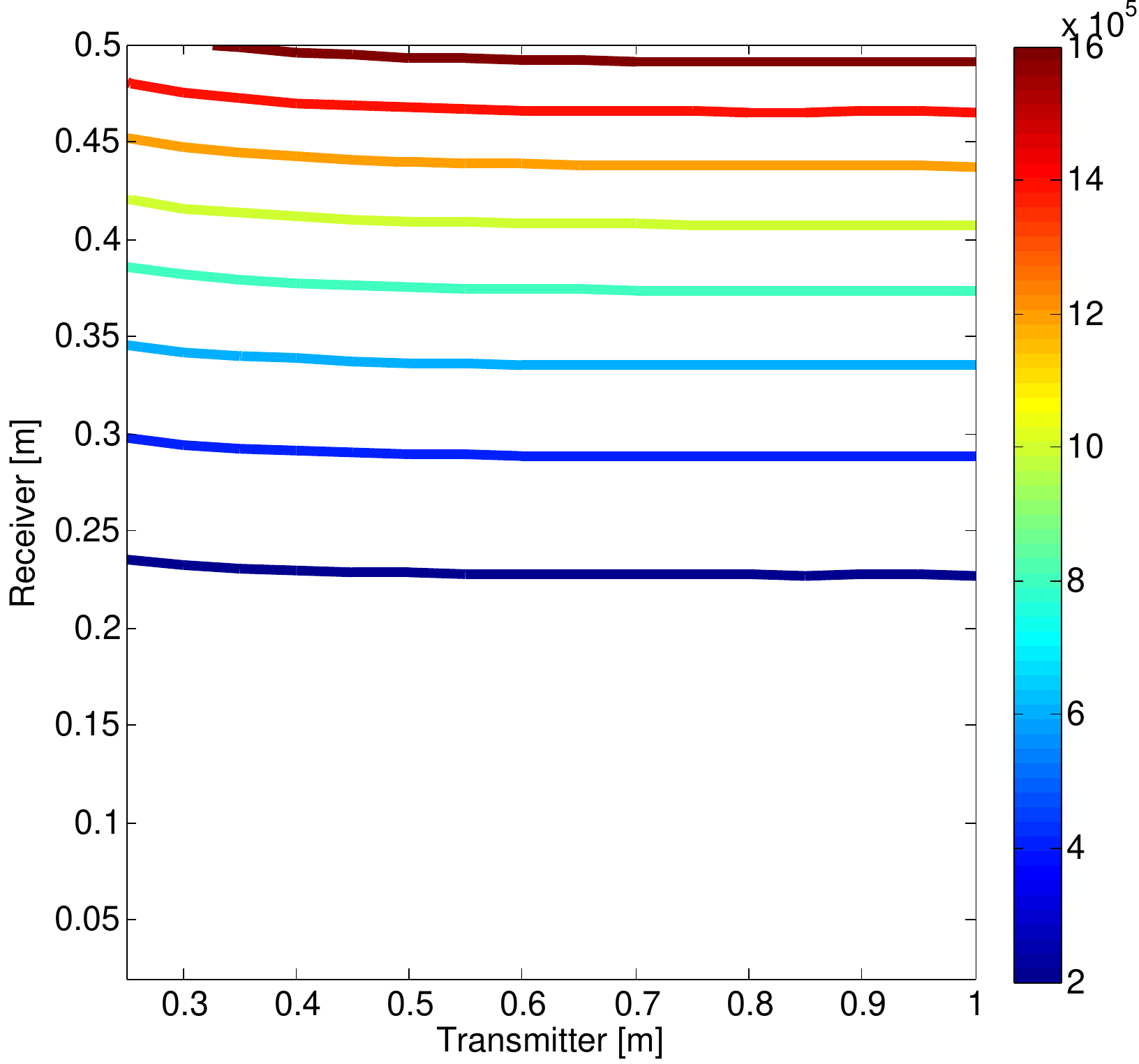}
 \caption[key entangled]{Estimated key per month with an entangled photon source for various telescope sizes, assuming half the passes are unobstructed by cloud cover. Top: downlink; bottom: uplink. A downlink with a satellite transmitter telescope of 10~cm and a receiver of 50~cm could be used to successfully exchange a key of 4.5~Mbit per month while an uplink with a 30~cm receiver telescope on the satellite and a ground transmitter of 25~cm could produce 0.4~Mbit per month. Again, the size of the ground transmitter in the uplink has little importance because atmospheric turbulence dominates diffraction. Conditions are as in previous figures, with a downlink wavelength of 670~nm, uplink wavelength of 785~nm, and source rate of 100~MHz.}
  \label{fig.key_entangled}
\end{figure}

\begin{figure}[tbp]
  \centering
 \includegraphics[width=0.49\linewidth]{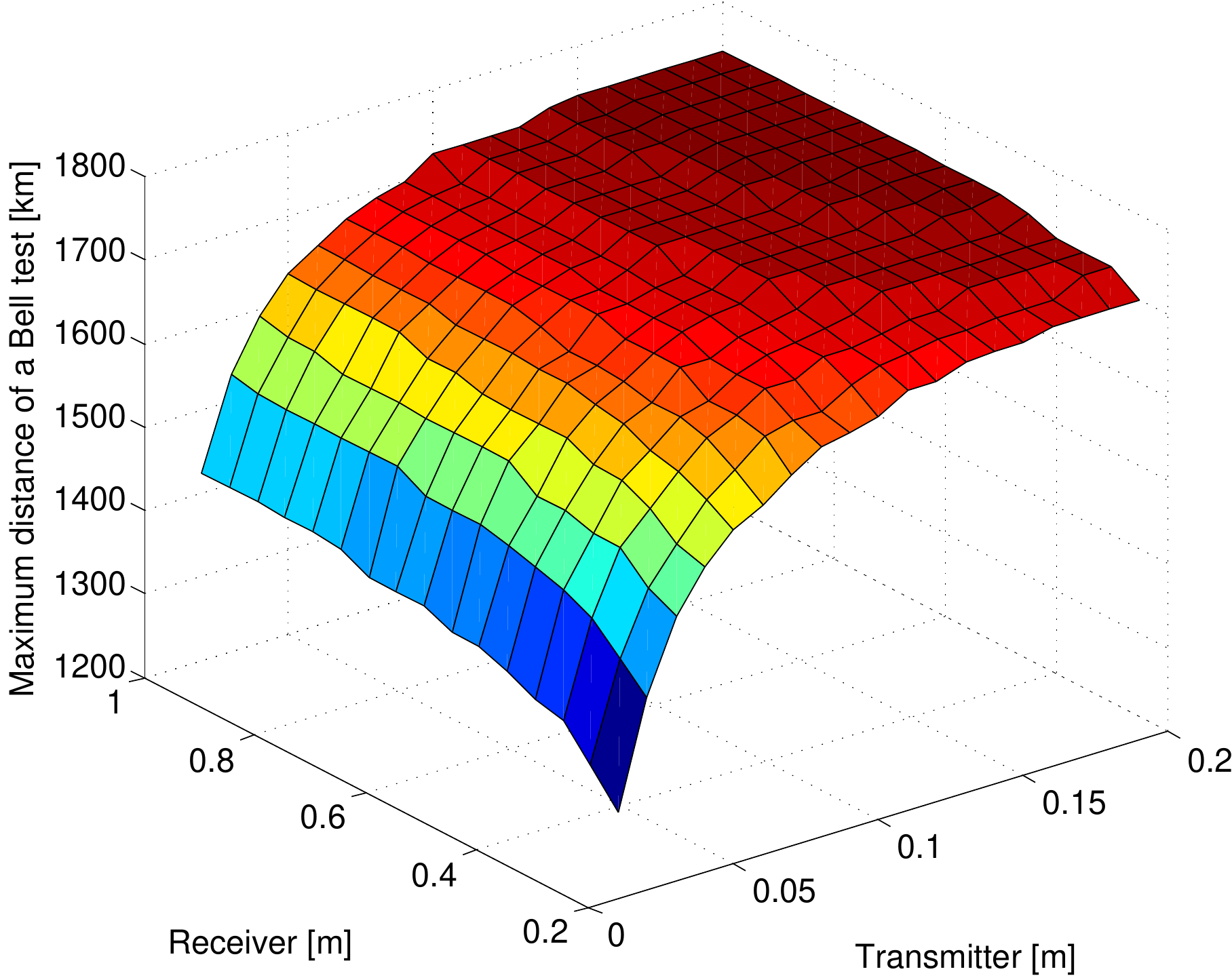}  
 \includegraphics[width=0.49\linewidth]{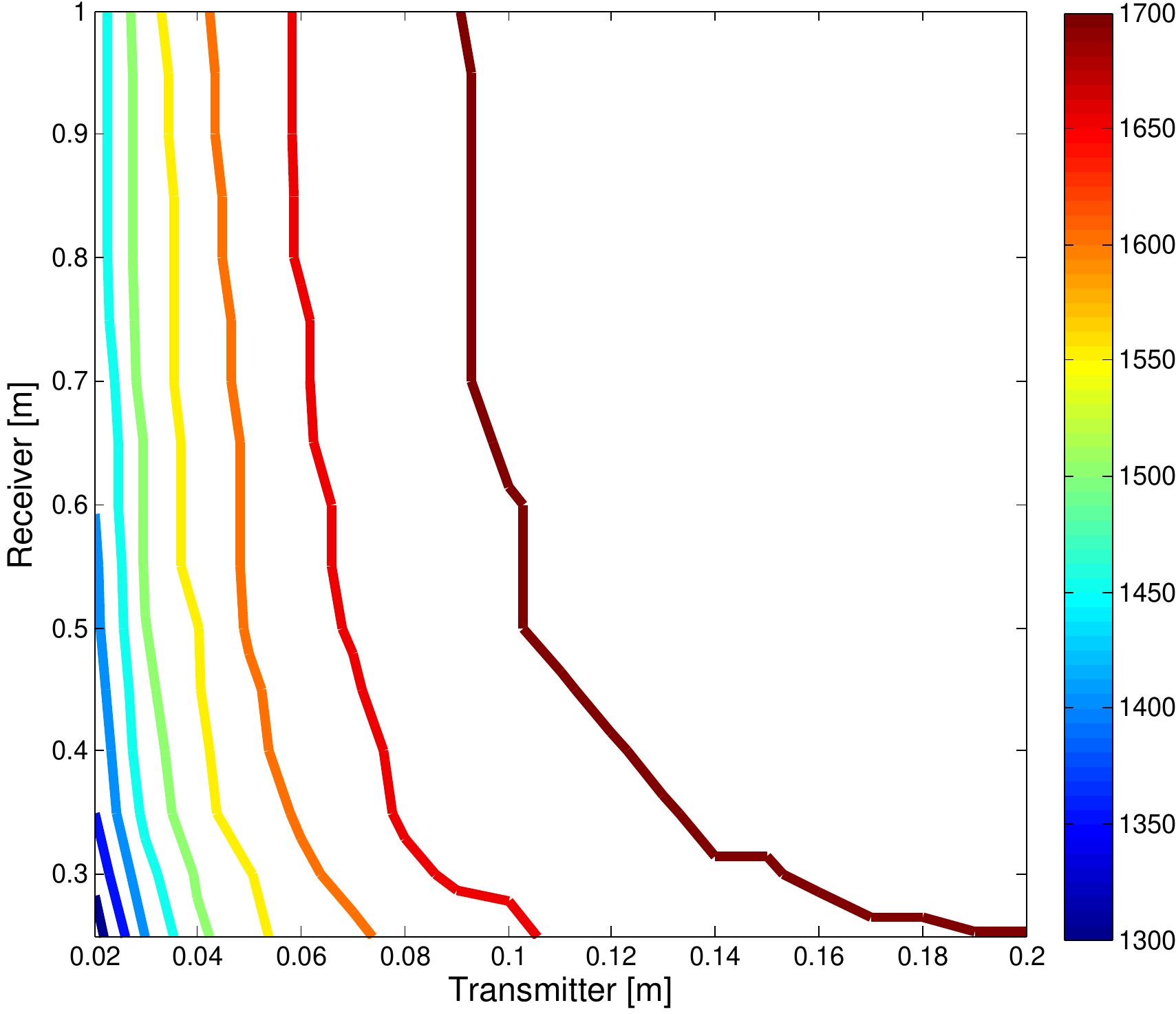} \\
 \includegraphics[width=0.49\linewidth]{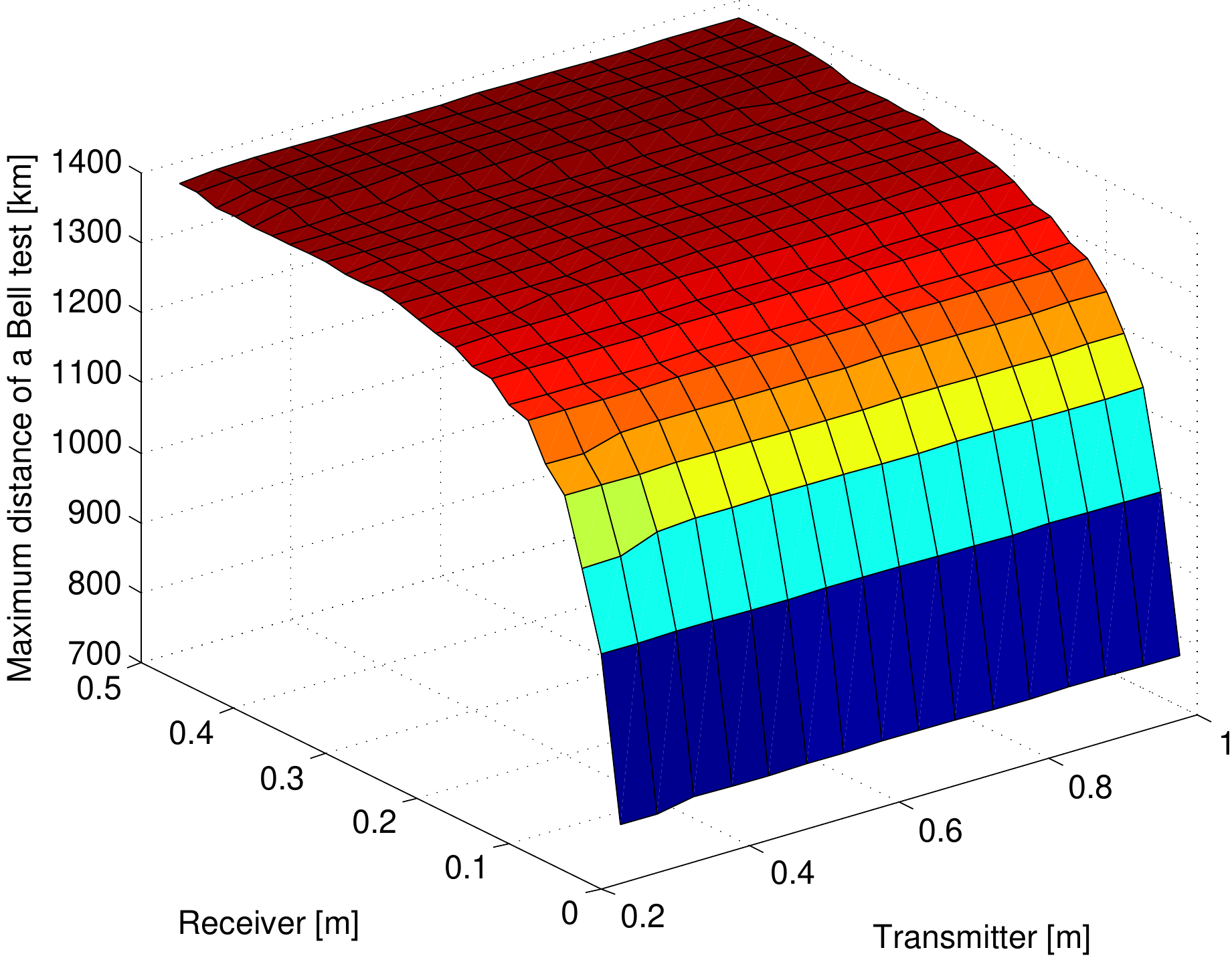}  
 \includegraphics[width=0.49\linewidth]{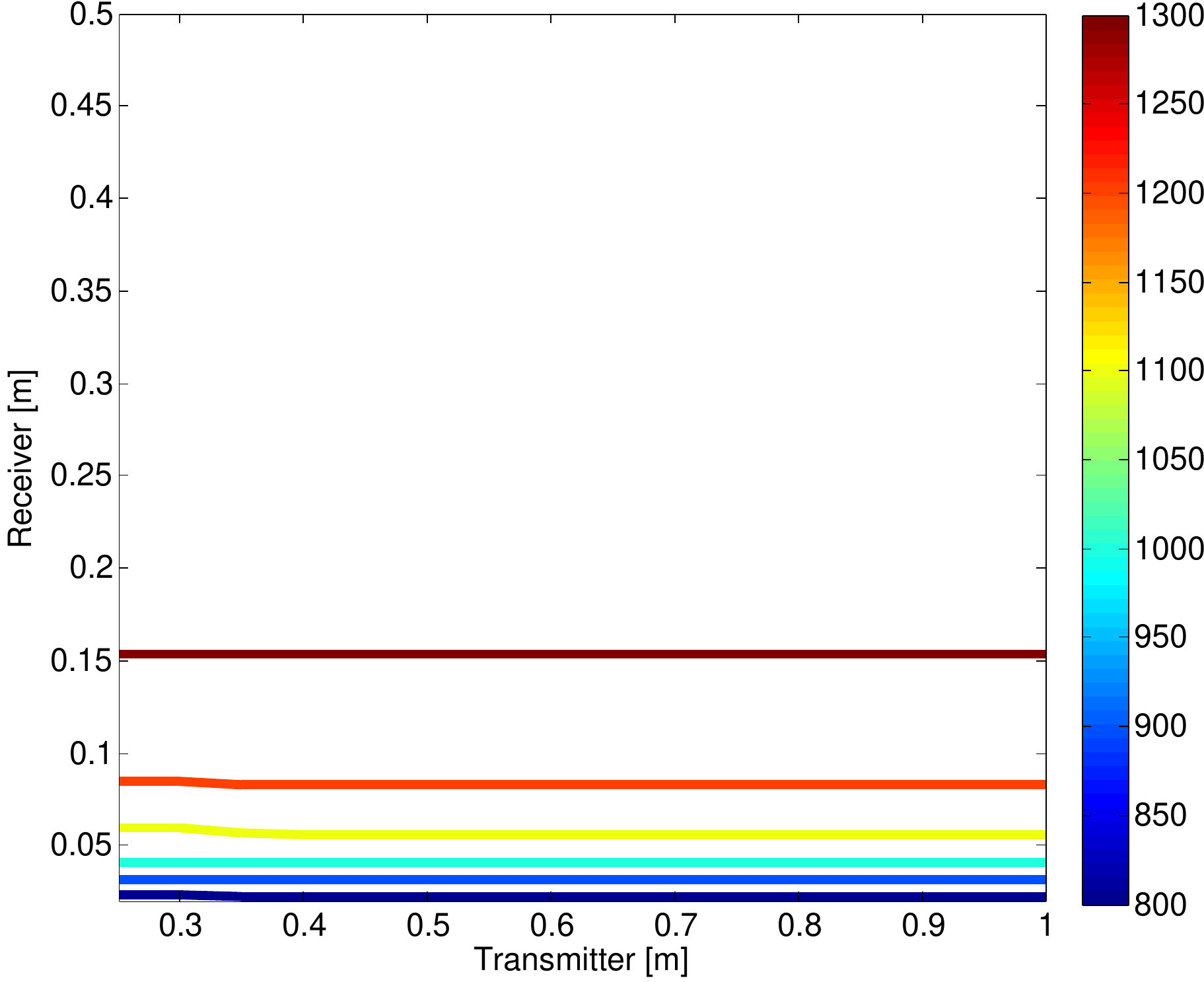}  
\caption{Maximum distance of a complete Bell test in a downlink (top) and an uplink (bottom) for various telescope sizes. A downlink with a satellite transmitter telescope of 10~cm and a receiver of 50~cm could be used to successfully violate the CHSH inequality at 1700~km, while an uplink with a 30~cm receiver telescope on the satellite and a ground transmitter of 25~cm could violate it at 1355~km. Jagged contours are an artefact of the finite sample of passes. Conditions are as in previous figures for an entangled photon source (100~MHz).}
  \label{fig.Bell}
\end{figure}

\begin{figure}[tbp]
  \centering
 \includegraphics[width=0.49\linewidth]{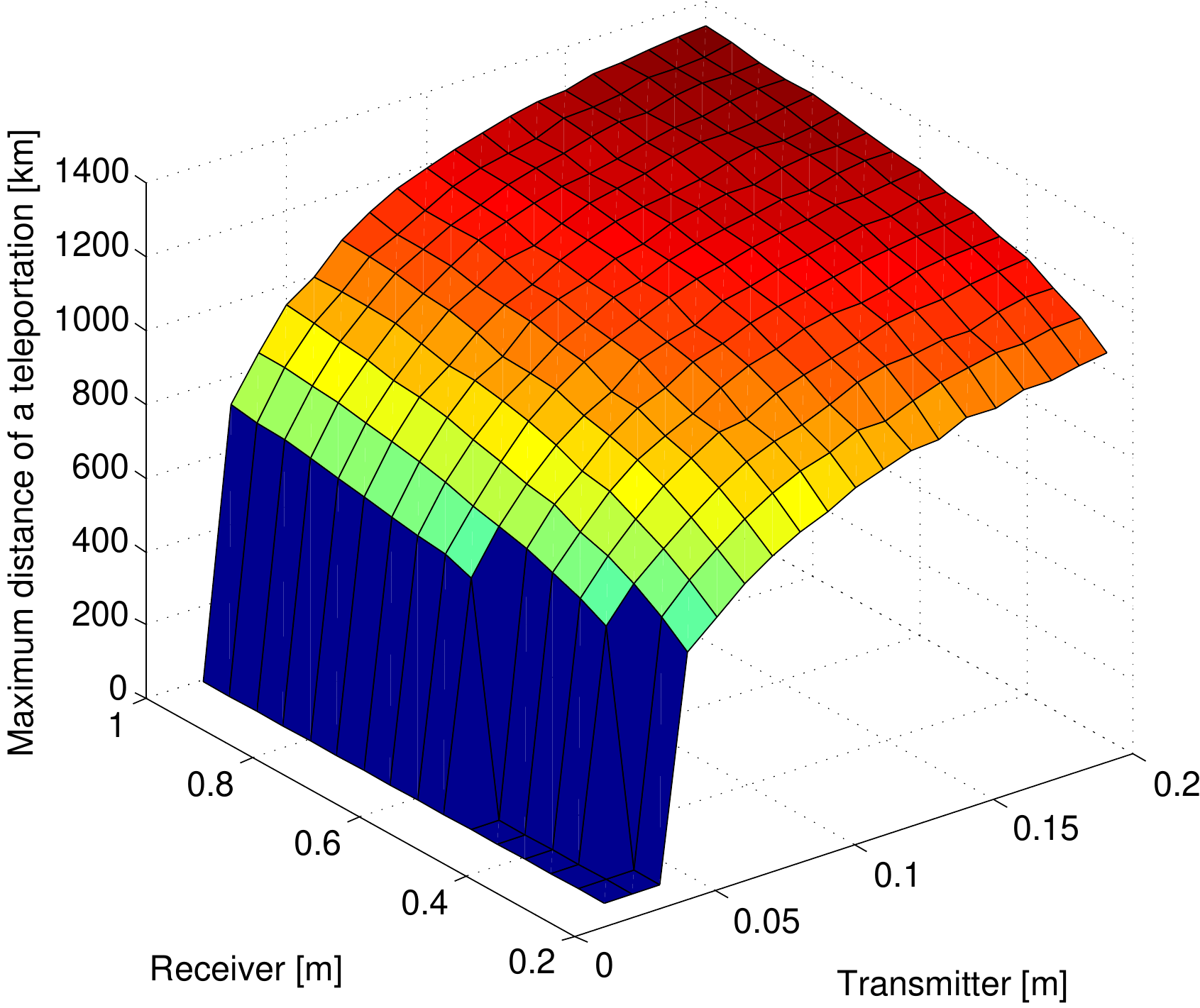}  
 \includegraphics[width=0.49\linewidth]{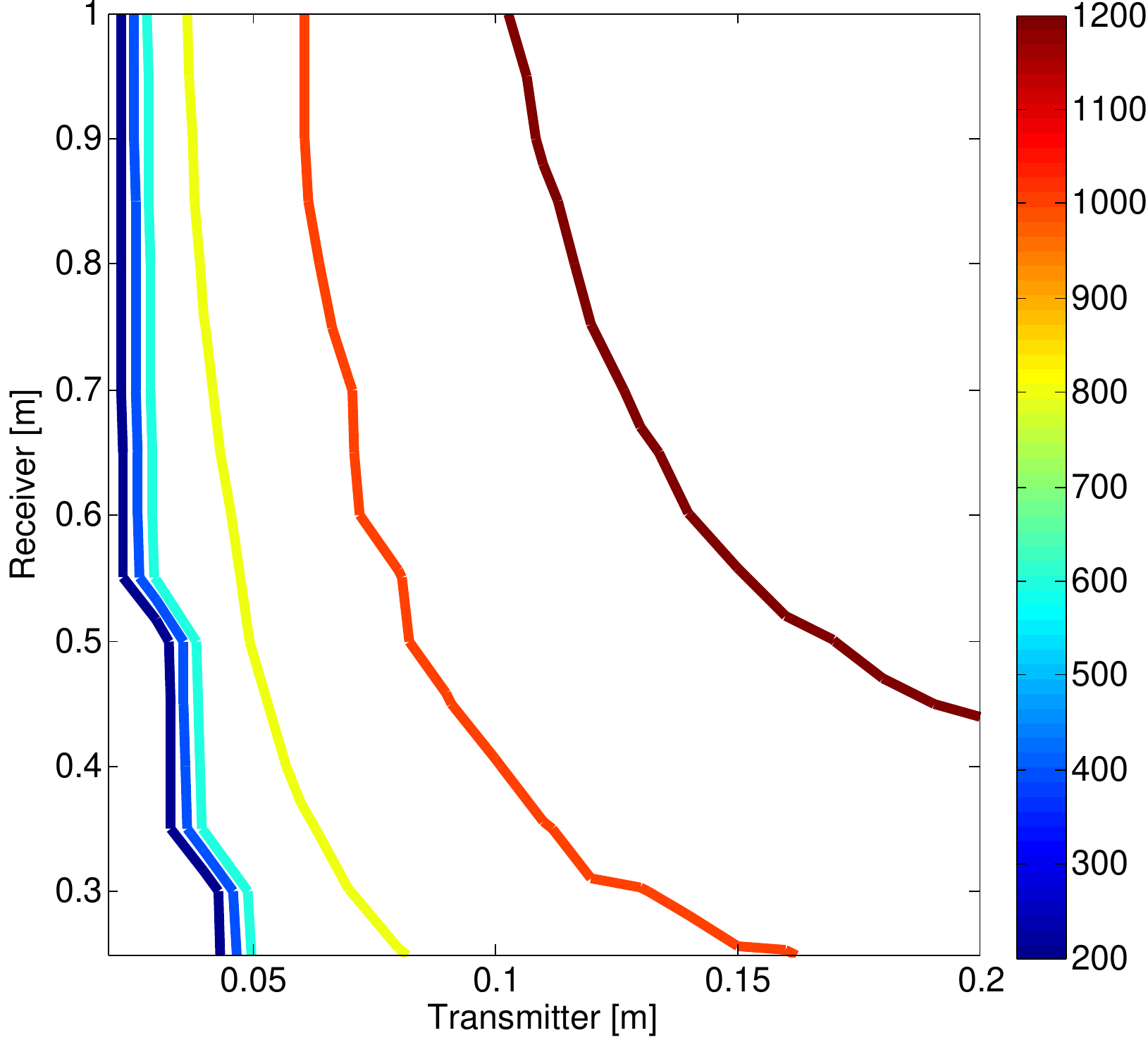} \\
 \includegraphics[width=0.49\linewidth]{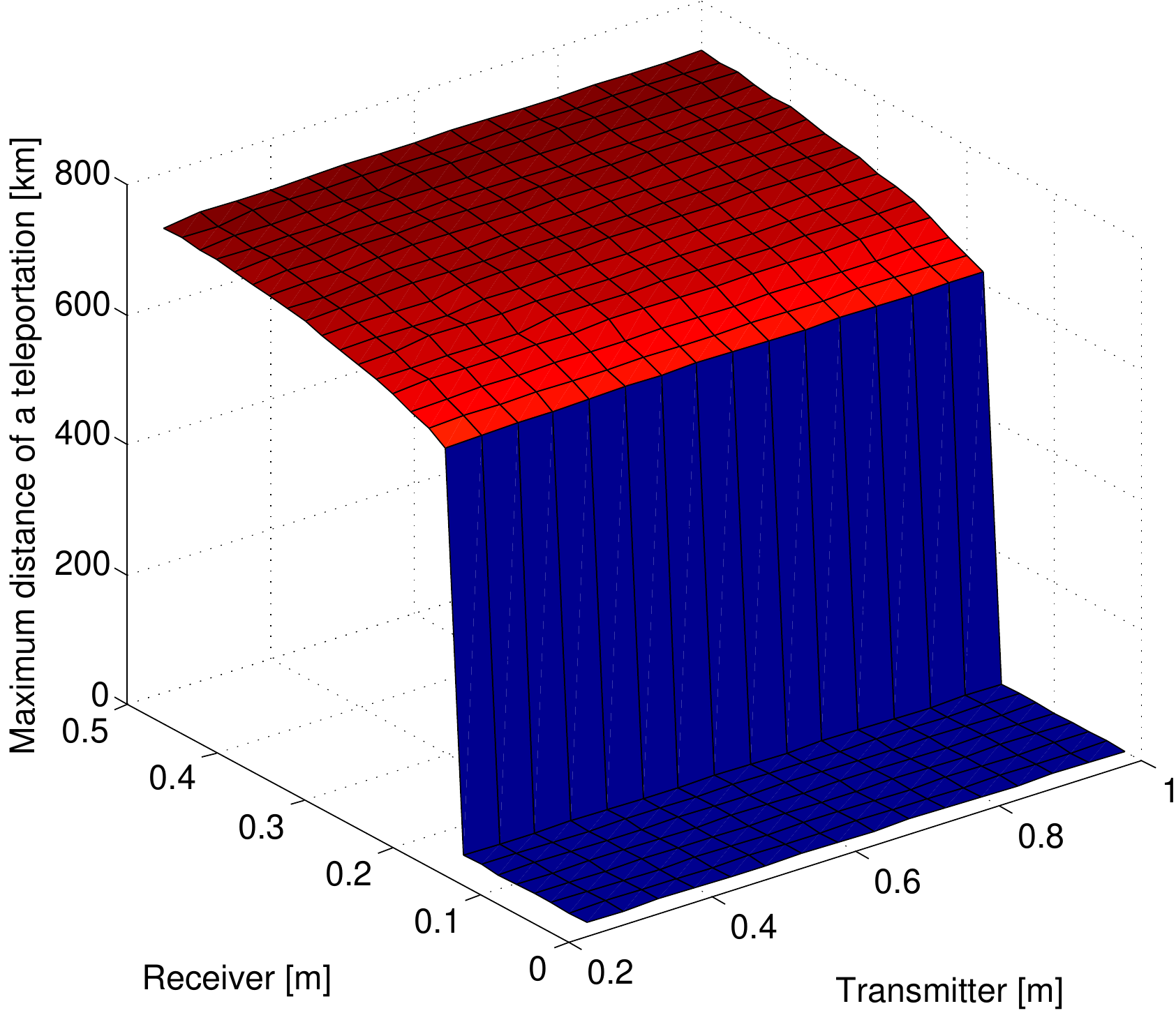}  
 \includegraphics[width=0.49\linewidth]{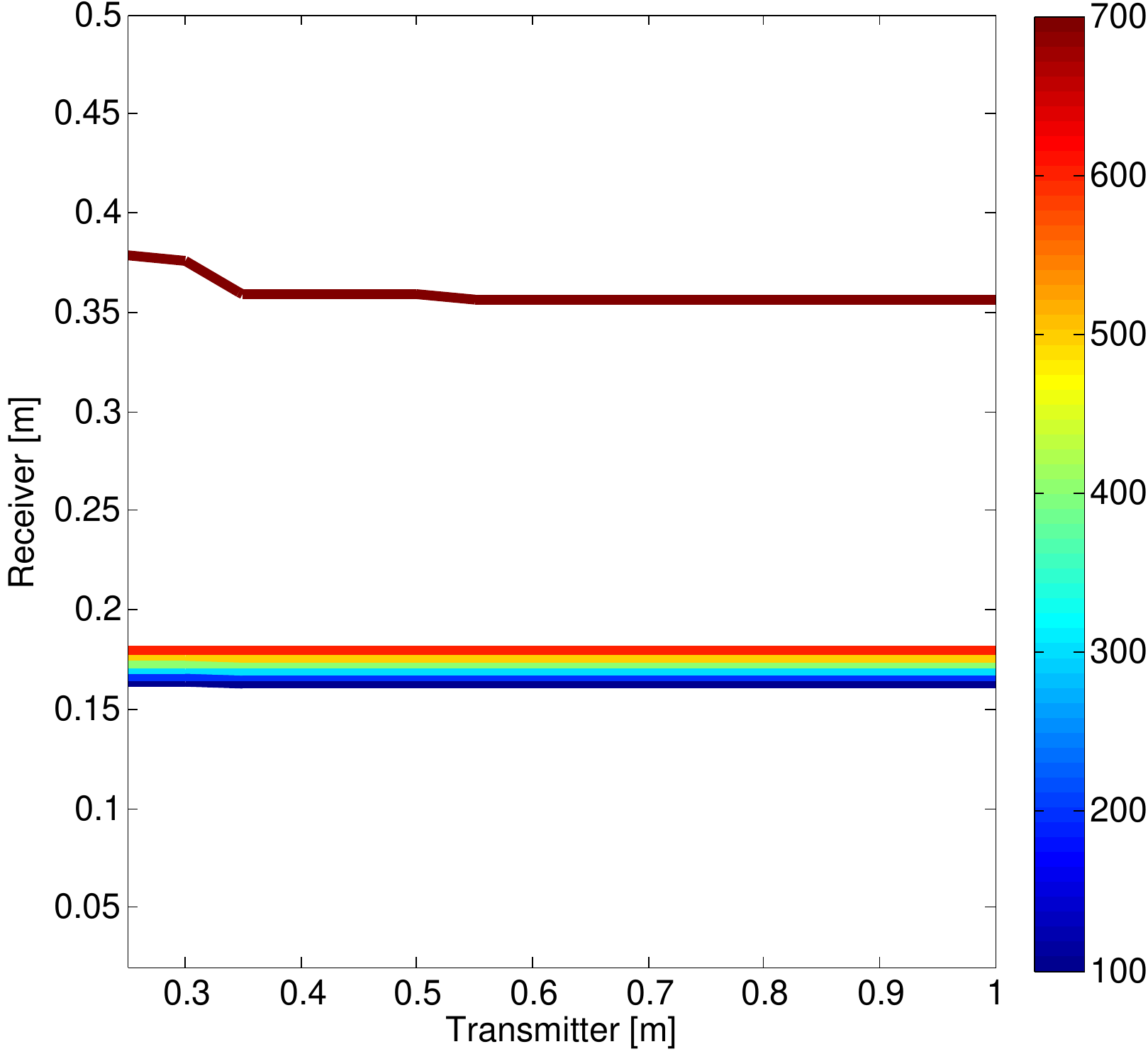}  
\caption{Maximum distance of a complete teleportation experiment in a downlink (top) and an uplink (bottom) for various telescope sizes. A downlink with a satellite transmitter telescope of 10~cm and a receiver of 50~cm could be used to successfully perform teleportation at 1050~km, while an uplink with a 30~cm receiver telescope on the satellite and a ground transmitter of 25~cm could perform it at 675~km. For small telescope sizes, teleportation cannot be performed with sufficient statistical certainty for any satellite pass studied. Again, jagged contours are due to the finite sample of passes. Conditions are as in previous figures for an entangled photon source (100~MHz).}
  \label{fig.teleport}
\end{figure}

With the results for loss (Appendix~\ref{app.LossCalcs}) and background counts (Appendix~\ref{app.BackgroundCalcs}), we calculate the QBER and raw key rate using the simulation of phonic quantum communication (Appendix~\ref{app.Detailed_QOsim}), allowing us to determine the length of secure key that can be obtained for each pass. Examples of the QBER and raw key rate for different passes are shown for a WCP source in Figure~\ref{fig.QBER_key_WCP} and for an entangled photon photon source in Figure~\ref{fig.QBER_key_Entangled}. A WCP source can have a higher repetition rate than an entangled photon source, on the order of GHz~\cite{DYDSS10,WCGYLZGH12} compared to MHz for the entangled photon source~\cite{SUFRMHPRKJZ09}. For our simulations, we use a source rate for the WCP of 3 times the rate of the entangled photon source. With this we can calculate the length of the secure key for each pass accounting for finite size effects. Figures~\ref{fig.key_WCP} and~\ref{fig.key_entangled} show the secure key that can be generated monthly for various transmitter and receiver aperture sizes. This assumes a cloud coverage of 50\%, as discussed in Section~\ref{sec.results}. The long-distance performance of a Bell test and a quantum teleportation are also obtained by analysing the success of each pass individually (requiring a $3\sigma$ violation of the classical limit). We then compare how close each successful pass comes to the ground station to determine the maximum nearest distance which completes a full bell test or quantum teleportation, with results shown in Figures~\ref{fig.Bell} and~\ref{fig.teleport}.

\section{MODTRAN parameters}\label{app.MODTRAN_parameters}

MODTRAN~\cite{Mod5} is a software package that calculates atmospheric transmittance, given a list of input parameters. These inputs are divided into ``cards''. Here we list the values we used for these inputs in our calculations. The descriptions are based on the descriptions in MODTRAN~5.2.1 user's manual.

\begin{table}[hbt]
\caption{Card 1: Main radiation transport driver.}\label{tab:card1}
\centering\begin{tabular}{p{2cm}p{1cm}p{9cm}} 
\hline
Name&Value&Description \\
\hline
MODTRN&M&MODTRAN band model  \\
SPEED&S&Slow speed Correlated-k option using 33 absorption coefficients (k values) per spectral bin (1~cm$^{-1}$ or 15~cm$^{-1}$) \\
LYMOLC&blank&Do not include auxiliary species with model atmosphere  \\
MODEL&2&Mid-Latitude Summer ($45^\circ$ North Latitude)  \\
ITYPE&3&Vertical or slant path to space or ground \\
IEMSCT&0&Program executes in spectral transmittance only mode  \\
IMULT&-1&Program executes with multiple scattering \\
I\_RD2C&0&Normal operation of program  \\
NOPRNT&0&Normal writing to tape6 and tape7 \\
TPTEMP&0&No surface emission if H2 is above ground  \\
SURREF&0.3&Albedo of the earth \\
\hline
\end{tabular}
\end{table}

\begin{table}[hbt]
\caption{Card 1A: Radiative transport driver cont'd.}\label{tab:card1a}
\centering\begin{tabular}{p{2cm}p{1cm}p{9cm}} 
\hline
Name&Value&Description \\
\hline
DIS&f&The less accurate but faster Isaac's two-stream algorithm is used  \\
DISAZM&f&Not using azimuth dependence with DISORT \\
DISALB&f&Not calculating the spectral spherical albedo of the atmosphere and diffuse transmittance for the line-of-sight and sun-to-ground paths \\
NSTR&8&Number of streams to be used by DISORT \\
SFWHM&0&Use default TOA solar data \\
CO2MX&365&CO$_2$ mixing ratio in ppmv \\
H2OSTR&0&Default vertical water vapor column character string \\
O3STR&0&Default vertical ozone column character string \\
C\_PROF&0&Do not scale default profiles \\
LSUNFL&f&The solar irradiance data to be used depends on the spectral resolution of the MODTRAN band model \\
LBMNAM&f&The default (1~cm$^{-1}$ bin) band model database files are to be used \\
LFLTNM&f&Do not read file name for user-defined instrument filter function from CARD 1A3 \\
H2OAER&f&Aerosol optical properties are not modified to reflect the changes from the original relative humidity profile arising from the scaling of the water column \\
SOLCON&0&Do not scale the TOA solar irradiance \\
CDASTM&blank&Use Angstrom Law description of boundary layer and tropospheric aerosol extinction data \\
NSSALB&0&Use reference aerosol spectral single scattering albedo values \\
\hline
\end{tabular}
\end{table}

\begin{table}[hbt]
\caption{Card 2: Main aerosol and cloud options.}\label{tab:card2}
\centering\begin{tabular}{p{2cm}p{1cm}p{9cm}} 
\hline
Name&Value&Description \\
\hline
APLUS&Blank&Don't use ``Aerosol Plus'' option \\
IHAZE&2&RURAL extinction, default VIS=5~km \\
CNOVAM&Blank&Don't use Navy Oceanic Vertical Aerosol Model (NOVAM) \\
ISEASN&0&Season determined by the value of MODEL \\
ARUSS&blank&Don't use user-defined aerosol optical properties \\
IVULCN&0&Background stratospheric profile and extinction \\
ICSTL&5&Air mass character (1--10, 1=open ocean, 10=stong Continental influence) \\
ICLD&0&No clouds or rain \\
IVSA&0&Army Vertical Structure Algorithm (VSA) not used \\
VIS&0&Uses the default meteorological range set by IHAZE \\
WSS&0&Default wind speeds are set according to the value of MODEL \\
RAINRT&0&Rain rate (mm/hr) \\
GNDALT&0&Altitude of surface relative to sea level (km) \\
\hline
\end{tabular}
\end{table}

\begin{table}[hbt]
\caption{Card 3: Line-of-sight geometry.}\label{tab:card3}
\centering\begin{tabular}{p{2cm}p{1cm}p{9cm}} 
\hline
Name&Value&Description \\
\hline
H1&0&Initial altitude (km) \\
H2&0&Final altitude, not used for ITYPE=3 \\
RANGE&0&Not used in this case for ITYPE=3 \\
BETA&0&Not used in this case for ITYPE=3 \\
RO&0&Default mid-latitude radius of the Earth (km) of 6371.23~km \\
LENN&0&Default \\
PHI&0&Zenith angle at H2 towards H1 \\
\hline
\end{tabular}
\end{table}

\begin{table}[hbt]
\caption{Card 4: Spectral range and resolution.}\label{tab:card4}
\centering\begin{tabular}{p{2cm}p{1cm}p{9cm}} 
\hline
Name&Value&Description \\
\hline
DV&0.1&Wavelength increment used for spectral outputs (in nm) \\
FWHM&2&Slit function Full Width at Half Maximum (in nm) \\
\hline
\end{tabular}
\end{table}

\clearpage
\bibliographystyle{iopart-num} 
\bibliography{Link_Bibliography}

\end{document}